\documentclass[aps,prb,times,twocolumn,amsmath,amssymb,superscriptaddress,floatfix,showpacs]{revtex4}

\usepackage{color} 
\usepackage{graphicx}
\usepackage{bbold} 
\usepackage{dsfont}
\usepackage{multirow}
\usepackage[section]{placeins}
\usepackage{color}
\usepackage{ulem}

\usepackage[caption=false]{subfig}
\usepackage{subfloat}

\usepackage{amssymb} 
\usepackage{amsmath} 
\usepackage{wasysym}
\usepackage[utf8]{inputenc} 

\begin{document}


\title{Of chain--based order and quantum spin liquids in dipolar spin ice}


\author{P. A. McClarty} 
\affiliation{ISIS Neutron and Muon Source, Rutherford-Appleton 
Laboratory, Harwell Campus, Oxfordshire}
\affiliation{Max-Planck-Institut f\"{u}r Physik komplexer Systeme, 
01187 Dresden, Germany}


\author{O. Sikora} 
\affiliation{Department of Physics, National Taiwan University, 
Taipei 10607, Taiwan}
\affiliation{Okinawa Institute for Science and Technology, 1919-1 
Tancha, Onna-son, Kunigami, Okinawa 904-0495, Japan}
\affiliation{H. H. Wills Physics Laboratory, University of Bristol, Bristol BS8 1TL, 
United Kingdom}


\author{R. Moessner} 
\affiliation{Max-Planck-Institut f\"{u}r Physik komplexer Systeme, 
01187 Dresden, Germany}


\author{K. Penc} 
\affiliation{Institute for Solid State Physics and Optics, Wigner Research 
Centre for Physics, Hungarian Academy of Sciences, H-1525 Budapest, 
P.O.B. 49, Hungary}


\author{F. Pollmann} 
\affiliation{Max-Planck-Institut f\"{u}r Physik komplexer Systeme, 
01187 Dresden, Germany}


\author{N.  Shannon} 
\affiliation{Okinawa Institute for Science and Technology, 1919-1 Tancha, 
Onna-son, Kunigami, Okinawa 904-0495, Japan}
\affiliation{H. H. Wills Physics Laboratory, University of Bristol, Bristol BS8 1TL, 
United Kingdom}


\date{\today} 


\begin{abstract}
Recent experiments on the spin--ice material Dy$_2$Ti$_2$O$_7$ suggest that 
the Pauling ``ice entropy'', characteristic of its classical Coulombic spin-liquid state, 
may be lost at low temperatures [D. Pomaranski et al., Nature Phys. 9, 353 (2013)].
However, despite nearly two decades of intensive study, the nature of the
equilibrium ground state of spin ice remains uncertain.
Here we explore how long-range dipolar interactions $D$, short-range exchange 
interactions, and quantum fluctuations combine to determine the ground 
state of dipolar spin ice.
We identify a new organisational principle, namely that ordered ground states are 
selected from a set of ``chain states'' in which dipolar interactions are 
exponentially screened.
Using both quantum and classical Monte Carlo simulation, we establish 
phase diagrams as a function of quantum tunneling $g$, and temperature $T$, 
and find that only a very small $g_c \ll D$ is needed to stabilize a 
quantum spin-liquid ground state.
We discuss the implications of these results for Dy$_2$Ti$_2$O$_7$.
\end{abstract}




\maketitle

\section{Introduction}
\label{section:introduction}


The search for materials which realize a spin-liquid state,
in which magnetic moments interact strongly, and yet fail to order, 
has become something of a {\it cause c\'el\`ebre}.\cite{fazekas74,lee08,balents10}
A rare three-dimensional example of a spin liquid is provided by 
the ``spin--ice'' materials, a family of rare--earth pycrochlore oxides 
exemplified by Ho$_2$Ti$_2$O$_7$ and Dy$_2$Ti$_2$O$_7$, 
which exhibit a ``Coulombic'' phase --- a classical spin liquid, 
exhibiting an emergent $U(1)$ gauge field,
whose excitations famously take the form of magnetic 
monopoles.\cite{bramwell01-Science294,castelnovo12}
The fate of this spin liquid at low temperatures is an important 
question, touching on the limits of our understanding of phase 
transitions,\cite{powell11} and the tantalising possibility of 
finding a quantum spin-liquid in three dimensions.
Nonetheless, after nearly two decades of intensive study, the nature of the 
quantum ground state of spin-ice materials remains a mystery.


\begin{figure}[h!]
\includegraphics[width=0.9\columnwidth]{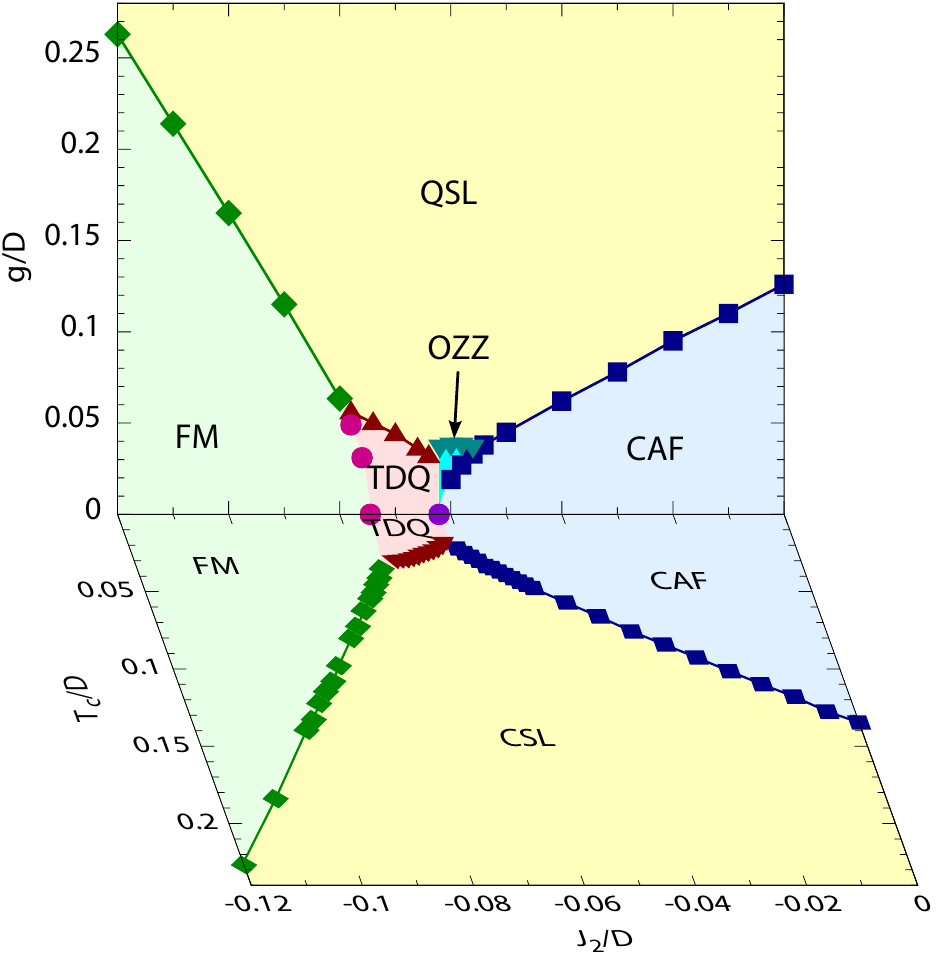}
\caption{
(Color online)  
Quantum and classical phase diagrams for a spin ice with long-range dipolar 
interactions $D$, as a function of second-neighbour exchange $J_2$.  
Quantum tunneling $g$, and temperature $T$, drive quantum (QSL)
and  classical (CSL) spin-liquid phases.
These compete with four distinct ordered phases based on ferromagnetically--polarised 
chains of spins, illustrated in Fig.~\ref{fig:ordered.phases}. 
Results are taken from quantum and classical Monte Carlo simulations
of $\mathcal{H_{\sf QDSI}}$~[Eq.~(\ref{eq:Hqdsi})], for a cubic cluster of 
128 spins, with exchange $J_k = 0$ for $k \ne 2$.}
\label{fig:combined.phase.diagram}
\end{figure} 


\begin{figure*}
%
\includegraphics[width=0.7\textwidth]{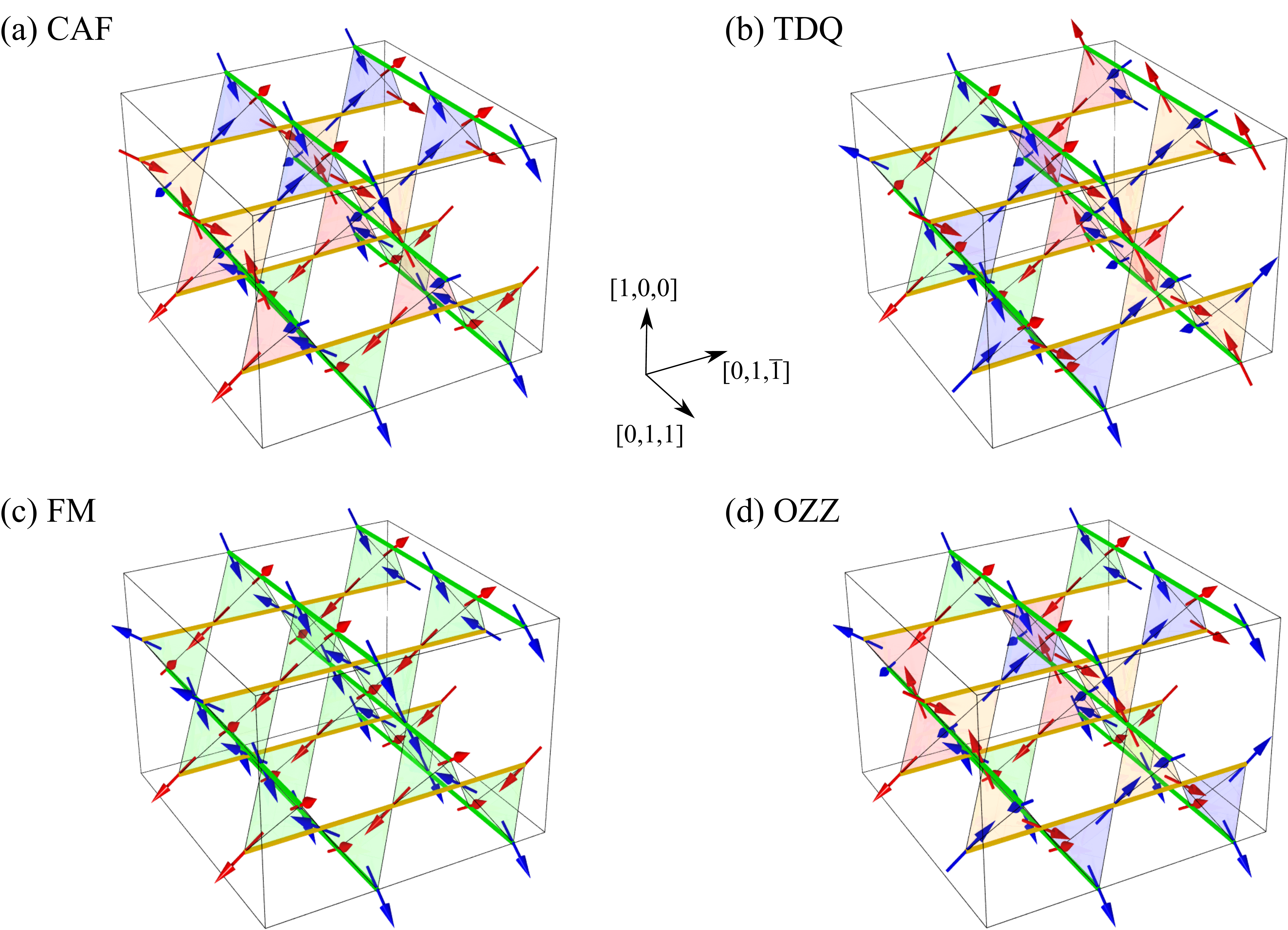}
\caption{ 
(Color online)  
Chain--based ordered ground states found in a dipolar spin--ice with 
competing exchange interactions.  
%
(a) cubic antiferromagnet (CAF); 
(b) tetragonal  double-Q state (TDQ);  
(c) ferromagnet (FM);
(d) orthorhombic ``zig-zag'' (OZZ) state stabilised by quantum fluctuations.
All states obey the ice rules, and are composed of chains of spins with
net ferromagnetic polarisation (green and yellow lines).
Tetrahedra of the same color have the same spin--configuration.   
Animated images of these ordered states can be found 
in the supplemental materials.\cite{supplemental}
}
\label{fig:ordered.phases}
\end{figure*} 


This question gains fresh urgency from recent experiments on the 
spin ice Dy$_2$Ti$_2$O$_7$,~\cite{pomaranski13} which suggest 
that the Pauling ice entropy, associated with an extensive number of 
states obeying the ``two-in, two-out'' ice rules,\cite{ramirez99,klemke11} 
is lost at the lowest temperatures.
Such a loss of entropy could herald the onset of a long-range ordered 
state~\cite{siddharthan99,siddharthan-arXiv,denHertog00,bramwell01-PRL87,melko01}, 
in which magnetic monopoles would be confined.  
Alternatively, it could signal the emergence of a three-dimensional 
{\it quantum} spin-liquid, in which monopoles would remain deconfined.
The theoretical possibility of such a spin-liquid has been widely 
discussed,~\cite{hermele04,banerjee08,savary12,shannon12,benton12,lee12,savary13,gingras14,hao-arXiv,kato-arXiv}
and is now well-established through quantum Monte Carlo simulations 
of models with anisotropic nearest-neighbour exchange.~\cite{banerjee08,shannon12,benton12,kato-arXiv}
These results have generated considerable excitement in the context of recent 
experiments on ``quantum spin ice'' 
systems such as Yb$_2$Ti$_2$O$_7$,~\cite{thompson11,ross11-PRX1,chang12},
Tb$_2$Ti$_2$O$_7$,~\cite{molovian07,fennell12,fennell14} and 
Pr$_2$Zr$_2$O$_7$.~\cite{kimura13}
However they leave unanswered the question of what happens in 
a realistic model of a  spin ice such as Dy$_2$Ti$_2$O$_7$.   
Moreover, the long equilibration time--scales encountered in both 
simulation~\cite{melko01} and experiment~\cite{pomaranski13} suggests
that it is difficult to access one low--energy spin configuration
from another.
It is therefore important to understand the nature of the different
low--energy spin configurations in a realistic model --- could a new organisational principle be in play ?


In this Article we address the question :
``What determines the equilibrium ground state of  
spin ice, once quantum effects are taken into account ?''
We start from a realistic model, directly motivated by experiment,
which treats both short-range exchange and long-range dipolar 
interactions, as well as quantum tunneling between different spin--ice configurations.
Our main theoretical results are summarized in the combined quantum and 
classical phase diagram 
Fig.~\ref{fig:combined.phase.diagram}, with illustrations of possible ordered 
ground states given in Fig.~\ref{fig:ordered.phases}.


We first consider the classical ground state of dipolar spin ice, 
in the absence of quantum fluctuations.
We find that long--range dipolar interactions are minimised by spin--configurations 
composed of chains of spins with net ferromagnetic polarisation.
We show that, within these ``chain states'', dipolar interactions are exponentially 
screened  and that all potential classical ground states can be described 
by a mapping onto an effective Ising model on a two-dimensional, anisotropic 
triangular lattice.
Within this mapping, the role of exchange interactions is to select between 
three different competing ordered ground states, a cubic antiferromagnet (CAF), 
a ferromagnet (FM) and tetragonal double--Q  (TDQ) state.
Classical Monte Carlo simulation is used to confirm this picture, 
and to assess the temperature at which the classical ground state 
``melts'' into a classical spin liquid (CSL), of the type observed in spin ice.


We then turn to the problem of determining the ground state of dipolar
spin ice in the presence of quantum fluctuations.
Using zero--temperature quantum Monte Carlo simulation, we establish
that even a small amount of quantum tunneling between different spin--ice 
configurations can ``melt'' chain states into a three-dimensional quantum 
spin-liquid (QSL) ground state.
For small tunneling, $g$, quantum fluctuations also stabilise a new,
ordered ``orthogonal zig--zag'' (OZZ) ground state, at the boundary between
CAF and TDQ states.


We conclude the Article with a discussion of the application of these
results to real materials, paying particular attention to Dy$_2$Ti$_2$O$_7$.
Based on published parameters [\onlinecite{yavorskii08}], 
we find that the ground state of Dy$_2$Ti$_2$O$_7$
should either be a quantum spin liquid, or an ordered CAF state, 
depending on the strength of quantum tunneling.  
We also provide estimates of the quantum tunneling needed to stabillize 
a quantum spin liquid in Dy$_2$Ti$_2$O$_7$, and a range of other materials.


The remainder of the Article is structured as follows~:


In Section~\ref{section:model} we define the models studied in this 
Article, first reviewing with the standard, classical, model for dipolar spin ice 
(DSI) with competing exchange interactions 
[Section~\ref{subsection:classical.model}], 
and then introduce a minimal model for quantum tunneling between 
different spin--ice configurations [Section~\ref{subsection:quantum.model}]. 


In Section~\ref{sec:classical.mean.field.theory} we use a 
mean--field theory 
to establish the 
ground state phase diagram for classical dipolar spin ice 
in the presence of competing second--neighbour exchange interaction $J_2$.


In Section~\ref{section:effective.Ising.model} we show how the ground state 
phase diagram for very general competing exchange interactions can be found 
from a mapping on to an effective, two--dimensional Ising model, 
describing exponentially--screened interactions between 
ferromagnetically---polarised 
chains of spins.


In Section~\ref{section:classical.monte.carlo} we use classical Monte Carlo 
simulation to establish a the finite--temperature phase diagram for dipolar spin 
ice in the presence of competing exchange interactions.


In Section~\ref{section:QMC} we use Green's function Monte Carlo
simulation (GFMC) to study the zero--temperature {\it quantum}
phase diagram of dipolar spin ice, taking into account quantum tunneling
between different spin--ice configurations, in the presence of competing
exchange interactions.


In Section~\ref{section:application.to.real.materials} we discuss the application
of these results to spin ice and quantum spin--ice materials, including 
Dy$_2$Ti$_2$O$_7$.


Finally, in Section~\ref{section:conclusions} we conclude with a 
summary of the results and discussion of some of the remaining
open issues.


The Article concludes with a number of technical appendices.


In Appendix~\ref{appendix:ewald.sum} the Ewald sum used to treat
long--range dipolar interactions is defined.


In Appendix~\ref{appendix:equivalence.J2.and.J3c} it is shown that
second--neighbour exchange $J_2$, and third--neighbour exchange 
along $[110]$ chains, $J_{3c}$, have the same effect when acting on 
spin--ice configurations.


In Appendix~\ref{section:CMC.technical} technical details
are given of classical Monte Carlo simulations.


In Appendix~\ref{section:QMC-technical} technical details
are given of quantum Monte Carlo simulations.


In Appendix~\ref{appendix:2nd.order.perturbation.theory.in.g} a
perturbarion theory is developed in the quantum tunneling between 
spin ice states 
and used to explore how the OZZ ground state emerges at the boundary 
between CAF and TDQ states.

\section{Model}
\label{section:model}

\subsection{The classical ``dipolar spin ice'' model}
\label{subsection:classical.model}


After almost twenty years of study, it is generally accepted that the 
finite--temperature properties of spin--ice materials are well-described by 
an effective Ising model with both short-range exchange and long--range 
dipolar interactions --- the so-called ``dipolar spin ice'' (DSI) model 
\cite{siddharthan99,siddharthan-arXiv,denHertog00,bramwell01-PRL87,melko01,yavorskii08}.  
The basic building blocks of this model are magnetic rare--earth ions, 
occupying the sites of a pyrochlore lattice.


This pyrochlore lattice is built of corner--sharing tetrahedra, and has 
the same cubic space group $Fd\overline{3}m$ as the diamond lattice.
It is convenient to represent this lattice in terms of its 4--site primitive
unit cell --- a tetrahedron.
The corresponding Bravais lattice is FCC, with sites
\begin{eqnarray}
 \mathbf{R}_{{\bf m}} &=  \frac{a}{2} (m_x,m_y,m_z) \quad , \quad m_{x,y,z} \in \mathbb{Z} \; ,
 \end{eqnarray}
where $a$ is the linear dimension of the chemical unit cell (which is cubic, and 
contains 16 magnetic ions), and \mbox{$m_x + m_y + m_z$} is an even integer.  
Magnetic ions then occupy sites belonging to one of the four 
sublattices $a$, $b$, $c$, $d$, with position 
\begin{subequations}
\begin{align}
 \mathbf{r}_i &=  \mathbf{R}_{{\bf m}_i} + \frac{a}{8} (1,1,1)  \; , &  i \in a \; ; \\
 \mathbf{r}_i &=  \mathbf{R}_{{\bf m}_i} + \frac{a}{8}(1,-1,-1) \; , &  i \in b \; ; \\
 \mathbf{r}_i &=  \mathbf{R}_{{\bf m}_i} + \frac{a}{8}(-1,1,-1) \; , &  i \in c \; ; \\
 \mathbf{r}_i &=  \mathbf{R}_{{\bf m}_i} + \frac{a}{8}(-1,-1,1) \; , &  i \in d \; . 
\end{align}
\label{eq:ri}
\end{subequations}


In spin ice, a cubic the crystal field 
lifts the degeneracy of the 4$f$ multiplets of the rare--earth ions, such that 
the ground state of each ion is a high--spin doublet.
This doublet acts like an Ising moment 
\begin{eqnarray}
\mathbf{M}_i =  2 \mu_{\sf eff} \mathsf{S}^z_i \mathbf{\hat{z}}_i \, ,
\label{eq:Ising-spin}
\end{eqnarray}
where 
\begin{eqnarray}
\mbox{$\mathsf{S}^z_i = \pm 1/2$} \, .
\end{eqnarray}
and the magnitude of the moment is given by 
\begin{eqnarray}
\mu_{\sf eff} =  g_{\sf L} \mu_B \langle \mathsf{J}^z \rangle \; .
\end{eqnarray}
The Ising moment on a given site is tied to a local easy--axis, 
parallel to the unit-vector $\mathbf{\hat{z}}_i$, where  
\begin{subequations}
\begin{align}
 \mathbf{\hat{z}}_i &= \frac{1}{\sqrt{3}}(1,1,1)\;, &\quad i\in a \;; \\
 \mathbf{\hat{z}}_i &= \frac{1}{\sqrt{3}}(1,-1,-1)\;, &\quad i\in b \;; \\
 \mathbf{\hat{z}}_i &= \frac{1}{\sqrt{3}}(-1,1,-1)\;, &\quad i\in c \;; \\
 \mathbf{\hat{z}}_i &= \frac{1}{\sqrt{3}}(-1,-1,1)\;, &\quad i\in d \; . 
\end{align}
\label{eq:local_axis}
\end{subequations}
It follows that the Ising spins $\mathbf{M}_i $ 
point into, or out of, the tetrahedron to which they belong [cf. Eqs.~(\ref{eq:local_axis}) 
and Eqs.~(\ref{eq:ri})].   


\begin{figure}[tb]
\includegraphics[width=0.4\columnwidth]{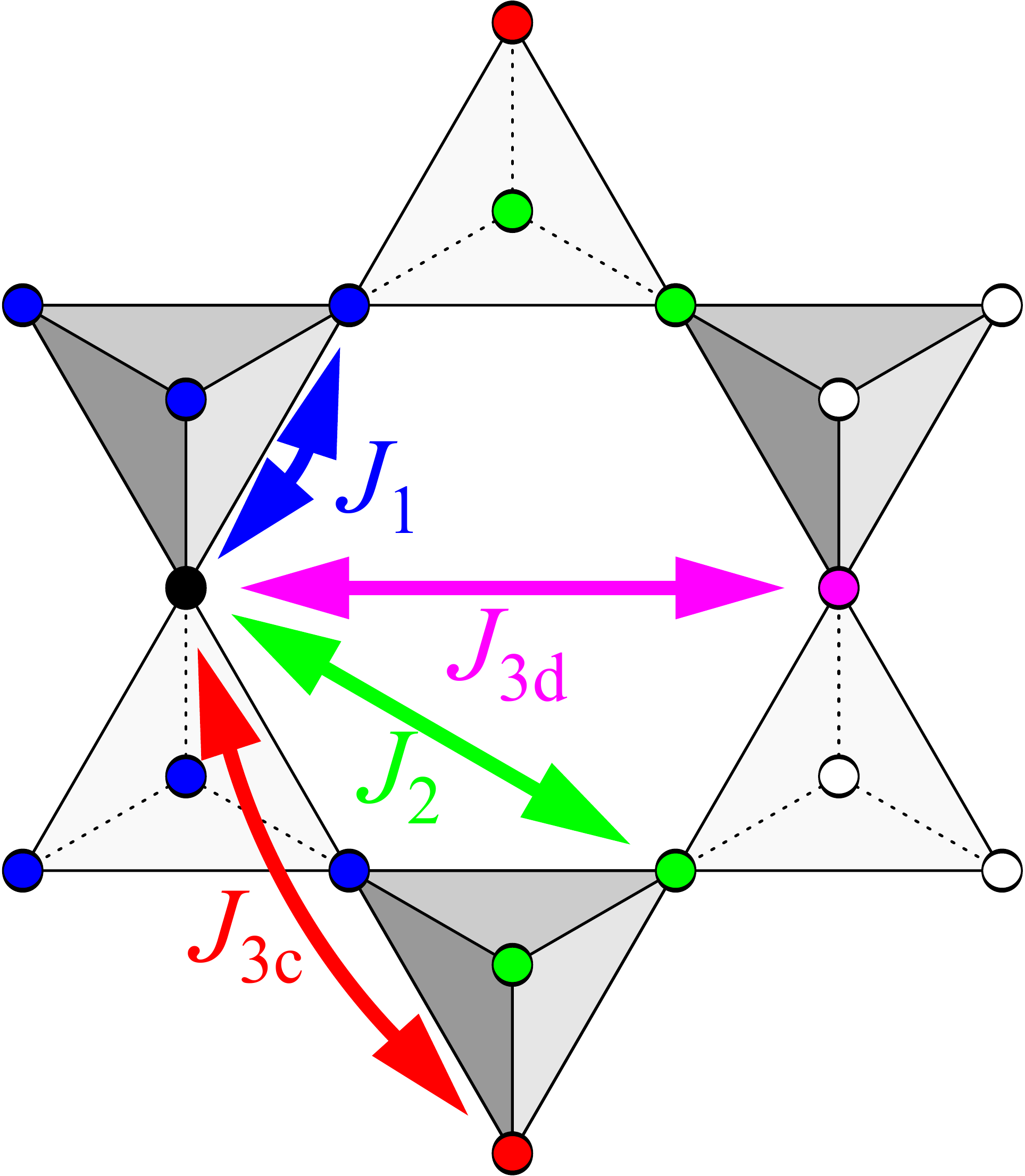}
\caption{(Color online)  
Exchange interactions up to \mbox{3$^{\text{rd}}$--neighbour} on the pyrochlore lattice.
The interactions $J_k$ appearing in ${\mathcal H}_{\sf exchange}$~[Eq.~(\ref{eq:H.exchange})]
are color-coded $J_1$ (blue), $J_2$ (green), $J_{3c}$ (red), and $J_{3d}$ (purple).
\label{fig:j2j3cj3dexchanges}
}
\end{figure}
 

The dipolar spin--ice model 
takes into account both dipolar and exchange interactions between
these Ising spins  
\begin{eqnarray}
\mathcal{H_{\sf DSI}} 
    =  \mathcal{H}_{\sf dipolar}
    + \mathcal{H}_{\sf exchange} \; .
\label{eq:Hdsi}
\end{eqnarray}
Dipolar interactions are long--ranged, and have the form 
\begin{eqnarray}
{\mathcal H}_{\sf dipolar} 
&=&  4 D \sum_{i<j} \left(\frac{r_1}{r_{ij}}\right)^3
  \left[
   \mathbf{\hat{z}}_i \cdot \mathbf{\hat{z}}_j  
   \right. \nonumber\\
&&   \qquad \left.  
    - 3  \left( \mathbf{\hat{z}}_i \cdot \hat{\mathbf{r}}_{ij} \right)
   \left( \mathbf{\hat{z}}_j \cdot \hat{\mathbf{r}}_{ij} \right)
  \right]
   \, \mathsf{S}^z_i  \, \mathsf{S}^z_j \,,
\label{eq:H.dipolar}
\end{eqnarray}
where $\mathbf{r}_{ij}$ is the vector connecting 
sites $i$ and $j$ (with $r_{ij} = |\mathbf{r}_{ij}|$ and  
$\hat{\mathbf{r}}_{ij} = \mathbf{r}_{ij}/r_{ij}$); 
\begin{eqnarray}
r_1 = \frac{a}{2\sqrt{2}}
\label{eq:r1}
\end{eqnarray}
is the distance between neighbouring 
magnetic ions; and 
\begin{eqnarray}
D = \frac{\mu_0 \mu_{\sf eff}^2}{16 \pi r_1^3}  
\label{eq:D}
\end{eqnarray}
is the strength of dipolar interactions at distance $r_1$.
To keep the definition of $D$ consistent with 
Refs.~[\onlinecite{siddharthan99,siddharthan-arXiv,denHertog00,bramwell01-PRL87,melko01,yavorskii08}], 
where spins have unit length \mbox{$\mathsf{S}^z_i = \pm 1$}, an overall factor of $4$ 
has been introduced in ${\mathcal H}_{\sf dipolar}$ [Eq.~(\ref{eq:H.dipolar})].
Dipolar interactions have an infinite range, so where we simulate finite--size clusters, 
with periodic boundary conditions we employ the Ewald resumption 
described in Appendix~\ref{appendix:ewald.sum}.


The dipolar spin--ice model also allows for competing exchange interactions 
\begin{eqnarray}
{\mathcal H}_{\sf exchange} 
&=& \sum_{k} 
   4 J_k \sum_{\langle ij \rangle_k} 
  \left( \mathbf{\hat{z}}_i \cdot \mathbf{\hat{z}}_j \right)
   \, \mathsf{S}^z_i  \, \mathsf{S}^z_j \,,
\label{eq:H.exchange}
\end{eqnarray}
where $k$ counts equivalent pairs of sites on the pyrochlore lattice 
and, once again, an overall factor of $4$ has been introduced in 
${\mathcal H}_{\sf exchange}$~[\ref{eq:H.exchange}] to keep the definition of $J_k$ 
consistent with 
Refs.~[\onlinecite{siddharthan99,siddharthan-arXiv,denHertog00,bramwell01-PRL87,melko01,yavorskii08}].
All possible exchange interactions up to \mbox{3$^{\text{rd}}$--neighbour},
including the two distinct forms of  \mbox{3$^{\text{rd}}$--neighbour} exchange 
$J_{3d}$ and $J_{3c}$, are illustrated in Fig.~\ref{fig:j2j3cj3dexchanges}.


\begin{figure*}
\includegraphics[width=0.6\textwidth]{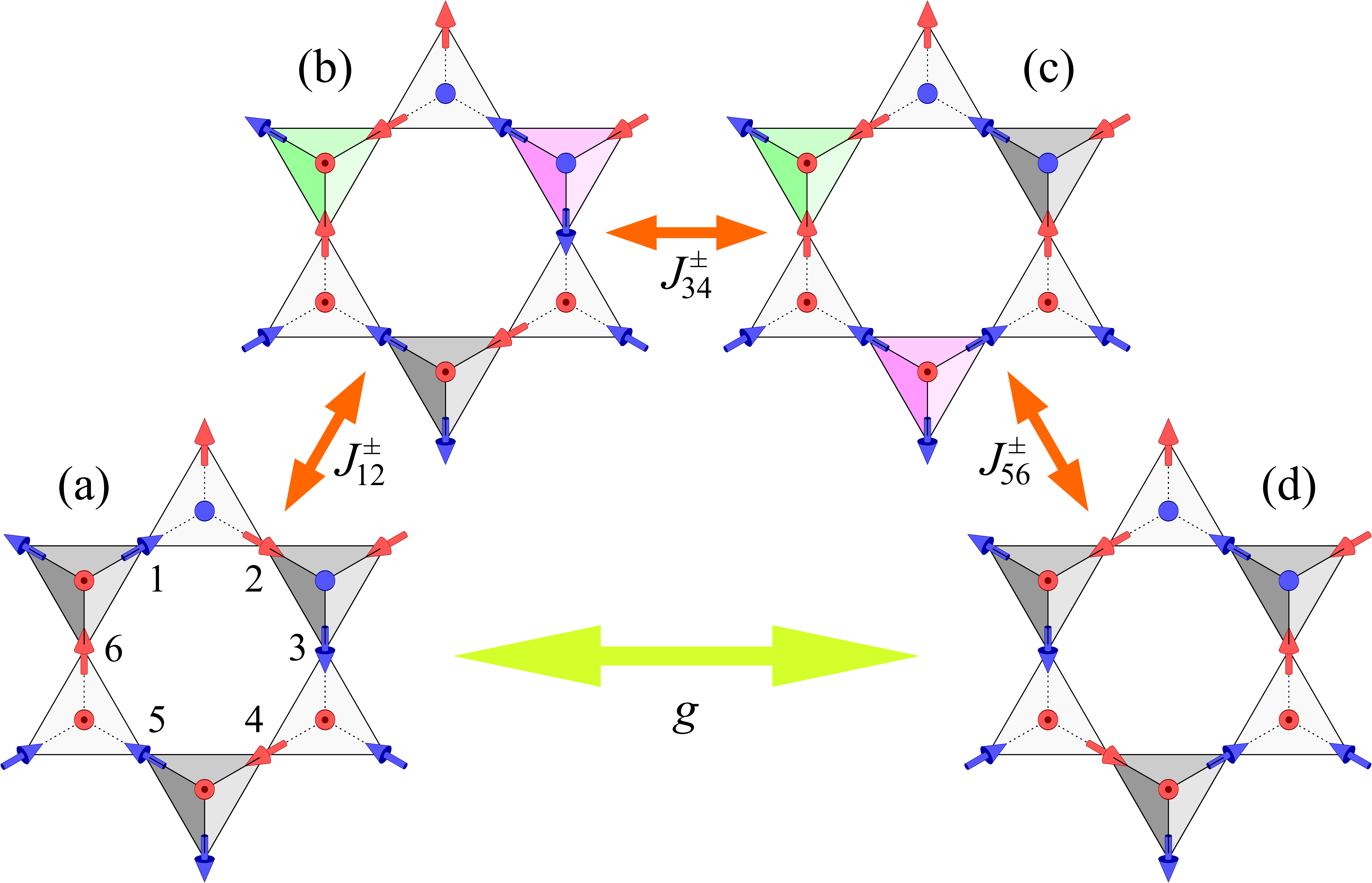}
\caption{
(Color online)  
An example of quantum tunnelling between two different spin configurations obeying the 
ice rules, mediated by the virtual excitation of a pair of magnetic monopoles.
(a) spin--configuration obeying the ice rules, containing a closed
loop of spins, numbered 1\ldots 6.  
(b) excited state containing a pair of magnetic monopoles 
(green and magnenta tetrahedra).
(c) degenerate excited state, in which one of the monopoles
has moved.   
(d) spin--configuration obeying the ice rules, in which the closed
loop of spin has been reversed.   
For the anisotropic exchange model ${\mathcal H}_{\sf xxz}$~[Eq.~(\ref{eq:xxzham})],
this process  corresponds to 3$^\text{rd}$--order degenerate perturbation in 
$J_\pm$, and leads to the tunneling amplitude $g = 12J_\pm^3/J_z^2$~[Eq.~(\ref{eq:gpertJpm3})].
\label{fig:tunneling}
}
\end{figure*}


The defining property of spin ice is that at low temperatures 
spin--configurations obey the ``ice rules'', which require 
that two spins point into, and two spins point out of, every 
tetrahedron on the lattice.
The simplest model leading to the ice rules contains only ferromagnetic 
exchange $J_1 < 0$, between nearest--neighbour Ising spins.~\cite{harris97}
In this case, all spin configurations obeying the ice rules are degenerate.


The presence of long--range dipolar interactions, and further--neighbour exchanges,
lifts this degeneracy, giving rise to the possibility of ordered ground states.     
However the differences in energy from dipolar interactions alone
are smaller than might be excepted, since dipolar interactions 
are ``self-screened'' \cite{gingras01, enjalran04, melko04} 
within spin-ice configurations, decaying as $1/r^5$.\cite{isakov05}
And, as discussed below, there exist a subset of spin-ice configurations,
the ``chain states'', in which dipolar interactions are even better screened,
with interactions decaying exponentially with distance.   

\subsection{Quantum tunneling between spin-ice states}
\label{subsection:quantum.model}

The minimal change in a spin ice, once quantum effects 
are taken into account, is the possibility of the system tunneling from
one spin--configuration obeying the ice rules to another.
Tunnelling matrix elements arise where it is possible to reverse 
closed loops of spins, with the shortest loop occurring on the 
hexagonal plaquette shown in Fig.~\ref{fig:j2j3cj3dexchanges}. 


The natural quantum generalisation of the dipolar spin--ice model is 
therefore 
\begin{eqnarray}
\mathcal{H_{\sf QDSI}} 
    =  \mathcal{H}_{\sf dipolar}
    + \mathcal{H}_{\sf exchange}
    + \mathcal{H}_{\sf tunneling} \,.
\label{eq:Hqdsi}
\end{eqnarray}
where 
\begin{equation}
\mathcal{H}_{\sf tunneling} =  
  -g \sum_{\hexagon} 
   |\! \circlearrowright \rangle\langle \circlearrowleft\! | + 
   |\! \circlearrowleft \rangle\langle \circlearrowright\! |  \,.
\label{eq:Htunneling}
\end{equation}
and the sum upon $\hexagon$ runs over the hexagonal plaquettes 
of the pyrochlore lattice.
In the absence of long--range dipolar or exchange interactions, quantum 
tunneling of the form $\mathcal{H}_{\sf tunneling}$ [Eq.~(\ref{eq:Htunneling})] 
is known to stabilize a quantum spin liquid described by 
a quantum $U(1)$ lattice gauge theory.~\cite{hermele04,banerjee08,shannon12,benton12}


Due to their relative smallness, it is hard even to estimate the strength
of quantum tunnelling in spin-ice materials such as Dy$_2$Ti$_2$O$_7$, and the 
microscopic aspects of the quantum dynamics are only beginning 
to be understood~\cite{rau-arXiv,tomasello-arXiv}.   
However, the form of the tunnelling matrix element $\mathcal{H}_{\sf tunneling}$ [Eq.~(\ref{eq:Htunneling})]  
is uniquely determined by the ice rules and the geometry of the pyrochlore lattice, 
so estimates of $g$ can be taken from any quantum model which supports 
a spin--ice ground state.


The simplest example is an anisotropic exchange model with 
interactions of ``XY'' type, 
\begin{eqnarray}
{\mathcal H}_{\sf xxz} 
= J_{zz} \sum_{\langle ij \rangle} 
\mathsf{S}_i^z \mathsf{S}_j^z 
- J_{\pm} \sum_{\langle ij \rangle}
(\mathsf{S}_i^+ \mathsf{S}_j^- + \mathsf{S}_i^- \mathsf{S}_j^+) \,,
\label{eq:xxzham}
\end{eqnarray}
with $\mathsf{S}_i^z$ promoted to a (pseudo) spin-1/2 operator such
that 
\begin{eqnarray}
[\mathsf{S}_i^+,\mathsf{S}_j^-] =  2 \mathsf{S}^z_i \delta_{ij} \, .
\end{eqnarray}
In this case $\mathcal{H}_{\sf tunneling}$~[Eq.~(\ref{eq:Htunneling})]
can be derived in degenerate perturbation theory about classical 
states obeying the ice rules. 
The tunneling process shown in Fig.~\ref{fig:tunneling} can be thought 
of as the spontaneous creation of a (virtual) pair of magnetic monopoles, 
which annihilate after one has traversed the hexagon, leading   
to an effective tunneling 
%
\begin{equation}
g = \frac{12 J_{\pm}^3}{J_{zz}^2} \,.
\label{eq:gpertJpm3}
\end{equation}


A more general starting point for describing a quantum spin ice is the 
anisotropic nearest-neighbour exchange model~\cite{onoda11,savary12-PRL108,lee12}
\begin{align}
  \mathcal{H}_{\sf S=1/2} 
  = & \sum_{\langle ij\rangle} \Big\{ J_{zz} \mathsf{S}_i^z \mathsf{S}_j^z - J_{\pm}
  (\mathsf{S}_i^+ \mathsf{S}_j^- + \mathsf{S}_i^- \mathsf{S}_j^+) 
\nonumber \\ & 
 + J_{\pm\pm} \left[\gamma_{ij} \mathsf{S}_i^+ \mathsf{S}_j^+ + \gamma_{ij}^*
    \mathsf{S}_i^-\mathsf{S}_j^-\right] 
\nonumber \\ & 
+ J_{z\pm}\left[ \mathsf{S}_i^z (\zeta_{ij} \mathsf{S}_j^+ + \zeta^*_{ij} \mathsf{S}_j^-) +
  {i\leftrightarrow j}\right]\Big\} \,,
  \label{eq:H-anisotropic-exchange}
\end{align}
where the sum $\langle ij\rangle$ runs over the nearest-neighbour bonds of the pyrochlore 
lattice; and $\gamma_{ij}$ and $\zeta_{ij} $ are $4 \times 4$ complex unimodular matrices 
encoding the rotations between the local axes $\mathbf{\hat{z}}_i$ and the cubic
axes of crystal.~\cite{curnoe08,ross11-PRX1}


The (pseudo) spin-1/2 model $\mathcal{H}_{\sf S=1/2}$~[Eq.~(\ref{eq:H-anisotropic-exchange})], 
has been shown to give a quantitative description of spin excitations in both the 
``quantum spin ice'' Yb$_2$Ti$_2$O$_7$~[\onlinecite{ross11-PRX1}] and quantum 
order-by-disorder system Er$_2$Ti$_2$O$_7$~[\onlinecite{savary12-PRL109}].
The parameterization of $\mathcal{H}_{\sf S=1/2}$ [Eq.~(\ref{eq:H-anisotropic-exchange})], 
and its mean-field phase diagram have been explored in 
Refs.~[\onlinecite{onoda11,savary12-PRL108,savary13,lee12}].
We will not develop this topic further here, but note that the additional terms, 
$J_{z\pm}$ and $J_{\pm\pm}$, can also contribute to the tunneling $g$, but 
do so in higher orders of perturbation theory than $J_{\pm}$ [Eq.~(\ref{eq:gpertJpm3})]. 

\subsection{Choice of parameters}
\label{subsection:model.parameters}

Like other spin ices, Dy$_2$Ti$_2$O$_7$ is believed to be well--described by the dipolar 
spin--ice model $\mathcal{H_{\sf DSI}}$ [Eq.~(\ref{eq:Hdsi})],  and the values of the 
parameters $D$ and $J_k$ have been estimated by Yavors'kii {\it et al.} in Ref.~[\onlinecite{yavorskii08}].  
In this case, the lattice constant $a = 10.124\ \AA$ [\onlinecite{fukazawa02}], and the 
Dy$^{3+}$ ions have a Land\'e factor $g_{\sf L} = 4/3$ associated with an Ising  
moment $\langle \mathsf{J}^z \rangle = 7.40\ \mu_B$.
It follows from Eq.~(\ref{eq:D}) that 
\begin{eqnarray}
D = 1.3224\ \text{K} \,. \qquad [\text{Dy$_2$Ti$_2$O$_7$}]
\label{eq:D.Dy2Ti2O7}
\end{eqnarray}
Competing exchange interactions were estimated 
on the basis of fits of classical Monte Carlo simulation to the structure 
factor $S({\bf q})$ measured in (diffuse) neutron scattering.   
Working within the simplifying assumption  
\begin{eqnarray}
J_{3c} = J_{3d} = J_3 \; ,  \qquad [\text{Ref.~\protect{\onlinecite{yavorskii08}}}] 
\end{eqnarray}
Yavors'kii {\it et al.}~[\onlinecite{yavorskii08}] find 
\begin{eqnarray}
J_1 =  &\ & 3.41\ \text{K} \,,  \nonumber\\
J_2 = &-&0.14\ \text{K} \,, \qquad [\text{Dy$_2$Ti$_2$O$_7$}] \\
J_3 =  & & 0.03\ \text{K} \,,  \nonumber
\label{eq:yavorskii.parameters}
\end{eqnarray}


For the purposes of this Article, we  work with parameters 
$D$ and $J_k$ chosen such that the net effect of the interactions in $\mathcal{H_{\sf DSI}}$ 
[Eq.~(\ref{eq:Hdsi})] is to enforce the ice--rules constraint.
We consider all possible exchanges 
up to \mbox{3$^{\text{rd}}$--neighbour}, as illustrated in Fig.~\ref{fig:j2j3cj3dexchanges}, 
maintaining the distinction $J_{3c} \ne  J_{3d}$.   
However, since $J_1$ plays no part in selecting ordered ground states
we set $J_1 \equiv 0$, 
except where needed for comparison with the finite--temperature
properties of real materials.


A further simplication arises since, within spin--configurations
obeying the ice rules, the effect of the \mbox{3$^{\text{rd}}$--neighbour} exchange $J_{3c}$ 
is simply to renormalise the \mbox{2$^{\text{nd}}$--neighbour} exchange, 
\begin{equation}
J_2 \to J_2 + 3 J_{3c}  \; , 
\label{eq:j2_j3c_equivalence}
\end{equation}
leaving only $J_2$ and $J_{3d}$ as independent parameters.   
This equivalence is proved in Appendix~\ref{appendix:equivalence.J2.and.J3c}.


Mindful of Dy$_2$Ti$_2$O$_7$ [cf. Eq.~(\ref{eq:yavorskii.parameters})], we will 
generally assume that $J_2$ is the leading form of exchange interaction.
And for the purposes of soft--spin mean--field theory [Sec.~\ref{sec:classical.mean.field.theory}], 
classical  Monte Carlo simulation [Sec.~\ref{section:classical.monte.carlo}], 
and quantum Monte Carlo simulation [Sec.~\ref{section:QMC}], we will generally 
consider ferromagnetic $J_2 < 0$, setting all other exchange interactions to zero.

\section{Mean--field ground states of dipolar spin ice}
\label{sec:classical.mean.field.theory}

Many of the properties of spin-ice materials~[\onlinecite{denHertog00,isakov05}] 
can be successfully described using a ``soft--spin'' mean field theory, 
in which the ``hard--spin'' constraint of fixed spin-length 
\begin{eqnarray}
({\sf S}^z_i)^2 = \frac{1}{4} \; ,
\label{eq:hard.spin.constraint}
\end{eqnarray}
is relaxed, and spins are treated as continuous variables.   


In what follows, we use such a soft--spin mean--field theory to explore
the classical ground state--phase diagram of $\mathcal{H_{\sf DSI}}$~[Eq.~(\ref{eq:Hdsi})].
We focus on the competition between long--range dipolar interactions $D$ and 
second--neighbour exchange $J_2$, and construct a mean--field phase diagram 
as a function of  $J_2/D$.



The starting point for our mean--field theory is the Fourier transform 
of the combined dipolar and exchange interactions,  $\mathcal{J}^{ab}_{\bf q}$, where the index 
\begin{eqnarray}
a,b =0,1,2, 3 
\end{eqnarray}
counts the 4 sites of the tetrahedron $i$ as defined in Eq.~(\ref{eq:ri}) (which is the primitive unit cell), with the local axis given by Eq.~(\ref{eq:local_axis}).
Following Reimers {\it et al.}~[\onlinecite{reimers91}], den Hertog {\it et al.}~[\onlinecite{denHertog00}], and 
Isakov {\it et al.}~[\onlinecite{isakov05}], we write 
\begin{eqnarray}
\mathcal{H_{\sf DSI}} \approx \overline{\mathcal H}_{\sf DSI} 
    = \sum_{\bf q}^{a,b} \mathcal{J}^{ab}_{\bf q}  m^a_{\bf q}  m^b_{\bf -q}  \; , 
    \label{eq:Hsoftspin}
\end{eqnarray}
where 
\begin{eqnarray}
m^a_{\bf q} = \frac{1}{\sqrt{N}} \sum_{i\in a} {\sf S}^z_{i} e^{i {\bf q} \cdot {\bf r}_{i}} \; ,
\label{eq:m}
\end{eqnarray}
and similar equations hold for sublattice $b$, $c$, $d$.
The contribution to $\mathcal{J}^{ab}_{\bf q}$ from long--range 
dipolar interactions is determined by an Ewald summation, as 
described in~Ref.~[\onlinecite{enjalran04}].  


\begin{figure}[tb]
\includegraphics[width=0.8\columnwidth]{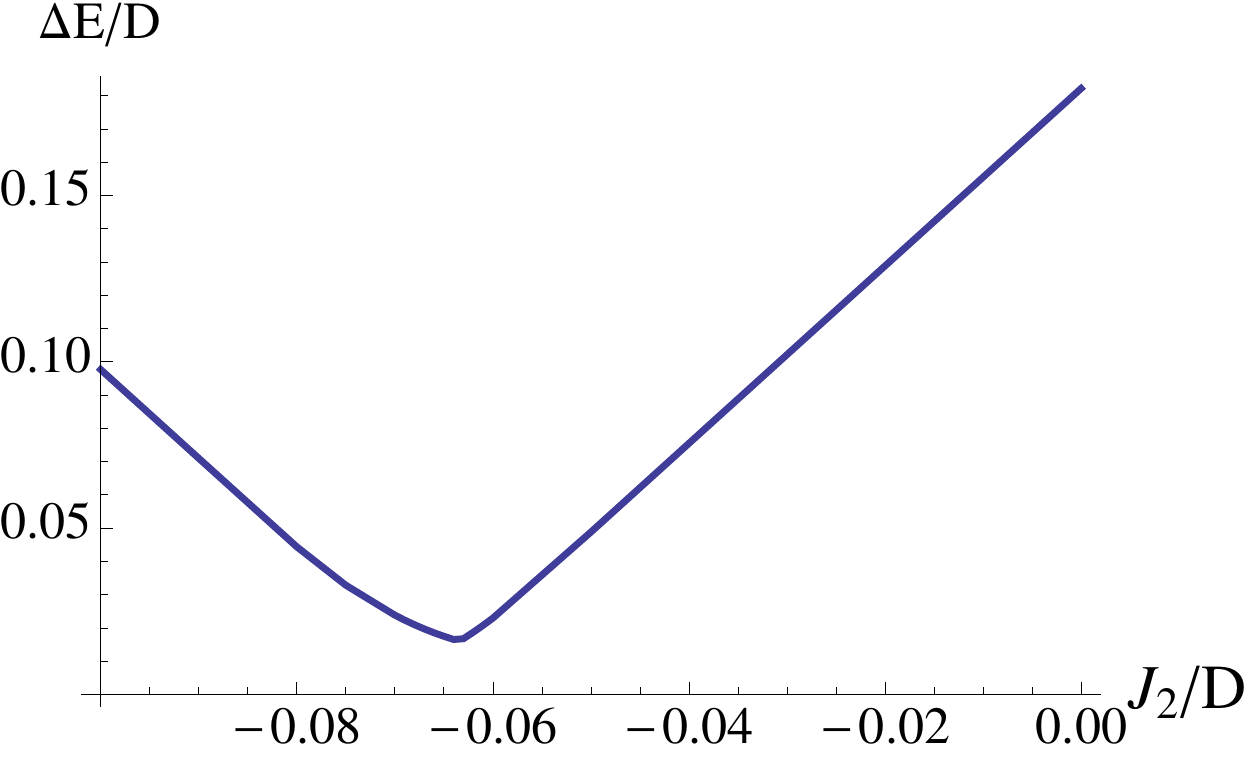}
\caption{(Color online) Band-width $\Delta E$ [Eq.~(\ref{eq:Delta.E})] of spin-ice states 
in the presence of long-range dipolar interactions $D$, as a function of second-neighbour 
exchange $J_2 < 0$, within the soft spin-mean field theory 
$\overline{\mathcal H}_{\sf DSI}$~[Eq.~(\ref{eq:Hsoftspin})].  
%
For small $|J_2|/D$, the competing exchange interaction 
leads to a reduction in the bandwidth of spin-ice states, which takes
on a minimum value for $J_2/D = -0.062$. 
\label{fig:kMFT_Bandwidth}
}
\end{figure}


The eigenvalues of the matrix $\mathcal{J}^{ab}_{\bf q} $, 
\begin{eqnarray}
\mathcal{J}_{\bf q} \cdot \boldsymbol{\mathcal{E}}_{\mathbf{q}}^{\mu}
   = \epsilon_{\mathbf{q}}^{\mu} \boldsymbol{\mathcal{E}}_{\mathbf{q}}^{\mu} \, ,
\label{eq:eigenvalue.equation}   
\end{eqnarray}
form four dispersing bands $\epsilon_{\mathbf{q}}^{\mu}$, labeled by $\mu$.    
%
The eigenvector 
$\boldsymbol{\mathcal{E}}_{\mathbf{q}_{\sf min}}^{\mu_{\sf min}}$
with the lowest eigenvalue(s)
\begin{eqnarray}
\mathcal{J}_{\bf q} \cdot  \boldsymbol{\mathcal{E}}_{\mathbf{q}_{\sf min}}^{\mu_{\sf min}}
   = \epsilon_{\mathbf{q}_{\sf min}}^{\mu_{\sf min}} \boldsymbol{\mathcal{E}}_{\mathbf{q}_{\sf min}}^{\mu_{\sf min}} \, .
\end{eqnarray}
%
is(are) a ground state of $\overline{\mathcal H}_{\sf DSI}$~[Eq.~(\ref{eq:Hsoftspin})].  
%
%
As long as the associated eigenvector  $\boldsymbol{\mathcal{E}}_{\mathbf{q}_{\sf min}}^{\mu_{\sf min}}$ 
satisfies the ``hard-spin'' constraint Eq.~(\ref{eq:hard.spin.constraint}), this state is also a valid 
ground state of original dipolar spin--ice model 
$\mathcal{H_{\sf DSI}}$~[Eq.~(\ref{eq:Hdsi})].  
Once this constraint is (re)imposed, the soft--spin approximation 
becomes equivalent to the well-known Luttinger--Tisza 
method.~\cite{luttinger46}


\begin{figure}[tb]
\includegraphics[width=0.75\columnwidth]{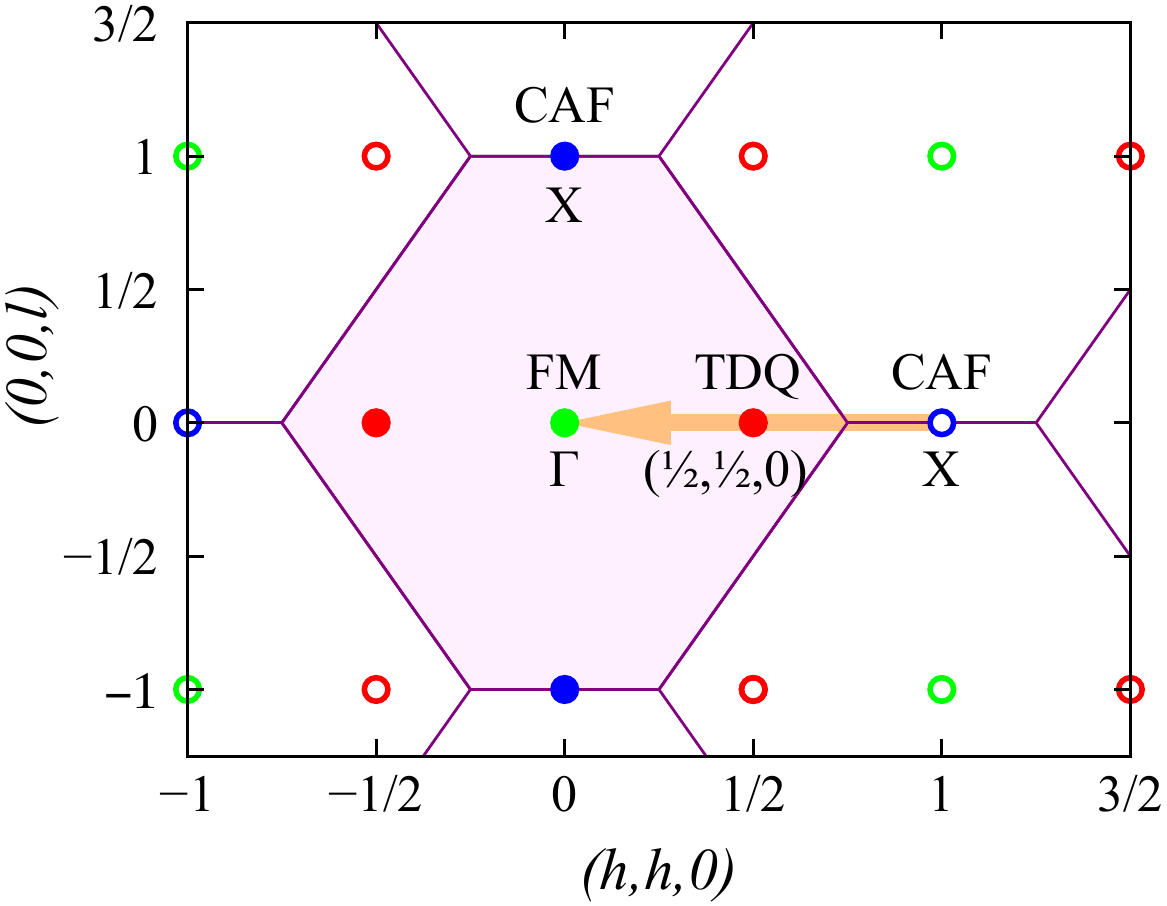}
\caption{(Color online)  
Wave vectors associated with ordered ground states 
in a dipolar spin ice described by $\mathcal H_{\sf DSI}$~[Eq.~(\ref{eq:Hdsi})].
The cubic antiferromagnet (CAF) has ordering vector 
${\bf Q}^{\sf CAF} = (0,0,1)$~[Eq.~(\ref{eq:Q.min.CAF})], 
and symmetry--related points [including the (1,1,0)], here labelled ``$X$'' (blue points).
The tetragonal double-Q state (TDQ) has is based on 
pairs of ordering vectors of the form
\mbox{${\bf Q}^{\sf TDQ} = \pm (1/2,1/2,0)$}~[Eq.~(\ref{eq:QTDQ})], shown with red points.  
The ferromagnet (FM) has ordering vector 
${\bf Q}^{\sf FM} = (0,0,0)$~[Eq.~(\ref{eq:Q.min.FM})], 
here labelled ``$\Gamma$'' (green points).
The evolution of the ordering vector within the soft--spin mean 
field--theory 
$\overline{\mathcal H}_{\sf DSI}$~[Eq.~(\ref{eq:Hsoftspin})]
is shown with an orange arrow.
\label{fig:ordering-vectors}
}
\end{figure} 


In the simplest model of a spin--ice, in which only nearest--neighbour interactions
are taken into account, there is no unique eigenvector 
$\boldsymbol{\mathcal{E}}_{\mathbf{q}_{\sf min}}^{\mu_{\sf min}}$
with a minimum energy.   
Instead the two lowest-lying 
eigenstates form ``flat'' bands with 
\begin{align}
\epsilon_{\mathbf{q}}^{\sf 1} =\epsilon_{\mathbf{q}}^{\sf 2} \equiv 0.
\end{align}
These bands describe the (extensively degenerate) set of spin 
configurations which obey the two-in two-out ``ice rules''~[\onlinecite{denHertog00,melko01,isakov05}]. 


The degeneracy of the spin--ice configurations is lifted by long--range dipolar
interactions, causing these flat bands to acquire a dispersion.
However dipolar interactions, despite being long-range, are effectively 
``self-screened'' within the spin-ice states,~\cite{denHertog00} a fact  
known as ``projective equivalence'' [\onlinecite{isakov05}].
The overall bandwidth of spin-ice states in the presence 
of dipolar interactions 
\begin{eqnarray}
\Delta E  = 
  \text{Max}(\epsilon_{\mathbf{q}}^{\sf 1},\epsilon_{\mathbf{q}}^{\sf 2}) 
   - \text{Min}(\epsilon_{\mathbf{q}}^{\sf 1},\epsilon_{\mathbf{q}}^{\sf 2})
   \approx 0.17 D 
   \label{eq:Delta.E}
\end{eqnarray}
and is significantly smaller than the bare scale of dipolar interactions $D$.
None the less, dipolar interactions do select an ordered ground state,
as described below.


We now turn to question of finding the ground state of
$\mathcal{H}_{\sf DSI}$~[Eq.~(\ref{eq:Hdsi})] as function of $J_2/D$.   
%
Within the soft-spin approximation   
$\overline{\mathcal H}_{\sf DSI}$~[Eq.~(\ref{eq:Hsoftspin})],
for $J_2 < 0$,  there are three distinct regimes, corresponding 
to different ordering vectors 
\begin{eqnarray}
{\bf q}_{\sf min} = \frac{2\pi}{a} \; {\bf Q}_{\sf min} \;,
\label{eq:Qmin}
\end{eqnarray}
where $a$ is the (cubic) lattice spacing, and ordering vectors
are measured relative to the usual, cubic, crystallographic coordinates.
We consider each of these regimes in turn, below.


\begin{figure}[t]
\includegraphics[width=0.8\columnwidth]{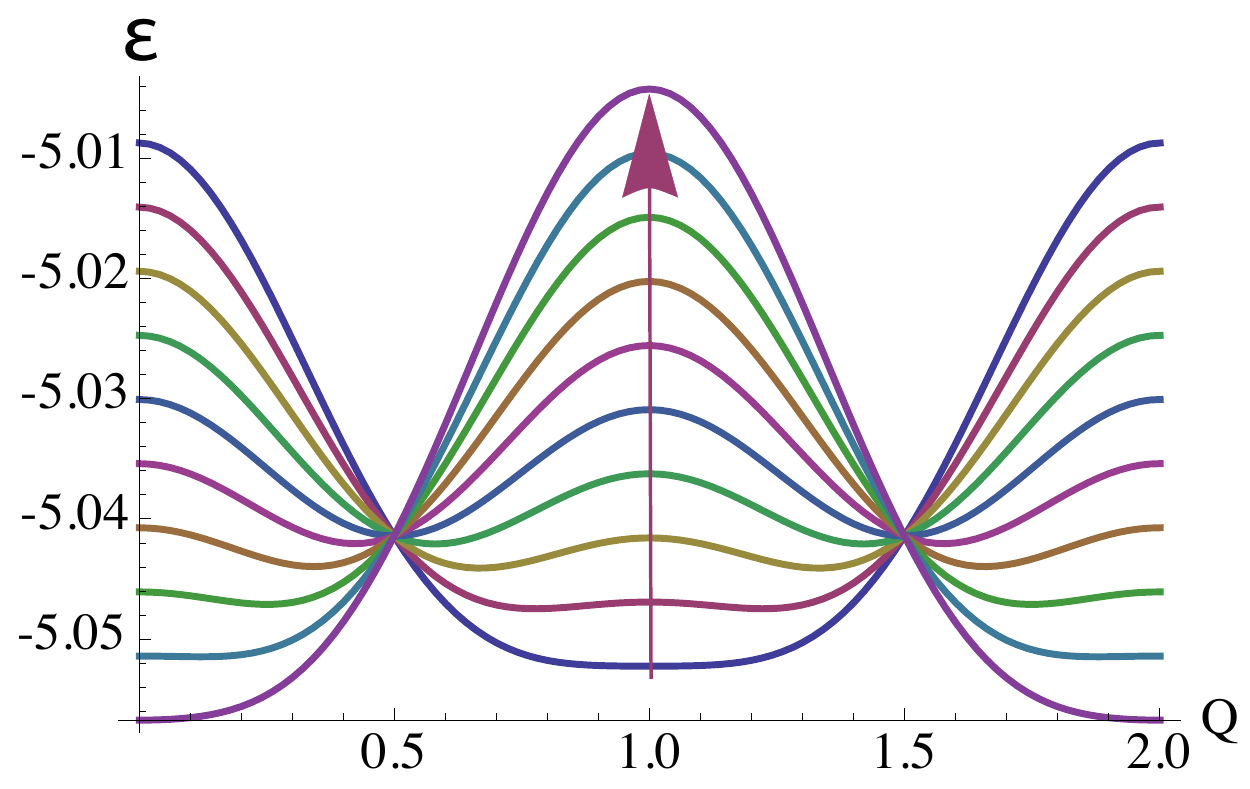}
\caption{ 
(Color online)
Evolution of the 
dispersion of lowest-lying eigenvalue of 
$\overline{\mathcal H}_{\sf DSI}$ [Eq.~(\ref{eq:Hsoftspin})],
as a function of second-neighbour exchange $J_2 < 0$, 
for wave vector ${\bf Q} = (Q,Q,0)$.  
%
For \mbox{$J_2/D = -0.057$}, (blue curve) the minimum of the dispersion 
is found at ${\bf Q}^{\sf CAF} = \left(1,1,0 \right)$.
For \mbox{$J_2/D = -0.08$}, (purple curve) the minimum of the dispersion 
is found at ${\bf Q}^{\sf FM} = \left(0,0,0 \right)$.
The direction of decreasing $J_2/D$ is shown with an arrow. The mean field energy at $Q=1/2$ and $3/2$ does not depend on $J_2/D$, this is the origin of the line crossings. %
}
\label{fig:band.evolution}
\end{figure} 


\begin{table*}[t]
\caption{ 
Chain--based ordered ground states of dipolar spin ice, 
$\mathcal{H_{\sf QDSI}}$~[Eq.~(\ref{eq:Hqdsi})].    
%
The ordering wave vector $\mathbf{Q}$, [cf.~Fig.~\ref{fig:ordering-vectors}], 
is measured in reciprocal lattice units 
[cf.~Eq.~(\ref{eq:Qmin})].   
Also listed are the corresponding spin eigenvector [cf.~Eq.~(\ref{eq:eigenvalue.equation})], 
the chain directions associated with each eignvector, and the degeneracy of the state.   
The cubic antiferromagnet (CAF), tetragonal double--Q (TDQ) and ferromagnet 
(FM) are all found in classical treatments of dipolar spin ice 
[Sec~\ref{sec:classical.mean.field.theory},\ref{section:effective.Ising.model},\ref{section:classical.monte.carlo}].  
In contrast, the orthorhombic zig-zag (OZZ) is only stabilised by quantum 
effects [Sec.~\ref{section:QMC}].  
All four ordered states are illustrated in Fig.~\ref{fig:ordered.phases}.
}
\centering
\begin{ruledtabular}
\begin{tabular}{c c c c r c }
state & ordering wavevector ${\bf Q}$ & eigenvectors $\boldsymbol{\mathcal{E}}_{\mathbf{Q}}$ & chain directions & degen. & figure \\ [0.5ex] 
\hline\vspace{0.1cm}
& $\begin{array}{c} (1,0,0) \end{array}$
& $\begin{array}{rl} \frac{1}{2}(-1)^j \left(1,-1,0,0 \right), & j=1,2 \\ \frac{1}{2}(-1)^{j'} \left(0,0,1,-1 \right), & j'=1,2 \end{array}$
& $\begin{array}{c}[0,1,1] \\{}[0,1,\bar{1}]\end{array} $ 
& 4 
& Fig.~\ref{fig:ordered.phases}(a) \\\cline{2-6}\vspace{1truemm}
CAF
& $\begin{array}{c} (0,1,0) \end{array}$ 
& $\begin{array}{rl} \frac{1}{2} (-1)^j \left(1,0,-1,0\right), & j=1,2\\ \frac{1}{2} (-1)^{j'}\left(0,1,0,-1\right), & j'=1,2\end{array}$   
& $\begin{array}{c} [1,0,1]\\{}[1,0,\bar{1}] \end{array}$ 
& 4 
& \\\cline{2-6}\vspace{1truemm}

& $\begin{array}{c} (0,0,1) \end{array}$
& $\begin{array}{rl} \frac{1}{2} (-1)^j \left(1,0,0,-1\right), & j=1,2 \\ \frac{1}{2} (-1)^{j'}\left(0,1,-1,0\right), & j'=1,2 \end{array}$
& $\begin{array}{c} [1,1,0]\\{} [1,\bar1,0] \end{array} $ 
& 4 
& \\
\hline\vspace{1truemm}
& $\begin{array}{c}\left(0,\frac{1}{2},-\frac{1}{2} \right)\\ \left(0,\frac{1}{2},\frac{1}{2} \right)\end{array}$ 
& $\begin{array}{rl}\frac{1}{\sqrt{2}}e^{i (2j+1)\pi/4}\left(1,-1,0,0\right), & j=1,2,3,4\\ \frac{1}{\sqrt{2}}e^{i (2j'+1)\pi/4}\left(0,0,1,-1\right), & j'=1,2,3,4\end{array}$ 
& $\begin{array}{c}[0,1,1] \\{}[0,1,\bar{1}]\end{array} $ 
& 16 
& Fig.~\ref{fig:ordered.phases}(b)     
\\\cline{2-6}\vspace{1truemm}
TDQ
& $\begin{array}{c}\left(\frac{1}{2},0,-\frac{1}{2} \right)\\ \left(\frac{1}{2},0,\frac{1}{2} \right)\end{array}$ 
& $\begin{array}{rl}\frac{1}{\sqrt{2}}e^{i (2j+1)\pi/4}\left(1,0,-1,0\right), & j=1,2,3,4\\ \frac{1}{\sqrt{2}}e^{i (2j'+1)\pi/4}\left(0,1,0,-1\right), & j'=1,2,3,4\end{array}$ 
& $\begin{array}{c} [1,0,1]\\{}[1,0,\bar{1}] \end{array}$ 
& 16 
& \\\cline{2-6}\vspace{1truemm}
& $\begin{array}{c}\left(\frac{1}{2},-\frac{1}{2},0 \right)\\ \left(\frac{1}{2},\frac{1}{2},0 \right)\end{array}$ 
& $\begin{array}{rl}\frac{1}{\sqrt{2}}e^{i (2j+1)\pi/4}\left(1,0,0,-1\right), & j=1,2,3,4\\ \frac{1}{\sqrt{2}} e^{i (2j'+1)\pi/4}\left(0,1,-1,0\right), & j'=1,2,3,4\end{array}$ 
& $\begin{array}{c} [1,1,0]\\{} [1,\bar1,0] \end{array} $ 
& 16 
& \\
\hline\vspace{0.1cm}
& $\begin{array}{c} (0,0,0)\\ {} \end{array}$ 
& $\begin{array}{c} \pm\frac{1}{2} \left(1,-1,-1,1\right)\\ {} \end{array}$  
& $\begin{array}{c} [0,1,1]\; \& \; [0,1,\bar1] \\  \text{} [1,0,1] \; \& \; [1,0,\bar{1}]  \end{array}$  
& $\begin{array}{c} 2 \\  {} \end{array}$ 
& $\begin{array}{c} \text{Fig.~\ref{fig:ordered.phases}(c)}\\{} \end{array}$ \\\cline{2-6} \vspace{1truemm}
FM  
& $\begin{array}{c} (0,0,0)\\ {} \end{array}$ 
& $\begin{array}{c} \pm\frac{1}{2} \left(1,-1,1,-1 \right)\\ {} \end{array}$  
& $\begin{array}{c} [0,1,1] \; \& \; [0,1,\bar1]\\ \text{}   [1,1,0]  \; \& \; [1,\bar{1},0] \end{array}$ 
& $\begin{array}{c} 2\\ {} \end{array}$ 
& \\\cline{2-6}\vspace{1truemm}
& $\begin{array}{c} (0,0,0)\\ {} \end{array}$ 
& $\begin{array}{c} \pm\frac{1}{2} \left(1,1,-1,-1 \right)\\ {} \end{array}$  
& $\begin{array}{c} [1,0,1] \; \& \; [1,0,\bar1]\\ \text{}  [ 1,1,0] \; \& \; [1,\bar{1},0] \end{array}$ 
& $\begin{array}{c} 2\\ {} \end{array}$ 
& \\
\hline\vspace{1truemm}
& $\begin{array}{c}\left(0,\frac{1}{2},-\frac{1}{2} \right)\\ \left(1,0,0 \right)\end{array}$ 
& $\begin{array}{rl}
\frac{1}{\sqrt{2}}e^{i (2j+1)\pi/4}\left(1,-1,0,0\right), & j=1,2,3,4 \\  \frac{1}{2}(-1)^{j'} \left(0,0,1,-1 \right), & j'=1,2
\end{array}$ 
& $\begin{array}{c}[0,1,1] \\{}[0,1,\bar{1}]\end{array} $ 
& 8
& Fig.~\ref{fig:ordered.phases}(d)     
\\\cline{2-6}\vspace{1truemm}
& $\begin{array}{c}\left(1,0,0 \right)\\\left(0,\frac{1}{2},\frac{1}{2} \right)\end{array}$ 
& $\begin{array}{rl}\frac{1}{2}(-1)^{j} \left(1,-1,0,0\right), & j=1,2\\  \frac{1}{\sqrt{2}}e^{i (2j'+1)\pi/4}\left(0,0,1,-1 \right), & j'=1,2,3,4\end{array}$ 
& $\begin{array}{c}[0,1,1] \\{}[0,1,\bar{1}]\end{array} $ 
& 8
&
\\\cline{2-6}\vspace{1truemm}
OZZ 
& $\begin{array}{c}\left(\frac{1}{2},0,-\frac{1}{2} \right)\\ \left(0,1,0 \right)\end{array}$ 
& $\begin{array}{rl}\frac{1}{\sqrt{2}}e^{i (2j+1)\pi/4}\left(1,0,-1,0\right), & j=1,2,3,4\\  \frac{1}{2}(-1)^{j'} \left(0,1,0,-1 \right), & j'=1,2\end{array}$ 
& $\begin{array}{c}[1,0,1] \\{}[1,0,\bar{1}]\end{array} $ 
& 8
&
\\\cline{2-6}\vspace{1truemm}
& $\begin{array}{c}\left(0,1,0 \right)\\\left(\frac{1}{2},0,\frac{1}{2} \right)\end{array}$ 
& $\begin{array}{rl}\frac{1}{2}(-1)^{j} \left(1,0,-1,0\right), & j=1,2\\  \frac{1}{\sqrt{2}}e^{i (2j'+1)\pi/4}\left(0,1,0,-1 \right), & j'=1,2,3,4\end{array}$ 
& $\begin{array}{c}[1,0,1] \\{}[1,0,\bar{1}]\end{array} $ 
& 8
&
\\\cline{2-6}\vspace{1truemm}
& $\begin{array}{c}\left(\frac{1}{2},-\frac{1}{2},0 \right)\\ \left(0,0,1 \right)\end{array}$ 
& $\begin{array}{rl}\frac{1}{\sqrt{2}}e^{i (2j+1)\pi/4}\left(1,0,0,-1\right), & j=1,2,3,4\\  \frac{1}{2}(-1)^{j'} \left(0,1,-1,0 \right), & j'=1,2\end{array}$ 
& $\begin{array}{c}[1,1,0] \\{}[1,\bar{1},0]\end{array} $ 
& 8
& 
\\\cline{2-6}\vspace{1truemm}
& $\begin{array}{c}\left(0,0,1 \right)\\\left(\frac{1}{2},\frac{1}{2},0 \right)\end{array}$ 
& $\begin{array}{rl}\frac{1}{2}(-1)^{j} \left(1,0,0,-1\right), &  j=1,2\\  \frac{1}{\sqrt{2}}e^{i (2j'+1)\pi/4}\left(0,1,-1,0 \right), & j'=1,2,3,4\end{array}$ 
& $\begin{array}{c}[1,1,0] \\{}[1,\bar{1},0]\end{array} $ 
& 8
&
\end{tabular}
\end{ruledtabular}
\label{table:ordered.states}
\end{table*}

\subsection{Cubic antiferromagnet (CAF)}
\label{subsection:CAF}

For purely dipolar interactions ${\mathcal H}_{\sf dipolar}$ [Eq.~(\ref{eq:H.dipolar})],  
the minimum of the lowest lying (nearly-flat) band $\epsilon_{\mathbf{q}}^{\mu}$ 
lies at 
\begin{eqnarray}
{\bf Q}^{\sf CAF} 
= 
   \left(0,0,1 \right)
\label{eq:Q.min.CAF}
\end{eqnarray}
--- the X point in Fig.~\ref{fig:ordering-vectors}, 
and two other wave vectors related by cubic symmetry.


The spectrum of $\overline{\mathcal H}_{\sf DSI}$~[Eq.~(\ref{eq:Hsoftspin})] 
is doubly--degenerate at these wavevectors, with associated eigenvectors   
\begin{eqnarray}
\boldsymbol{\mathcal{E}}_{\mathbf{Q}^{\sf CAF}}^{1^\prime} &=& \frac{1}{\sqrt{2}} \left(1,0,0,-1 \right) \, ,
\nonumber\\
\boldsymbol{\mathcal{E}}_{\mathbf{Q}^{\sf CAF}}^{2^\prime} &=& \frac{1}{\sqrt{2}} \left(0,1,-1,0 \right) \, .
\label{eq.CAF.ev}
\end{eqnarray} 
These eigenvectors satisfy the hard--spin constraint Eq.~(\ref{eq:hard.spin.constraint}), 
and correspond to the CAF --- an antiferromagnet ground state with 
cubic symmetry, studied extensively by Melko {\it et al.} [\onlinecite{melko01}]. 
Competing second-neighbour exchange, $J_2 < 0$, leads a 
reduction in the bandwidth of spin-ice configurations $\Delta E$, 
as illustrated in Fig.~\ref{fig:kMFT_Bandwidth}.
However the CAF remains a mean-field ground state for 
\begin{equation}
J_{2}/D > -0.057(1) \; , 
\label{eq.MF.phase.boundary.CAF.TDQ}
\end{equation} 
where the bracket indicates the uncertainty in the final digit.


\begin{figure}[t]
\includegraphics[width=0.85\columnwidth]{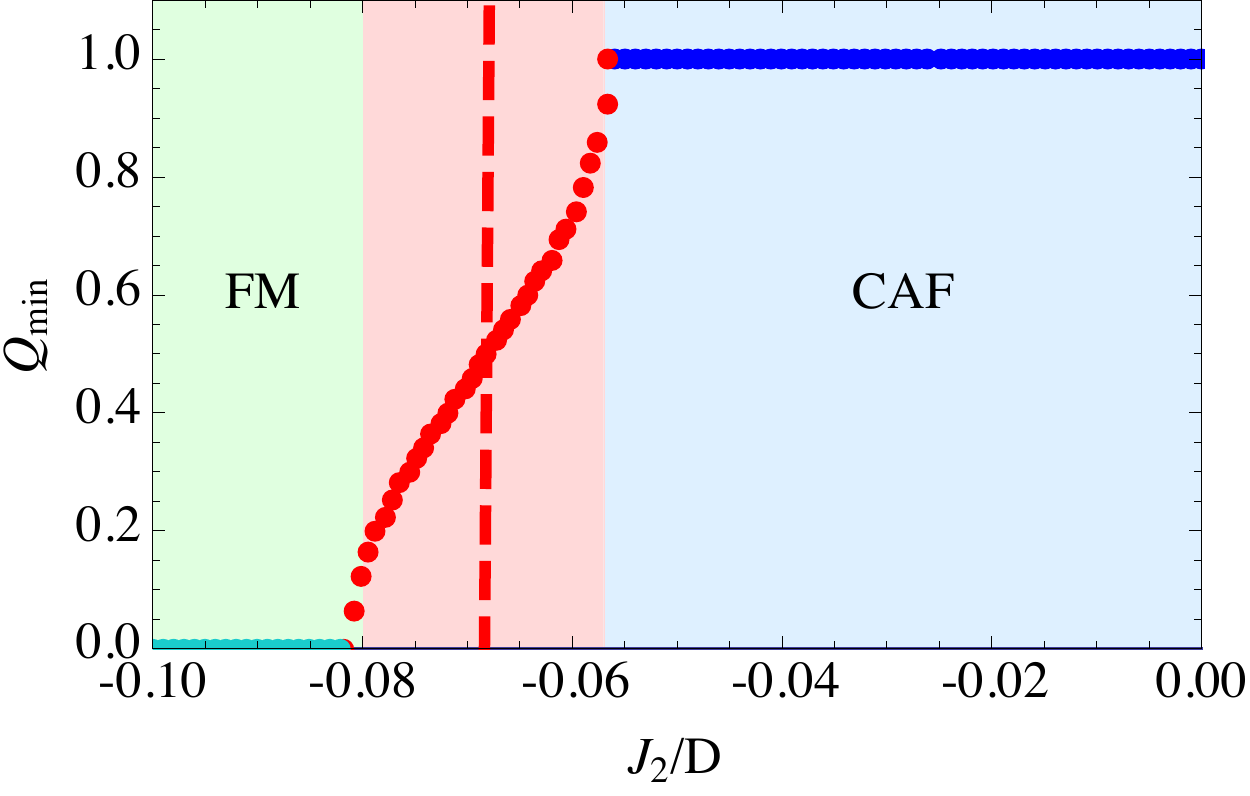}
\caption{(Color online)  
Evolution of the mean--field ordering wave vector, 
\mbox{${\bf Q}_{\sf min} = (Q,Q,0)$} 
as a function of second--neighbour exchange \mbox{$J_2 < 0$}, 
within a dipolar spin ice described by the soft--spin mean--field theory 
$\overline{\mathcal H}_{\sf DSI}$~[Eq.~(\ref{eq:Hsoftspin})].
${\bf Q}_{\sf min}$ 
interpolates smoothly from \mbox{${\bf Q}^{\sf CAF} = (1,1,0)$} 
[labelled $X$ in Fig.~(\ref{fig:ordering-vectors})],   
through incommensurate values, to 
\mbox{${\bf Q}^{\sf FM} = (0,0,0)$}, 
[labelled $\Gamma$ in Fig.~(\ref{fig:ordering-vectors})].
%
The wave vectors associated with the tetragonal double-Q (TDQ) 
state, including \mbox{${\bf Q}^{\sf TDQ} = (1/2,1/2,0)$} 
[cf.~Table~\ref{table:ordered.states}], occur for  $J_2/D =  -0.068$ 
(red dashed line), within the incommensurate region.
}
\label{fig:evolution-of-Qmin}
\end{figure}


The CAF ground state is illustrated in Fig.~\ref{fig:ordered.phases}(a).
(An equivalent animated figure is provided in the supplemental materials).
It has the same 16--site cubic unit cell as the pyrochlore lattice.
Once time--reversal symmetry is taken into account, each of the three possible 
ordering vectors ${\bf Q}^{\sf CAF}$ contributes four possible ground states, 
leading to an overall 12--fold degeneracy.


In Fig.~\ref{fig:ordered.phases}(a), the CAF is shown within a tetragonal 32--site 
cell, aligned with the $[110]$ and $[1\overline{1}0]$ axes of the lattice.
Plotted in this way, it becomes clear that the CAF is built of alternating chains
of 
spins (coloured blue and red, respectively), running parallel
to the $[110]$ and $[1\overline{1}0]$ axes (green and yellow lines, respectively), corresponding to the $\boldsymbol{\mathcal{E}}_{\mathbf{Q}^{\sf CAF}}^{1^\prime}$ and $\boldsymbol{\mathcal{E}}_{\mathbf{Q}^{\sf CAF}}^{2^\prime}$ eigenvectors [Eqs.~(\ref{eq.CAF.ev})], respectively.
Each of these chains has a net ferromagnetic polarisation.
However, the polarisation of the chains rotates between the different $[001]$ 
planes of the lattice, to give a state with no net magnetisation.

\subsection{Incommensurate states and tetragonal double--Q (TDQ) order} 
\label{subsection:TDQ}

In the intermediate parameter range 
\begin{equation}
- 0.080 < J_{2} /D < - 0.057 
\label{eq:J2.for.TDQ}
\end{equation}
the dispersion of lowest-lying eigenvalue of 
$\overline{\mathcal H}_{\sf DSI}$, Eq.~(\ref{eq:Hsoftspin}),  
evolves smoothly from a band with a minimum at 
${\bf Q}^{\sf CAF}$ [Eq.~(\ref{eq:Q.min.CAF})]
to band with a minimum at 
\begin{eqnarray}
{\bf Q}^{\sf FM} = (0,0,0) \, , 
\label{eq:Q.min.FM}
\end{eqnarray}
as illustrated in Fig.~\ref{fig:band.evolution}.
The corresponding mean-field ordering wave vector, $\mathbf{Q}_{\sf min}$, 
interpolates between ${\bf Q}^{\sf CAF}$
and ${\bf Q}^{\sf FM}$, following the path shown in Fig.~\ref{fig:ordering-vectors}.


In general, the eigenvectors $\boldsymbol{\mathcal{E}}_{\mathbf{Q}_{\sf min}}^\mu$ in
this range of $J_2/D$ [Eq.~(\ref{eq:J2.for.TDQ})] do {\it not} satisfy the hard--spin 
constraint Eq.~(\ref{eq:hard.spin.constraint}).   
However a special case, occurring for 
\begin{eqnarray}
J_2/ D = - 0.068(1) \; , 
\end{eqnarray}
is the commensurate wavevector 
\begin{eqnarray}
{\bf Q}^{\sf TDQ} 
&=& 
\left( \frac{1}{2}, \frac{1}{2}, 0 \right) \; .
\label{eq:QTDQ}
\end{eqnarray}
In this case, it {\it is} possible to construct to linear combinations 
of pairs of the 12 eigenvectors $\boldsymbol{\mathcal{E}}_{\mathbf{Q}^{\sf TDQ}}^\mu$, 
listed in Table~\ref{table:ordered.states},  which {\it do} satisfy the hard-spin constraint.   
These correspond to the 48--fold degenerate, tetragonal, double--Q state (TDQ) 
illustrated in Fig.~\ref{fig:ordered.phases}(b).
(An equivalent animated figure is provided in the supplemental materials).
  

Close examination of Fig.~\ref{fig:ordered.phases}(b) reveals that the TDQ state, 
like the CAF, is built of alternating chains of 
spins, running parallel to the $[110]$  and $[1\overline{1}0]$ axes.  
Each alternating chain has a net ferromagnetic polarisation.
However the sense of this polarisation alternates between neighbouring chains, 
to give a state with no net magnetisation.


To rule out the possibility of other mean--field ground states in this 
parameter range, we have carried out a search of all possible 
multiple--Q states of the form 
\begin{eqnarray}  
\Psi^a (\mathbf{r}_{i,a}) 
   &=& \sum_{\eta}  
\left[  
z^a_\eta e^{ i \mathbf{q}_{\sf min}^\eta \cdot \mathbf{r}_{i,a} } 
+ \text{c.c.}
 \right] \; .
\label{eq:TL-order_parameter}
\end{eqnarray}
where $z^a_\eta$ 
is a 4--component vector proportional to  
$\boldsymbol{\mathcal{E}}^\mu_{\mathbf{Q}_{\sf min}^\eta}$,
and the sum $\sum_{\eta}$ runs over the six distinct mean--field 
ordering wave vectors 
given in Table~\ref{table:ordered.states}.
We find that the only solutions $\Psi^a (\mathbf{r}_{i,a})$
which satisfy the hard--spin constraint [Eq.~(\ref{eq:hard.spin.constraint})], 
are those corresponding to the TDQ states.

\subsection{Ferromagnet (FM)}
\label{subsection:FM}

Finally, for parameters 
\begin{eqnarray} 
J_{2}/D < -0.080(1) 
\label{eq.MF.phase.boundary.TDQ.FM}
\end{eqnarray}  
we find $\mathbf{Q}_{\sf min}$ equal to ${\bf Q}^{\sf FM}$ [Eq.~(\ref{eq:Q.min.FM})], 
and eigenvectors have unique solutions of the simple ``two--in, two--out'' form 
\begin{eqnarray}  
\boldsymbol{\mathcal{E}}_{\mathbf{Q}^{\sf FM}}^{\mu_{\sf min}} &=& \frac{1}{2} \left(1,1,-1,-1 \right) \; .
\end{eqnarray}  
There are three such eigenvectors up to time reversal and they are listed in Table~\ref{table:ordered.states}. 
These eigenvectors trivially satisfy the hard-spin constraint 
Eq.~(\ref{eq:hard.spin.constraint}), and correspond to a simple 
ferromagnet (FM) in which all tetrahedra have the same spin 
configuration.
Since there are six possible ``two--in, two--out'' spin configurations
for a single tetrahedron, the FM is six--fold degenerate.


The FM state is illustrated in Fig.~\ref{fig:ordered.phases}(c).
(An equivalent animated figure is provided in the supplemental materials).
Once again, the FM can be seen to be built of alternating chains of 
spins, running parallel to the $[110]$ and $[1\overline{1}0]$ axes.  
However, unlike the CAF or TDQ state, all chains parallel to $[110]$ or $[1\overline{1}0]$ 
have the same polarisation, and as a result the FM has a net magnetization parallel 
to the $[100]$ axis.

\section{Mapping to an effective triangular--lattice Ising model}
\label{section:effective.Ising.model}

The mean--field treatment of dipolar spin ice, 
developed in Sec.~\ref{sec:classical.mean.field.theory}, reveals three different 
ordered ground states as a function of second neighbour exchange $J_2 < 0$ ---
a cubic antiferromagnet (CAF), a tetragonal double--Q (TDQ) state, and 
a cubic ferromagnet (FM).
These three ordered states have a striking common feature --- they are all built 
of alternating chains of 
spins.


Numerical simulations, described in Sec.~\ref{section:classical.monte.carlo},  
confirm that the CAF, TDQ and FM states are indeed the classical ground states of 
$\mathcal{H_{\sf DSI}}$~[Eq.~(\ref{eq:Hdsi})] for $ J_2 < 0$.
However neither these simulations, nor the mean--field theory, explain why 
ordered ground states should be built of alternating chains of spins.
Moreover, the fact that three different ground states are found within such a small 
range of $J_2/D$ [cf. Fig.~\ref{fig:evolution-of-Qmin}] suggests that ground state order 
might also be very sensitive to third neighbour exchanges $J_{3c}$ and $J_{3d}$, 
not treated in Sec.~\ref{sec:classical.mean.field.theory}.


Taken together, these results suggest that a new ordering principle is at 
work in dipolar spin ice at low temperatures.
In what follows we identify this ordering principle, showing how long--range dipolar 
interactions between alternating chains of spins can be described by an 
effective Ising model on an anisotropic triangular lattice, with only weak, 
short--ranged interactions.
The extreme sensitivity of the ground state dipolar spin--ice to competing 
exchange interactions is shown to follow from the exponential--screening 
of dipolar interactions within such ``chain states''.


We develop, below, the classical, ground-state phase diagram of this Ising model, and show 
how it can be used to determine the  ordered phases of a dipolar spin ice 
with competing further-neighbour exchange. 

\subsection{Effective Ising model} 
\label{subsection:madlung.sum}

Spin ice is not the only material where long--range interactions arise
within an ice--like manifold of states.
Another example, famously studied by Anderson 
is the charged ordered system magnetite, Fe$_3$O$_4$.
In a seminal paper,~\cite{anderson56} Anderson argued that Fe$^{2+}$ 
and Fe$^{3+}$ ions, occupying the sites of a pyrochlore lattice in magnetite, 
could be equated with the hydrogen bonds in water ice.
The tendency to charge order means that they are subject to the same 
``ice rule'', namely that there should be exactly two Fe$^{2+}$ and 
two Fe$^{3+}$ in every tetrahedron in the lattice.
The degeneracy of these ice--like, locally charge--ordered states is lifted
by long--range Coulomb interactions between the Fe$^{2+}$ and Fe$^{3+}$ 
ions.~\cite{anderson56,mcclarty14}
At first sight, evaluating the effect of these long--range interactions 
is a very challenging problem.   
However, as Anderson realised, the particular geometry of pyrochlore 
lattice leads to a significant simplification.


The pyrochlore lattice can be broken down into sites on two sets of 
chains, running parallel to $[110]$ and $[1\overline{1}0]$, with a tetrahedron
at every point where two perpendicular chains cross.
States satisfying the ``ice rule''
can be constructed by populating these chains with alternating 
Fe$^{2+}$ and Fe$^{3+}$ ions.
These chains of alternating charges are charge--neutral objects 
(relative to the average valence of Fe$^{2.5+}$), and so interact
only weakly.
%
Moreover, it follows from the symmetry of the lattice that interactions
between perpendicular chains vanish.
What remains are two, independent, sets of weakly--interacting chains,
whose low--energy states can be described by an Ising variable on a triangular lattice.
The two states of the Ising variable stand for the two possible states of the ferromagnetic chains.


\begin{figure}[tb]
\subfloat[\label{fig:chain}]{\includegraphics[width=0.7\columnwidth]{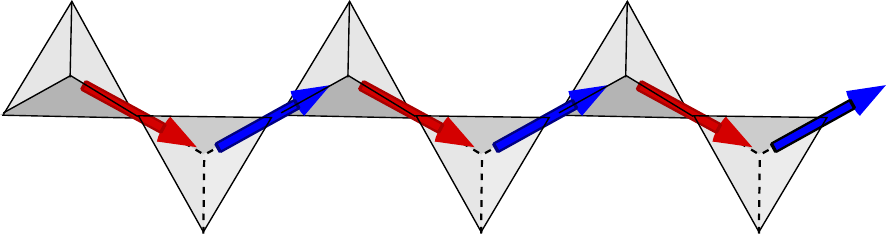}} \\
\subfloat[\label{fig:FM-to-ANNNI}]{\includegraphics[width=0.7\columnwidth]{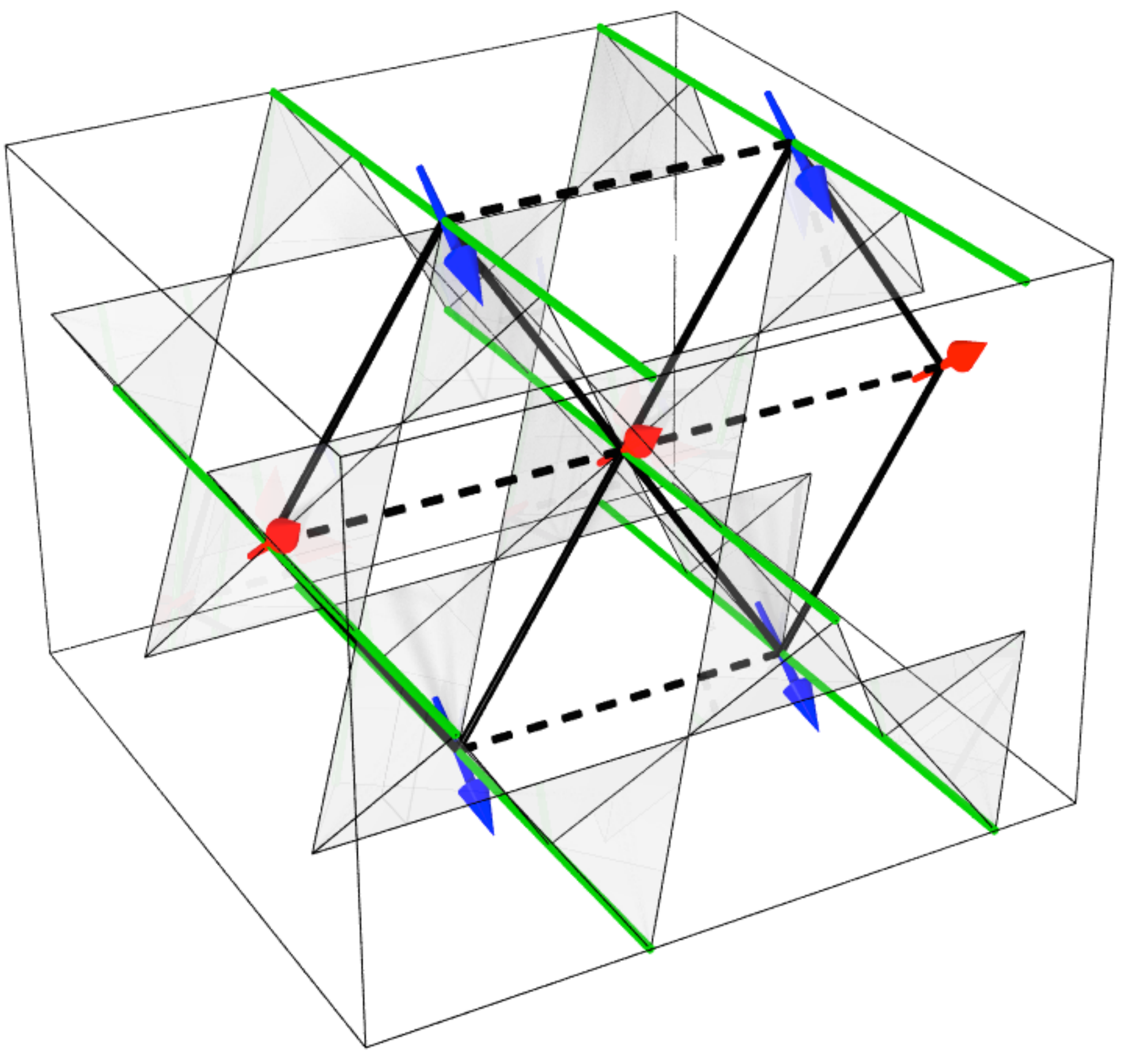}} \\ 
\caption{(Color online) 
Mapping  from spin-ice states to the anisotropic Ising model on a triangular lattice.  
(a) Alternating chain of 
spins, with net ferromagnetic magnetisation.  
(b) Pyrochlore lattice, showing how chains parallel to $[110]$ form a triangular lattice. 
}
\label{fig:Ising-model-mapping}
\end{figure}


All of the same considerations apply in spin-ice, where spins interact 
through long--range dipolar interactions, and the alternating charges 
are replaced by alternating 
spins (blue and red arrows in Fig.~\ref{fig:Ising-model-mapping}), to form
ferromagnetic chains.
We consider the two sets of chains parallel to the $[1,1,0]$ and 
$[1,\overline{1},0]$ directions. 
In units of 
\begin{eqnarray}
r_1^\prime = \frac{a}{4} = \frac{r_1}{\sqrt{2}} 
\label{eq:r1.prime}
\end{eqnarray}
[cf.~Eq.~(\ref{eq:r1})], the coordinates of the spins on these chains 
are given by 
\begin{eqnarray}
&(n+l,-n+l, 2m)      &\quad (\text{chain} \parallel [1,1,0]) \, , \nonumber\\
&(n+1+l,n-l,2m+1)  &\quad (\text{chain} \parallel [1,\overline{1},0]) \, ,
\end{eqnarray}
where 
\begin{eqnarray}
l = -\infty  \ldots -2, -1, 0, 1, 2, \ldots \infty  
\end{eqnarray}
counts the spins on a given chain, while the integers 
$m$ and $n$ (such that $m+n$ is even) determine the chain in question and at the same time, for $l=0$, define the sites of an anisotropic triangular lattice 
(solid and dashed black lines in Fig.~\ref{fig:Ising-model-mapping}), 
with coordinates 
\begin{eqnarray}
\boldsymbol{\delta} &=& (\delta_1,\delta_2) 
   \equiv (\delta_1,-\delta_1,\sqrt{2} \delta_2) 
   = (n,-n,2m) \nonumber\\
   && \qquad  \qquad \qquad \qquad (\text{chain} \parallel [1,1,0]) \, .
\label{eq:triangular.lattice.of.chains}
\end{eqnarray}
In units of $r_1$ [Eq.~(\ref{eq:r1})], projecting onto a $(1,1,0)$ plane, 
these correspond to a lattice with primitive lattice vectors 
\begin{eqnarray}
\mathbf{a} = (2,0) \quad , \quad \mathbf{b} = (1,\sqrt{2}) \; .
\end{eqnarray}
An exactly equivalent triangular lattice can be assigned to 
chains parallel to $[1,\overline{1},0]$.  
We note that the local easy--axis of the spins within 
each of these triangular lattices points in one of two directions, 
and is the same for all even (odd) $n$ --- cf. Fig.~\ref{fig:Ising-model-mapping}.


Following Anderson,~\cite{anderson56} we now consider the specific
case of states composed of alternating chains of 
spins, running parallel to $[1,1,0]$ and $[1,\overline{1},0]$, with net ferromagnetic polarisation.
Such states automatically satisfy the ``ice rules'', and so are candidates
as ground states in a spin ice.
Moreover, dipolar interactions between orthogonal ferromagnetic spin--chains 
vanish by symmetry (by analogy to the charge problem mentioned above), while interactions between parallel chains 
can be described by an effective Ising model
\begin{eqnarray}
{\mathcal H}^{\sf 2D}_{\sf Ising} 
&=& \frac{1}{2} \sum_{\boldsymbol{\rho},\boldsymbol{\delta}} 
   K_{\boldsymbol{\delta}} 
       \, \sigma_{\boldsymbol{\rho}}  \ \sigma_{\boldsymbol{\rho}+\boldsymbol{\delta}} \,,
\label{eq:HIsing}
\end{eqnarray}
where the sum $\sum_{\boldsymbol{\rho},\boldsymbol{\delta}}$ runs over all 
pairs of sites within the triangular lattice defined by 
Eq.~(\ref{eq:triangular.lattice.of.chains}), and 
\begin{eqnarray}
\sigma_{\boldsymbol{\rho}} \equiv 2 \sf \mathsf{S}^z_{\boldsymbol{\rho}} = \pm 1 \; ,
\end{eqnarray}
is the Ising variable characterising the state of a given ferromagnetic chain.  


What remains is to determine the strength of the interaction 
$K_{\boldsymbol{\delta}}$ [Eq.~(\ref{eq:HIsing})]
between parallel chains. 
These will have contributions from both long--range dipolar interaction 
${\mathcal H}_{\sf dipolar}$~[Eq.~(\ref{eq:H.dipolar})], and exchange 
interactions ${\mathcal H}_{\sf exchange}$~[Eq.~(\ref{eq:H.exchange})].    
Just as in the problem of charge--order,~\cite{anderson56} the contribution of the long range dipolar interactions can be calculated
through a Madelung sum.
We start by considering the dipolar interaction between a test spin 
at $\boldsymbol{\delta} = (0,0)$ and a chain $\parallel [1,1,0]$, at position 
$\boldsymbol{\delta} = (\delta_1,\delta_2)$
\begin{align}
K_{\boldsymbol{\delta}} = 
 \frac{D}{3} \sum_{l=-\infty}^{\infty} 
 &\left[ 
 (-1)^l  
  \frac{\left(\delta_1^2 -2\delta_2^2 + l^2\right)}
  {\left(\delta_1^2 + \delta_2^2 + l^2\right)^{5/2}} 
\right. 
\nonumber\\  &\quad  
\left. 
  + 2 (-1)^{\delta_1} 
  \frac{\left(\delta_1^2 + \delta_2^2 -2 l^2 \right)}
  {\left(\delta_1^2 + \delta_2^2 + l^2\right)^{5/2}} 
  \right] \, , 
\label{eq:K.delta.as.sum}  
\end{align}
where the coordinates of the sites on the chain are given by 
\begin{align}
(n+l,-n+l, 2m) = (\delta_1+l,\delta_1-l ,\sqrt{2} \delta_2) \; , \nonumber
\end{align}
The term with alternating sign comes from the alternating spin components perpendicular 
to the chain, while the uniform term comes from the spin components parallel to the chain.


\begin{table}[t]
\caption{
Interactions $K_{\boldsymbol{\delta}}$  of the extended Ising model 
${\mathcal H}^{\sf 2D}_{\sf Ising}$~[Eq.~(\ref{eq:HIsing})], written
in terms of the microscopic parameters of 
$\mathcal{H_{\sf DSI}} $~[Eq.~(\ref{eq:Hdsi})].
The contribution of the long-range dipolar interactions 
$\mathcal{H}_{\sf dipolar}$~[Eq.~(\ref{eq:H.dipolar})] 
shows exponential decay as a function of distance $|{\boldsymbol{\delta}}|$. 
}
\centering
\begin{ruledtabular}
\begin{tabular}{l c l c}
$K_{\boldsymbol{\delta}}$  & 
$|\boldsymbol{\delta}|$ \quad & 
\qquad  ${\mathcal H}_{\sf dipolar}$ & 
\qquad  ${\mathcal H}_{\sf exchange} $ \\ [0.5ex] 
\hline\vspace{-0.25cm}\\
$K_{(1, \sqrt{2})}$ & $\sqrt{3}$ & $ -0.0227426 D $ & $- J_2/3 - J_{3c} - J_{3d}$  \\
$K_{(2,0)}$ & $2$ & $ \phantom{-}0.0021957 D$ & $ J_{3d}$  \\ [1ex]
$K_{(0,2\sqrt{2})}$  & $2\sqrt{2}$ & $-0.0008443 D$ & \\
$K_{(3,\sqrt{2})}$ & $\sqrt{11} $ & $-0.0000178 D$ & \\ 
$K_{(2,-2\sqrt{2})}$  & $2\sqrt{3}$ & $-0.0000649 D$ & \\ 
$K_{(1,3\sqrt{2})}$ & $\sqrt{19}$ & $-0.0000051 D$ & \\
$K_{(4,0)}$  & $4$ & $ \phantom{-}0.0000013 D $\\
\end{tabular}
\end{ruledtabular}
\label{table:Kdelta}
\end{table}


Evaluating the sum in Eq.~(\ref{eq:K.delta.as.sum}) numerically, we find 
that the  interchain couplings $K_{\boldsymbol{\delta}}$ are very small, 
and decay very rapidly, with the first few interactions given by 
\begin{align}
K_{(1, \sqrt{2})} &= -0.0227 D  \; ,
 \nonumber\\
\label{eq:Kdelta}
K_{(2,0)} &=  \phantom{-}0.0022 D \; ,
\\
K_{(0,2\sqrt{2})} &=   -0.0008 D \nonumber \; .
\end{align}
Interactions 
up to 7$^\text{th}$--neighbour, including the contribution of ${\mathcal H}_{\sf exchange}$ 
[Eq.~(\ref{eq:H.exchange})], are listed in Table~\ref{table:Kdelta}.

\begin{figure}[tb]
\includegraphics[width=0.9\columnwidth]{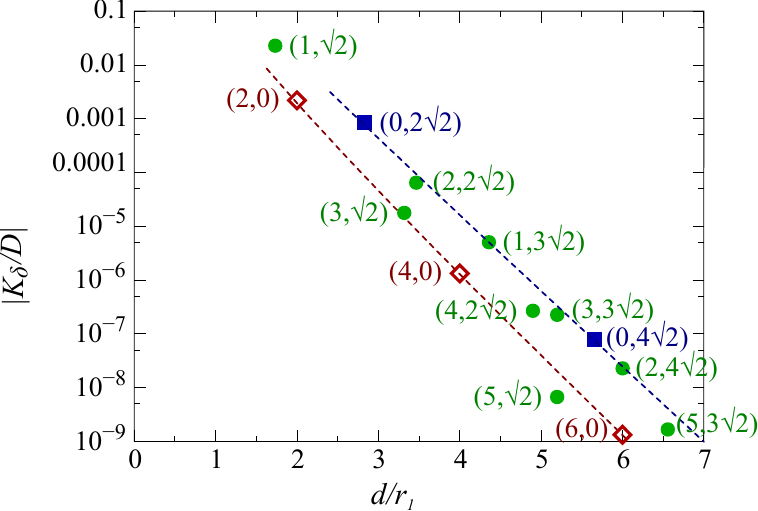}
\caption{(Color online) 
Exponential decay of dipolar contributions to the interchain interaction 
$K_{\boldsymbol{\delta}}$, as a function of the distance between the 
chains $d/r_1=\sqrt{\delta_1^2+\delta_2^2}$, where 
$\boldsymbol{\delta} = (\delta_1,\delta_2)$, and $r_1$ is defined by Eq.~(\ref{eq:r1}).  
Interactions are measured in units of $D$ [cf. Eq.~(\ref{eq:D})].  
The family of exchanges $K_{(0,\delta_2)}$ are shown with 
blue squares, while those for $K_{(\delta_1,0)}$ are 
shown with red diamonds.  
Dashed lines of the same color show the corresponding asymptotic 
expressions Eqs.~(\ref{eq:as_k_0_delta}) and (\ref{eq:as_k_delta_0}). 
Exchanges for general $(\delta_1,\delta_2)$ are plotted as green circles.
Interactions denoted with solid symbols are ferrromagnetic ($K_{\boldsymbol{\delta}} < 0$); 
interactions denoted with open symbols are antiferrromagnetic ($K_{\boldsymbol{\delta}} > 0$). 
\label{fig:Kdecay}
}
\end{figure}

In fact, $K_{\boldsymbol{\delta}}$ decays {\it exponentially}
with distance, as can be seen from  Fig.~\ref{fig:Kdecay}, 
where interactions are plotted for the two main lattice 
directions, $(0,\delta_2)$ and $(\delta_1,0)$.
The origin of this exponential decay lies in the alternation of the spins, 
and can be understood by converting the sums on $l$ in 
Eq.~(\ref{eq:K.delta.as.sum}) into integrals, 
using Fourier representations of the Dirac delta function~: 
\begin{subequations}
\begin{align}
\sum_{l=-\infty}^{\infty} f(l)
&=
\sum_{q=-\infty}^\infty 
\int\limits_{-\infty}^{\infty} d l  \;
f(l) \cos  2q l \pi \;,
\\
 \sum_{l=-\infty}^{\infty} 
 (-1)^l   f(l)
&=
\sum_{q=-\infty}^\infty 
\int\limits_{-\infty}^{\infty} d l \; 
f(l) \cos (2q+1)l \pi \;.
\end{align}
\end{subequations}
Doing so, we obtain 
\begin{align}
K_{\boldsymbol{\delta}} = & \frac{D}{3} 
\sum_{q=-\infty}^\infty 
\int\limits_{-\infty}^{\infty} d l 
\left[ 
 \frac{\left(\delta_1^2 -2\delta_2^2 + l^2\right)}
{\left(\delta_1^2 + \delta_2^2 + l^2\right)^{5/2}}  \cos (2q+1)l \pi
\right.\nonumber\\&\left.
+ 2(-1)^{\delta_1}
\frac{\left(\delta_1^2 + \delta_2^2 -2 l^2 \right)}
{\left(\delta_1^2 + \delta_2^2 + l^2\right)^{5/2}} \cos 2q l \pi
\right]\, .
\label{eq:intermediate.result}
\end{align}
The leading contribution to $K_{\boldsymbol{\delta}}$ 
comes from the first term in Eq.~(\ref{eq:intermediate.result}) with $q=-1$ and $q=0$, which 
decays exponentially with distance~:
\begin{align}
\frac{K_{\boldsymbol{\delta}}}{D} \approx & \frac{2}{3} \int_{-\infty}^{\infty} d l  \,
 \frac{\left(\delta_1^2 - 2\delta_2^2 + l^2\right)}
{\left(\delta_1^2 + \delta_2^2 + l^2\right)^{5/2}}  
\cos \pi l 
\nonumber\\ =& 
\frac{4 \pi}{3 \delta} K_1(\pi\delta) - \frac{4 \pi^2 \delta_2^2}{3 \delta^2} K_2(\pi\delta)
\nonumber\\
 \approx &  -\frac{2 \sqrt{2}}{3} \left[
   \pi^2 \left(\frac{\delta_2}{\delta}\right)^2 \delta^{-1/2}
  - \pi \delta^{-3/2}
  + \cdots
   \right] e^{-\pi \delta} \; , 
  \label{eq:sup_dip_decay}
\end{align}
where $K_1(x)$ and $K_2(x)$ are modified Bessel functions 
of the second kind and 
\begin{align}
\delta=\sqrt{\delta_1^2 + \delta_2^2} \; . 
\end{align}
The neglected integrals decay as $e^{-2\pi\delta}$ or faster with the distance (more precisely, the integral with $\cos p l \pi $ decays as $e^{-p \pi \delta }$). 


\begin{figure*}[t]
\includegraphics[width=0.99\columnwidth]{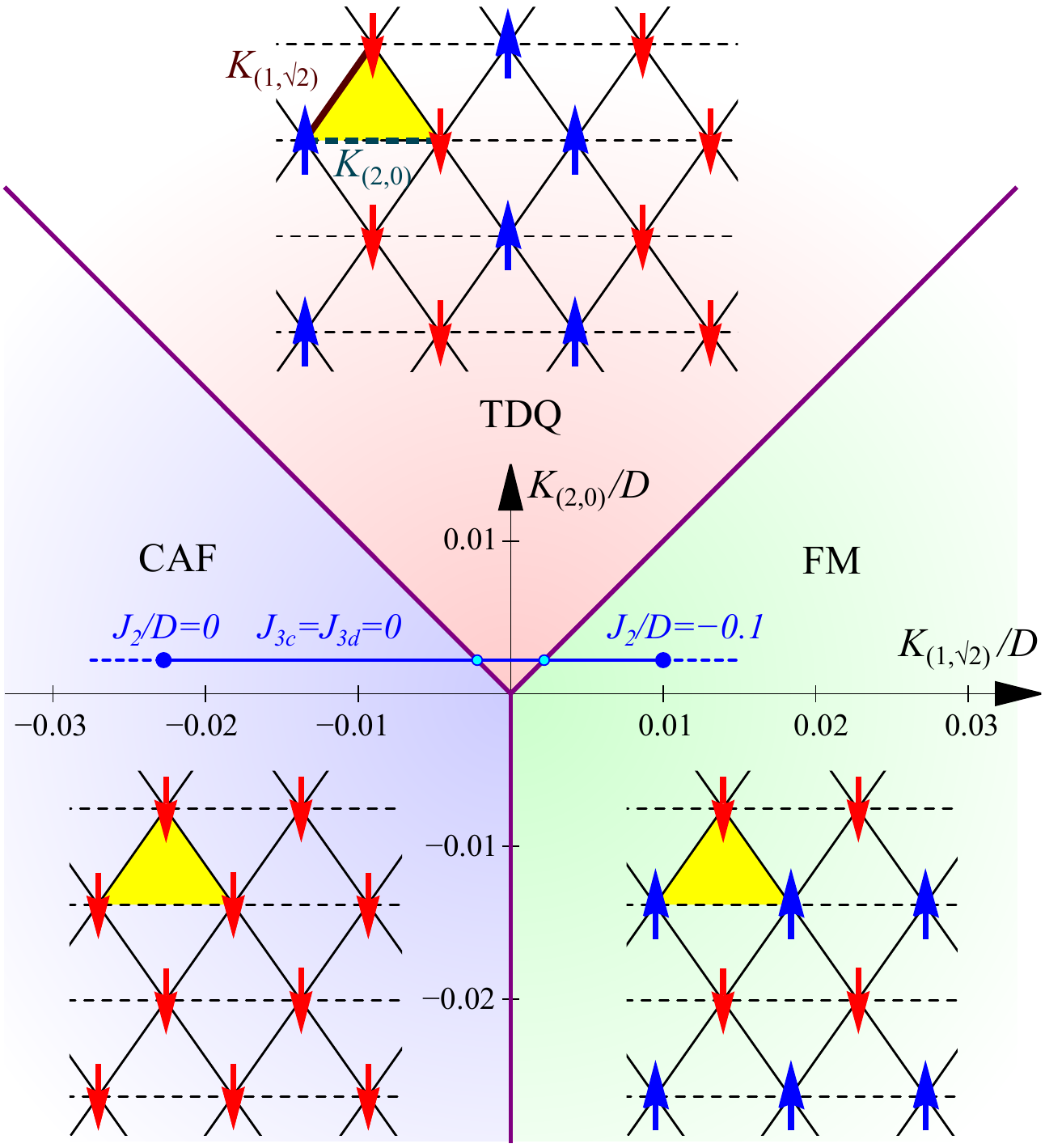}
\caption{
Phase diagram of the extended Ising model 
${\mathcal H}^{\sf 2D}_{\sf Ising}$~[Eq.~(\ref{eq:HIsing})], 
as a function of the leading interchain interactions $K_{(1, \sqrt{2})}$ and $K_{(2,0)}$.  
In this model, each Ising spin $\sigma_{{\bf r}}$ corresponds to a chain of alternating 
spins in a dipolar spin--ice described by $\mathcal{H_{\sf DSI}}$ 
[Eq.~(\ref{eq:Hdsi})], with parameters $K_{\boldsymbol{\delta}}$ given 
in Table~\ref{table:Kdelta}.  
Three ordered ground states are found, a cubic antiferromagnet (CAF), 
a tetragonal double--Q state (TDQ) and a ferromagnet (FM), illustrated 
in Fig.~\ref{fig:ordered.phases}.
The parameters estimated by Ya'vorskii {\it et al.} [\onlinecite{yavorskii08}]
place Dy$_2$Ti$_2$O$_7$ in the CAF phase.  
The parameters considered in soft--spin mean--field theory 
[Sec.~\ref{sec:classical.mean.field.theory}], classical  Monte Carlo simulation 
[Sec.~\ref{section:classical.monte.carlo}], and quantum Monte Carlo simulation 
[Sec.~\ref{section:QMC}], are shown with blue line.
}
\label{fig:Ising-phase-diagram}
\end{figure*}


It follows that the assymptotic form of $K_{\boldsymbol{\delta}}$ at large distances 
is given by 
\begin{eqnarray} 
K_{(0,\delta_2)}/D  &\approx& - \frac{\pi^2}{3} 2^{3/2} \delta_2^{-1/2}
 e^{-\pi \delta_2} \; , 
 \label{eq:as_k_0_delta} \\
K_{(\delta_1,0)}/D  &\approx& \phantom{-} \frac{\pi}{3} 2^{3/2} 
\delta_1^{-3/2} e^{- \pi \delta_1} \,.
\label{eq:as_k_delta_0} 
\end{eqnarray}
These functions are plotted as dashed lines in Fig.~\ref{fig:Kdecay}.

\subsection{Ground--state phase diagram}
\label{subsection:Ising.phase.diagram}

Finding the ground state of the dipolar spin--ice model, 
$\mathcal{H_{\sf DSI}}$~[Eq.~(\ref{eq:Hdsi})], is a daunting 
task, combining the geometric frustration of the pyrochlore lattice, 
with long--range interactions and competing exchanges.~\cite{siddharthan-arXiv,melko01,yavorskii08}
In contrast, finding the ground state of the effective two--dimensional Ising model 
${\mathcal H}^{\sf 2D}_{\sf Ising}$~[Eq.~(\ref{eq:HIsing})], describing chain states, 
is relatively easy.
In this case, all interactions are short--ranged, and the frustration of the triangular lattice 
\cite{Wannier1950} is lifted by the anisotropy of the leading interactions, 
$K_{(1,\sqrt{2})}$ and $K_{(2,0)}$.  
However since dipolar interactions are suppressed by two orders of magnitude within 
chain state --- cf. Table~\ref{table:Kdelta} --- the behaviour of the model is {\it very} 
sensitive to competing exchange.


Since nearest--neighbour interactions dominate, the ground---state phase diagram of 
${\mathcal H}^{\sf 2D}_{\sf Ising}$~[Eq.~(\ref{eq:HIsing})] can be found by 
examining spin--configurations on the elementary unit of the lattice, a triangle.
The resulting phase diagram is shown in Fig.~\ref{fig:Ising-phase-diagram}, 
with the parameter set considered in Sec.~\ref{sec:classical.mean.field.theory}
shown as a blue line.   
This phase diagram contains the same three ordered ``chain states'' as are found 
in mean--field theory \mbox{[cf. Table~\ref{table:ordered.states}]}~:
\begin{enumerate}

\item  A cubic antiferromagnet (CAF), 
with energy per triangle
\begin{eqnarray} 
E^{\sf CAF}_\triangle = 2 K_{(1,\sqrt{2})} + K_{(2,0)} \; .
\end{eqnarray}

\item A tetragonal, double-q state (TDQ) with energy per triangle
\begin{eqnarray} 
E^{\sf TDQ}_\triangle = -K_{(2,0)} \; .
\end{eqnarray}

\item A cubic ferromagnet (FM) with energy per triangle
\begin{eqnarray} 
E^{\sf FM}_\triangle  = -2 K_{(1,\sqrt{2})}+K_{(2,0)} \;.
\end{eqnarray}
%

\end{enumerate}
While the CAF and FM are selected uniquely by the nearest--neighbour interactions 
$K_{(1,\sqrt{2})}$ and $K_{(2,0)}$, the TDQ state is selected from a larger family of 
degenerate states by ferromagnetic $K_{(0,2\sqrt{2})}$ [cf. Table~\ref{table:Kdelta}].


The effective Ising model ${\mathcal H}^{\sf 2D}_{\sf Ising}$~Eq.~(\ref{eq:HIsing}), 
has much in common with the anisotropic next--nearest nieghbour Ising (ANNNI) model,
famous for supporting a ``Devil's staircase'' of ordered states.\cite{bak82, selke88} 
And while the ground state phase diagram, Fig.~\ref{fig:Ising-phase-diagram}, 
is dominated by three ordered states, additional degeneracies 
arise on the boundaries between the CAF and the TDQ state, 
\begin{eqnarray}
K_{(2,0)} = -K_{(1,\sqrt{2})} > 0 \; ,
\end{eqnarray}
and on the boundary between the TDQ state and the FM, 
\begin{eqnarray}
K_{(2,0)}=K_{(1,\sqrt{2})}> 0 \; .
\end{eqnarray}
An example of one these degenerate ground states is the orthorhombic ``zig--zag'' state 
(OZZ) shown in Fig.~\ref{fig:ordered.phases}(d), which is 
found on the boundary between the CAF and the TDQ state.
Overall, these additional degeneracies are essentially the same as those found 
in the Ising model on an anisotropic triangular lattice.~\cite{dublenych13}


\begin{figure}[tb]
\includegraphics[width=0.95\columnwidth]{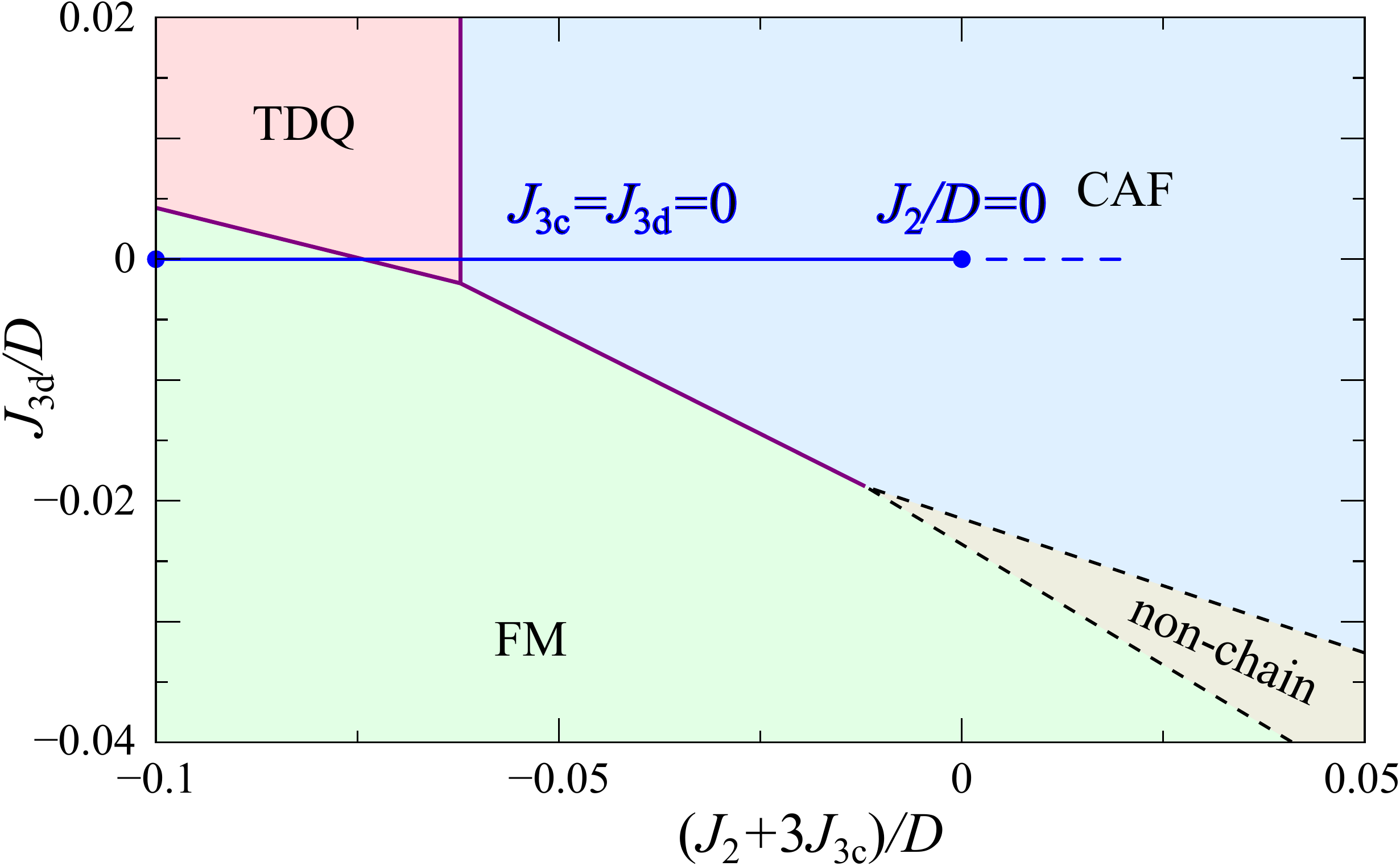}
\caption{(Color online) 
Classical ground-state phase diagram of a dipolar spin ice described by 
$\mathcal{H_{\sf DSI}}$~[Eq.~(\ref{eq:Hdsi})], showing the breakdown
of the chain picture of sufficiently strong, ferromagnetic third--neighbour
interactions $J_{3d}$.  
The TDQ, CAF and FM states are composed of ``chains'' 
of ferromagnetically polarised spins.
For $J_{3d} < -0.018 D$, the FM and CAF phases are separated by a small
region of non-chain states. 
Results are taken from a zero-temperature Monte-Carlo search 
of spin-ice configurations, for a cubic cluster of 128 sites. 
The parameters considered in soft--spin mean--field theory 
[Sec.~\ref{sec:classical.mean.field.theory}], classical  Monte Carlo simulation 
[Sec.~\ref{section:classical.monte.carlo}], and quantum Monte Carlo simulation 
[Sec.~\ref{section:QMC}], are indicated with a blue line.
\label{fig:phase-diagram-J3d}
}
\end{figure}


For purpose of comparison with the mean--field theory of 
$\mathcal{H_{\sf DSI}}$~[Eq.~(\ref{eq:Hdsi})] developed in 
Sec~\ref{sec:classical.mean.field.theory}, and the numerical simulations 
described in Sec.~\ref{section:classical.monte.carlo} and Sec.~\ref{section:QMC}, 
it is interesting to express the phase boundaries found 
from ${\mathcal H}^{\sf 2D}_{\sf Ising}$~[Eq.~(\ref{eq:HIsing})] in terms of 
second--neighbour exchange~$J_2$, setting all $J_{k \ne 2} \equiv 0$.
Taking into account all $K_{\boldsymbol{\delta}}$ up to 7$^{th}$--nieghbour 
[cf. Table~\ref{table:Kdelta}],we find that the transition between the CAF 
and TDQ occurs for
\begin{equation}
   J_2/D = -0.0621  \; ,
   \label{eq:J2.CAF.TDQ}
 \end{equation}
while the transition between the TDQ and the FM occurs for
 \begin{equation}
   J_2/D = -0.0745 \; .
   \label{eq:J2.TDQ.FM}
 \end{equation}
These results are consistent with the results of 
classical Monte Carlo simulation, described in
Sec.~\ref{section:classical.monte.carlo}, 
and in excellent agreement with the numerical values 
from zero--temperature quantum Monte Carlo simulation, 
described in Sec.~\ref{section:QMC}, below.
Mean field theory, on the other hand, is seen to over-estimate the 
stability of the TDQ phase, giving values of 
\mbox{$J_2/D = -0.57$}~[Eq.~(\ref{eq.MF.phase.boundary.CAF.TDQ})] 
and \mbox{$J_2/D = -0.80$}~[Eq.~(\ref{eq.MF.phase.boundary.TDQ.FM})].


In the light of the recent experiments by Pomaranski  {\it et al.}~[\onlinecite{pomaranski13}], 
it is also interesting to ask what interchain couplings might arise in the spin--ice Dy$_2$Ti$_2$O$_7$.
Taking values for exchange and dipolar interactions from Yavorskii {\it et al.}
[\onlinecite{yavorskii08}], we find
\begin{eqnarray}
K_{(1,\sqrt{2})}/D &=& -0.025   \,, \nonumber\\
K_{(2, 0)}/D &=& \phantom{-}0.020  \,, \qquad [\text{Dy$_2$Ti$_2$O$_7$}] \\
K_{(0,2\sqrt{2})}/D &=& \phantom{-}0.001  \,. \nonumber
\end{eqnarray}
These parameters suggest that the classical ground state of Dy$_2$Ti$_2$O$_7$ 
would be a CAF --- cf. Fig.~\ref{fig:Ising-phase-diagram}.
We return to this point in Section~\ref{section:application.to.real.materials}, below.

\subsection{Breakdown of the of chain--state picture}
\label{subsection:sanity.check}

``Chain states''' provide an extremely efficient way of minimising dipolar interactions
${\mathcal H}_{\sf dipolar}$ [Eq.~(\ref{eq:H.dipolar})], but do not necessarily minimise
the exchange interactions ${\mathcal H}_{\sf exchange}$ [Eq.~(\ref{eq:H.exchange})].
Given this, it is natural to ask how strong competing exchange interactions need to be 
to invalidate the ``chain picture''.
This proves to be a somewhat subtle question.


Exchange interactions up to third neighbour (cf.~Fig.~\ref{fig:j2j3cj3dexchanges}) 
can be grouped in three classes.
First--neighbour interactions $J_1$ help determine the stability of the spin--ice
manifold, but play no role in selecting an ordered ground state.
Second--neighbour interactions $J_2$ can be combined with third--neighbour 
interactions $J_{3c}$ [see Appendix~\ref{appendix:equivalence.J2.and.J3c}], 
to give a combined interaction $J_2 + 3 J_{3c}$.
This combined interaction selects between different chain states, and does
not by itself lead to any breakdown of the chain picture.
Third--neighbour interactions $J_{3d}$ also selects between different 
chain states, but can also lead to a breakdown of the chain picture if 
ferromagnetic, and sufficiently strong.


To asses the impact of $J_{3d}$, we performed a numerical search 
for ground states of cubic clusters of 128, 432 and 1024 sites, 
using a zero-temperature Monte Carlo ``worm'' algorithm.
Results for an 128-site cluster are shown in Fig.~\ref{fig:phase-diagram-J3d}.
Apart from a small window of parameters for $J_{3d} < 0$, the ground state is 
dominated by the chain-based TDQ, CAF and FM states discussed above.
%
The precise range of parameters for which non-chain states occur was found to 
depend on the geometry of the cluster.
We note that {\it no} non-chain states were found for $J_{3d} > -0.018D$, in 
{\it any} cluster.
  
\section{Classical Monte Carlo simulation}
\label{section:classical.monte.carlo}


The classical ground--states of dipolar spin ice 
are based on alternating chains of spins [Sec.~\ref{sec:classical.mean.field.theory}], 
a fact which can be understood through the mapping onto 
an effective Ising model [Sec.~\ref{section:effective.Ising.model}].
However, at finite temperature, a spin--ice can gain an extensive ``ice entropy''
by fluctuating between different spin--ice configurations.~\cite{harris97,ramirez99}
As a result, chain--based ordered ground states will give way to a classical 
spin liquid (CSL).

\begin{figure}[t]
\includegraphics[width=0.9\columnwidth]{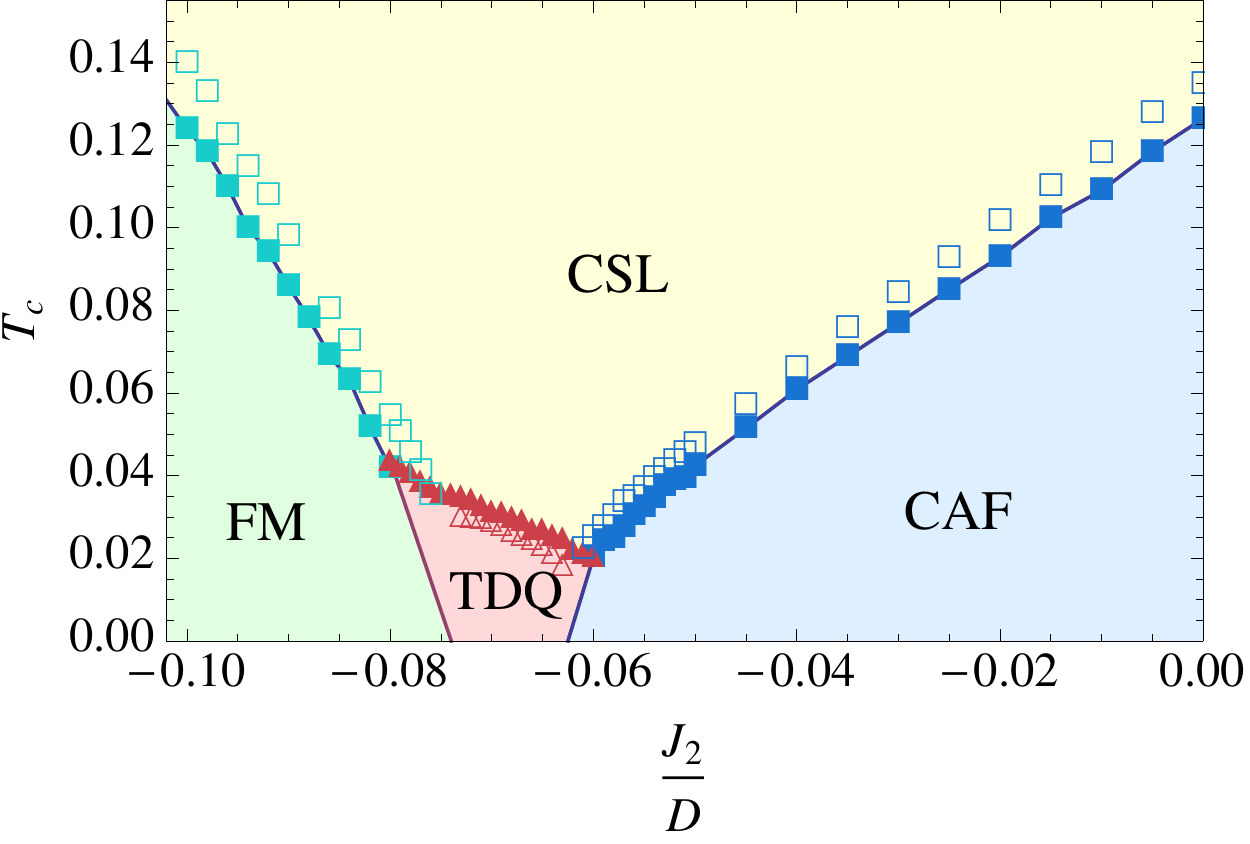}
\caption{(Color online) 
Finite-temperature phase diagram of spin ice with long-range dipolar interactions, 
as a function of competing 2$^{\text{nd}}$-neighbour exchange $J_2/D$.
Results are taken from classical Monte Carlo simulation of $\mathcal{H}_{\sf DSI}$ [Eq.~(\ref{eq:Hdsi})] 
for cubic clusters of 128 (filled symbols) and 1024 spins (open symbols). 
The error in the estimate of $T_c$ is set by the interval between consecutive temperatures 
in simulations using parallel tempering.   
\label{fig:phase.diagram.classical.MC}
}
\end{figure}

To learn more about the nature of this transition, and 
whether thermal fluctuations stabilise new ordered states, we have performed 
classical Monte Carlo simulations of $\mathcal{H_{\sf DSI}}$~[Eq.~(\ref{eq:Hqdsi})].
Simulations were carried out for cubic clusters of 128 and 1024 spins, using 
the worm algorithm and parallel--tempering 
methods described in Appendix~\ref{section:CMC.technical}, 
for \mbox{2$^{\text{nd}}$--neighbour} interaction $J_2/D$ spanning the 
cubic antiferromagnet (CAF), tetragonal double-Q (TDQ)
and ferromagnetic (FM) ground states [cf. Fig.~\ref{fig:Ising-phase-diagram}].
All other exchange interactions $J_{k \ne 2}$ were set to zero.
The results of these simulations are summarised in Fig.~\ref{fig:phase.diagram.classical.MC}.


\begin{figure*}[t]
%
\begin{tabular}{c}
\subfloat[$J_2/D=0.000$ \label{Sq.classical.a}]{\includegraphics[width=0.5\columnwidth]{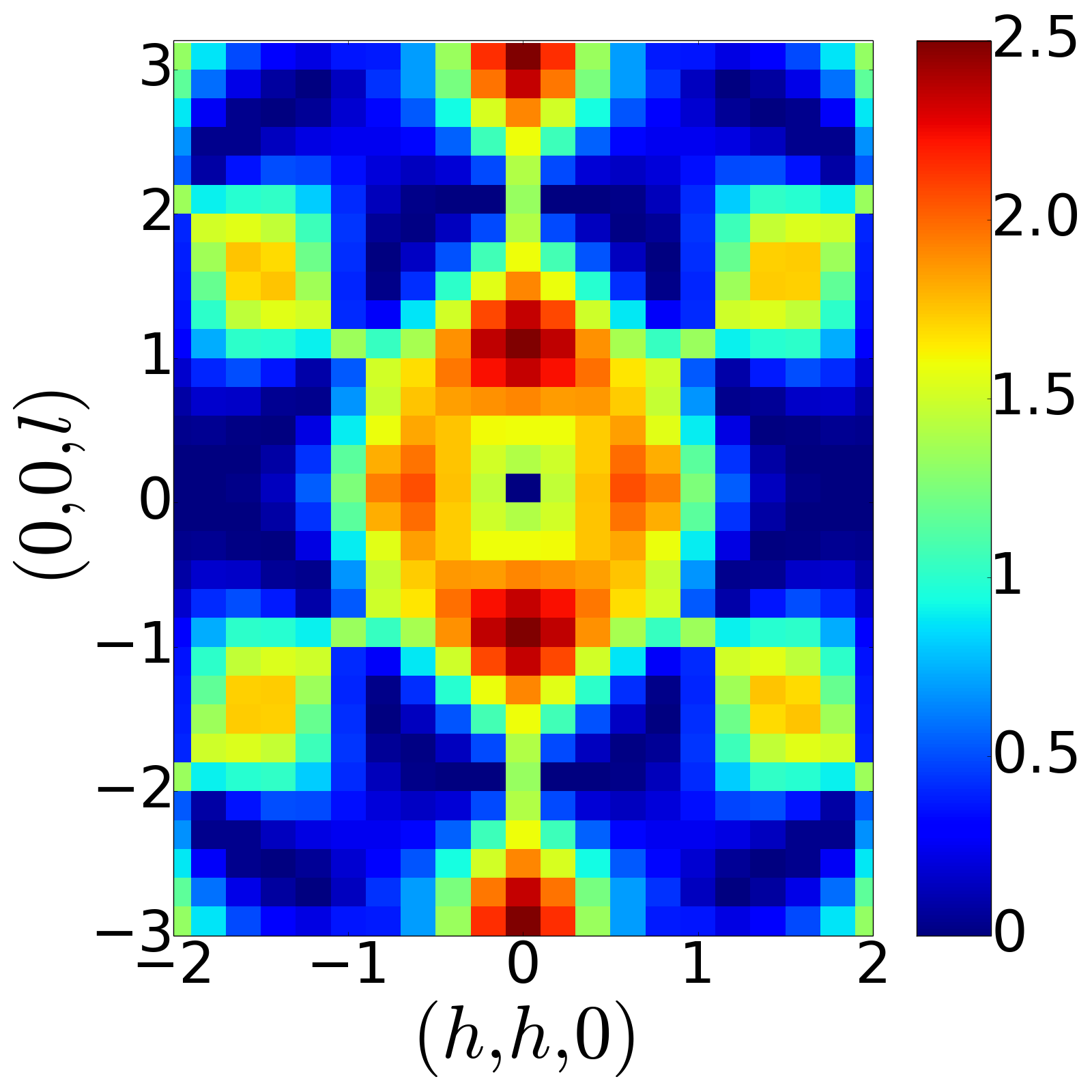}} 
\hspace{1cm}
\subfloat[$J_2/D=-0.070$ \label{Sq.classical.b}]{\includegraphics[width=0.5\columnwidth]{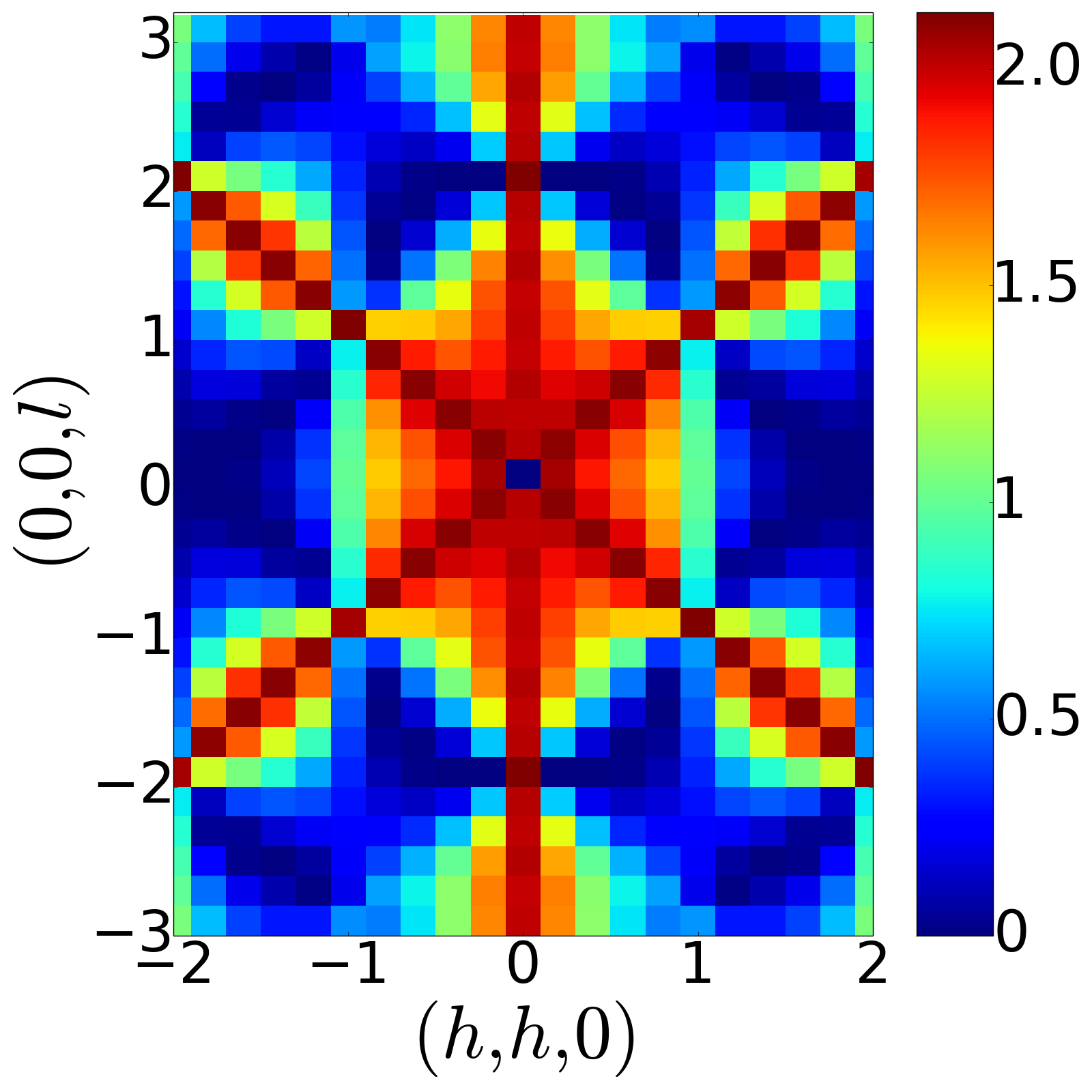}} 
\hspace{1cm}
\subfloat[$J_2/D=-0.090$ \label{Sq.classical.c}]{\includegraphics[width=0.5\columnwidth]{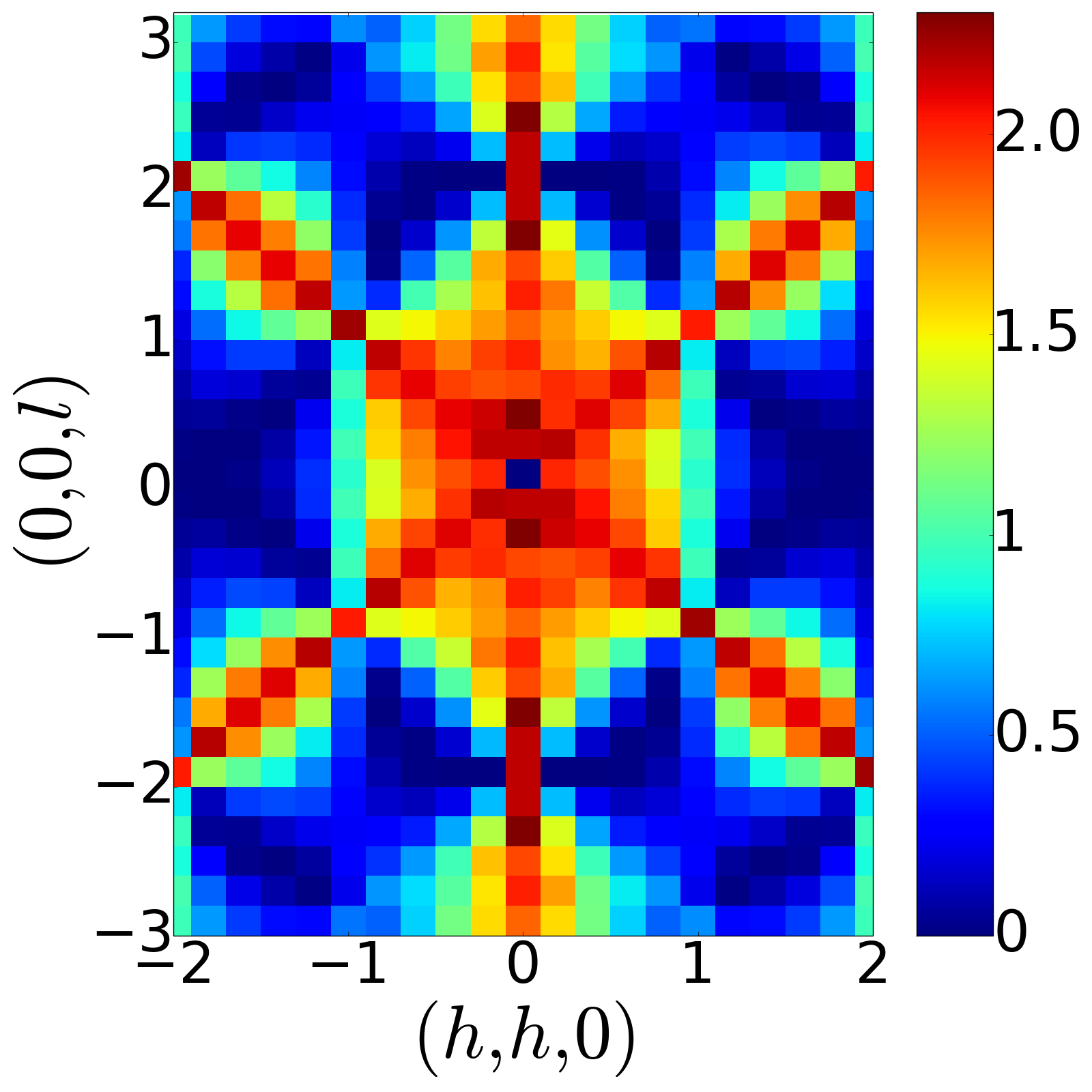}}  \\
\end{tabular} 
\caption{(Color online)  
Equal-time structure factor $S^{\sf SF} (\mathbf{q})$ [Eq.~(\ref{eq:def.Sq.SF})] 
for a dipolar spin ice with competing further-neighbour exchange $J_2$, 
as found from classical Monte Carlo simulation of 
$\mathcal{H_{\sf DSI}}$~[Eq.~(\ref{eq:Hdsi})].
(a) $S^{\sf SF} ({\bf q})$ for $J_2 = 0$.   
(b) $S^{\sf SF} ({\bf q})$ for $J_2 = - 0.07 D$.   
(c) $S^{\sf SF} ({\bf q})$ for $J_2 = -0.09 D$.   
Simulations were performed for a cubic cluster of $2000$ spins, 
for parameters spanning the CAF, TDQ and FM ground states, 
for a temperature $T = 0.5 D$ within the disordered spin-ice regime, 
with $J_{k \ne 2} \equiv 0$.  
$S^{\sf SF} (\mathbf{q})$ is shown in the $[hhl]$ plane, in the spin-flip channel 
measured by Fennell {\it et al.}~[\onlinecite{fennell09}]. 
}
\label{fig:Sq.classical}
\end{figure*} 

\subsection {Classical spin liquid}

The finite--temperature phase diagram of dipolar spin ice is dominated by 
a classical spin liquid (CSL), shown in yellow in Fig.~\ref{fig:phase.diagram.classical.MC}.
This CSL has the character of a classical Coulombic phase, described 
by a $U(1)$ lattice gauge theory.~\cite{henley10}  
For $J_2/D = 0$, simulations reproduce known results for a purely dipolar spin 
ice,~\cite{melko01} with the transition into the spin liquid occuring for $T_c/D \approx 0.12$. 
For ferromagnetic $J_2$, this transition temperature is at first suppressed, 
reaching a minimum value of $T_c/D \approx 0.02$ for $J_2/D \approx -0.06$.
For stronger ferromagnetic $J_2$, there is a rise in $T_c/D$.
These are the same trends as are observed in the overall band-width of spin-ice states 
[Fig.~\ref{fig:kMFT_Bandwidth}], within the mean--field theory described 
in Sec.~\ref{sec:classical.mean.field.theory}.


Spin correlations within the CSL phase are dipolar,\cite{isakov04,henley05}
leading to singular ``pinch--points'' 
\begin{eqnarray}
S (\mathbf{q}) \sim \left[ \delta_{\alpha\beta} - \frac{q^\alpha q^\beta}{q^2} \right] 
\end{eqnarray}
in the spin structure factor.
Pinch--points of exactly this form have been observed in neutron scattering 
experiments on the spin ice Ho$_2$Ti$_2$O$_7$ by Fennell {\it et al.} 
[\onlinecite{fennell09}].


To characterise the CSL found in the presence of competing exchange interactions, 
we have used classical Monte Carlo simulation to calculate the (equal--time) 
structure factor 
\begin{eqnarray}
S^{\alpha\beta} (\mathbf{q}) = \sum_{i, j=1}^{4} 
    \langle 
      S^{\alpha}_i(-{\bf q})  S^{\beta}_j ({\bf q}) 
   \rangle \; , 
   \label{eq:def.Sq}
\end{eqnarray}
where $i,j$ run over the sites of a tetrahedron and the spin 
$S^{\alpha}_i$ is considered in frame of the cubic crystal axes 
$\alpha = \{ x, y, z\}$.
We consider in particular the spin--flip component of scattering, 
for neutrons polarised $\parallel [1\bar10]$, as measured by 
Fennell {\it et al.} [\onlinecite{fennell09}]~:
\begin{eqnarray}
S^{\sf SF} (\mathbf{q}) 
   &=& \sum_{i, j=1}^{4} 
    \langle 
        \left[ {\bf S}_i (-{\bf q}) \cdot {\bf u} ({\bf q})  \right]
        \left[ {\bf S}_j ({\bf q}) \cdot {\bf u} ({\bf q}) \right]
   \rangle \; , \nonumber\\
 &&   {\bf u} ({\bf q}) = \hat{\bf n} \times {\bf q}, \quad 
   \hat{\bf n} = (1,\bar1,0)/\sqrt{2} \; .
   \label{eq:def.Sq.SF}
\end{eqnarray}


Simulation results for $S^{\sf SF} (\mathbf{q})$ in the CSL phase 
are shown in Fig.~\ref{fig:Sq.classical}, for $\mathbf{q}$ in the $[hhl]$ 
plane, and a range of values of $J_2/D$ 
spanning the three classical ordered ground states.   
For $J_2/D = 0$, these simulations reproduce known
results for dipolar spin ice, with pinch--points clearly visible at a subset of reciprocal
lattice vectors, {\it e.g.} $\mathbf{q} = (1,1,2)$ [cf. Fig.~\ref{fig:Sq.classical}(a)].   
For ferromagnetic $J_2/D < 0$, there is a progressive 
redistribution of spectral weight within the $[hhl]$ plane [cf. Fig.~\ref{fig:Sq.classical}(b, c)],
None the less, pinch--points remain clearly defined.

\subsection {Ordered phases}

At low temperatures, in the absence of quantum tunnelling, 
long--range dipolar interactions drive dipolar spin ice into a state with chain--based order.
Classical Monte Carlo simulation of $\mathcal{H_{\sf DSI}}$~[Eq.~(\ref{eq:Hqdsi})]
reveals the same three, chain--based ordered phases as are found 
in mean--field theory [Sec.~\ref{sec:classical.mean.field.theory}], and through 
mapping onto an effective Ising model [Sec.~\ref{section:effective.Ising.model}]~:
a cubic antiferromagnet (CAF), a tetragonal double-Q (TDQ) state
and a cubic ferromagnetic (FM) [cf. Fig.~\ref{fig:ordered.phases}].
Transition temperatures for the transition from the classical spin liquid (CSL) 
into each of these ordered states can be extracted from the susceptibility
associated with the appropriate order parameter [cf. Eq.~(\ref{eq:TL-order_parameter})].


The results of this analysis, for clusters of 128 and 1024 spins, are summarised in the 
finite--temperature phase diagram, Fig.~\ref{fig:Ising-phase-diagram}, where 
the error on the estimated ordering temperature $T_c$ is indicated by the size of the point.
All phase transitions are found to be first--order, with finite-size corrections to 
$T_c$ of order $10\% $ between the 128-site cluster and the 1024-site cluster.
Classical Monte Carlo simulations do {\it not} reveal any new phases on 
the (degenerate) phase boundaries between the CAF and the TDQ, 
or the TDQ and the FM~\footnote{
We have explored the possibility that boundary states fan into finite temperature 
phases using self-consistent mean field theory in real space for an 128--site cluster.
Such a conventional real space mean field theory successfully captures the finite 
temperature phases of the original 3D ANNNI model.\cite{selke88} 
We find that the mean-field theory confirms the picture obtained from Monte Carlo 
simulations in particular that no further phases arise between the CAF and TDQ states}. 

\section{Quantum Monte Carlo simulation}
\label{section:QMC}

Just as thermal fluctuations stabilize a classical spin liquid (CSL), so quantum tunneling 
might be expected to stabilize a quantum spin liquid (QSL), of the type previously studied 
in idealised models of a quantum spin-ice with nearest-neighbour interactions.\cite{hermele04,banerjee08,savary12,shannon12,benton12,lee12,savary13,gingras14}
There is also the possibility that quantum fluctuations might stabilise new ordered
states, not found in classical dipolar spin ice.
To address these questions, we have carried out extensive quantum Monte Carlo (QMC) 
simulations of  $\mathcal{H_{\sf QDSI}}$~[\ref{eq:Hqdsi}].


\begin{figure}[]
\includegraphics[width=0.9\columnwidth]{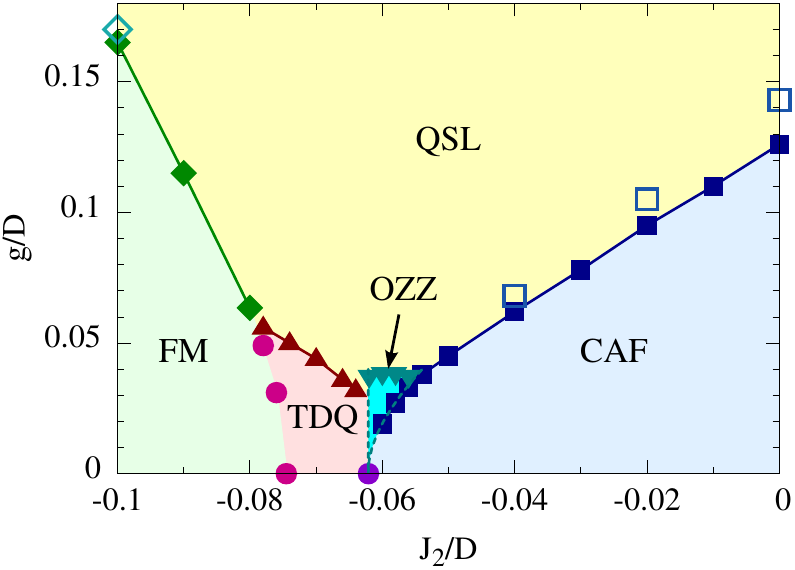}
\caption{(Color online) 
Quantum ground-state phase diagram of a dipolar spin ice, as a function of competing 
\mbox{2$^{\rm nd}$-neighbour} exchange $J_2/D$, and quantum tunneling $g/D$.
Results are taken from Green's function Monte Carlo (GFMC) 
simulation of $\mathcal{H}_{\sf QDSI}$~[Eq.~(\ref{eq:Hqdsi})], 
for cubic clusters of 128 sites (solid symbols) and 1024 spins (open symbols), 
with $J_{k \ne 2} = 0$. 
Phase boundaries for $g/D=0$ were determined from the solution of the extended
Ising model described in Sec.~\ref{section:effective.Ising.model}.   
Dashed lines bordering the OZZ state are taken from the degenerate perturbation
theory described in Appendix~\ref{appendix:2nd.order.perturbation.theory.in.g}. 
\label{fig:phase.diagram.QMC}
}
\end{figure}


\begin{figure*}[t]
\begin{tabular}{c}
\subfloat[$J_2/D=0.00$ \label{Sq.quantum.a}]{\includegraphics[width=0.5\columnwidth]{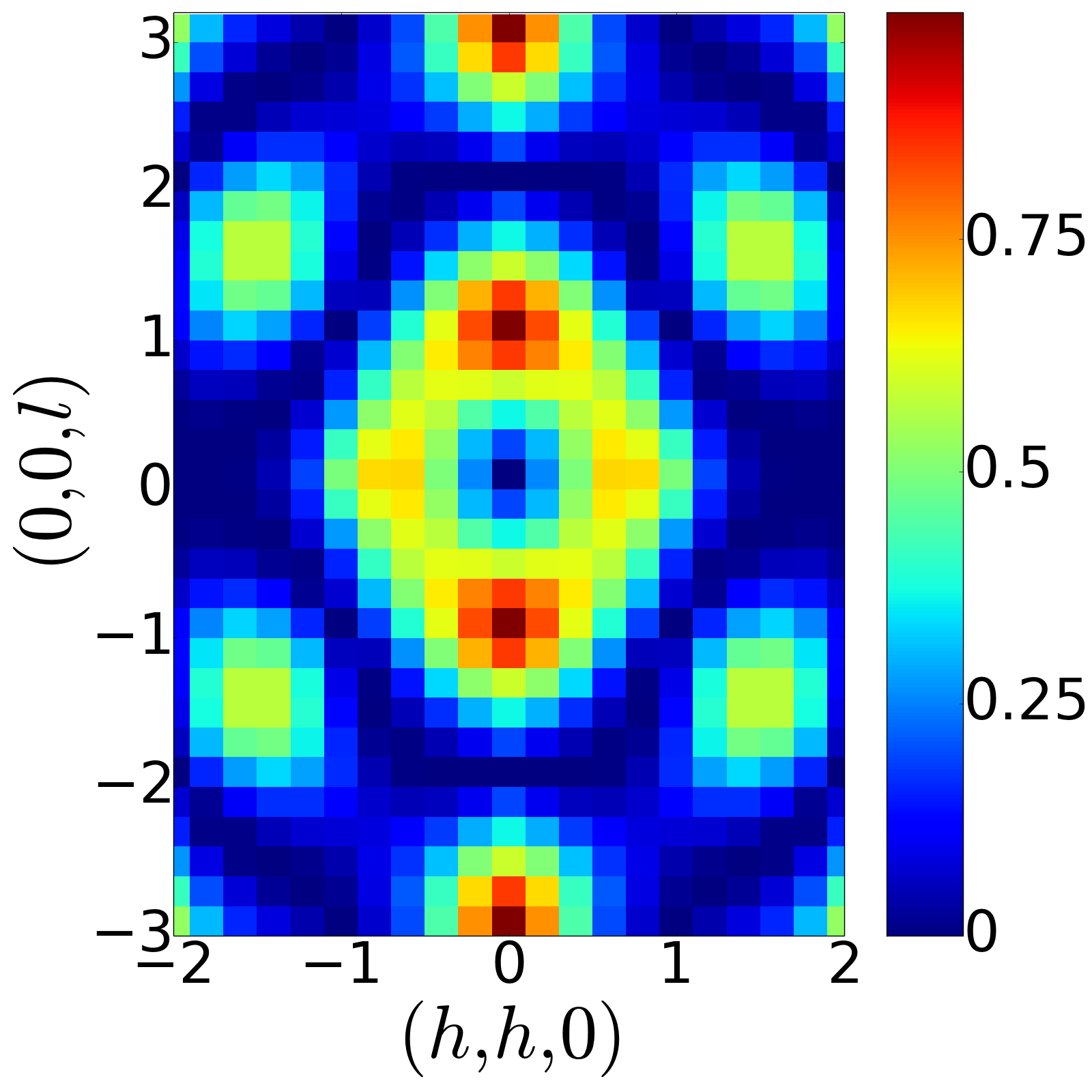}} 
\hspace{1cm}
\subfloat[$J_2/D=-0.07$ \label{Sq.quantum.b}]{\includegraphics[width=0.5\columnwidth]{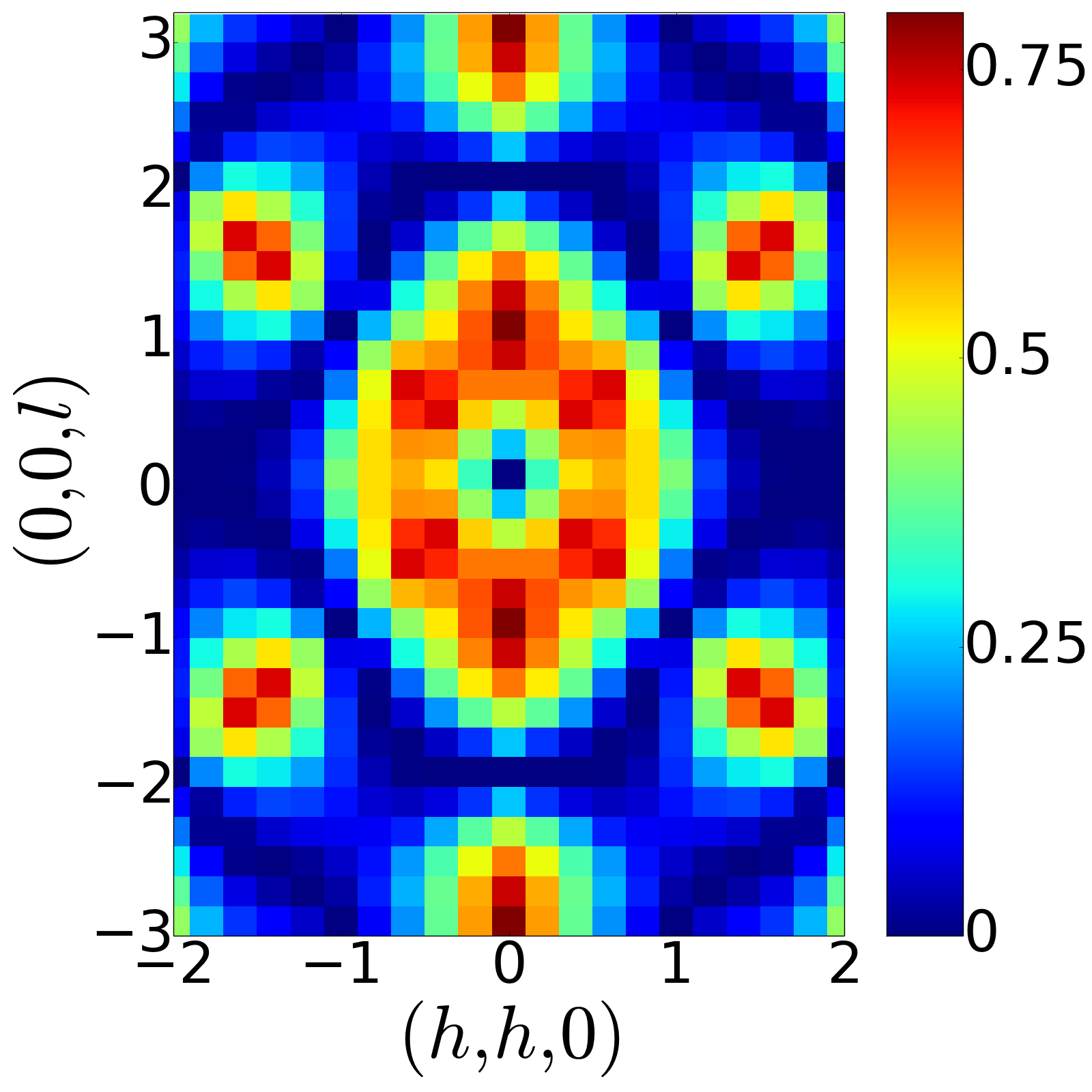}} 
\hspace{1cm}
\subfloat[$J_2/D=-0.09$ \label{Sq.quantum.c}]{\includegraphics[width=0.5\columnwidth]{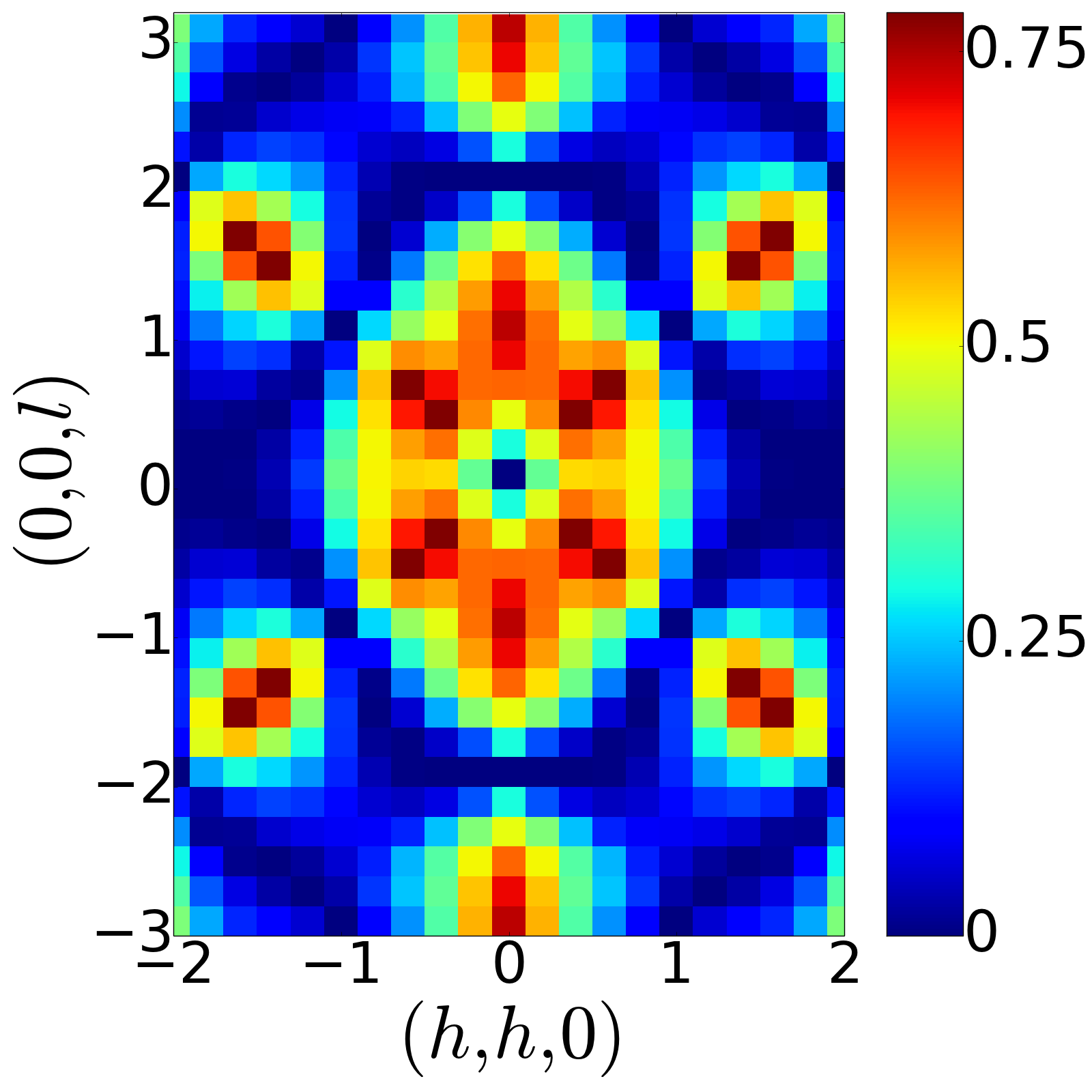}}  \\
\end{tabular} 
\caption{ 
Equal-time structure factor $S^{\sf SF} (\mathbf{q})$ [Eq.~(\ref{eq:def.Sq.SF})] 
for a dipolar spin ice with competing further-neighbour exchange $J_2$, 
as found from Green's function Monte Carlo (GFMC) 
simulation of $\mathcal{H}_{\sf QDSI}$~[Eq.~(\ref{eq:Hqdsi})].
(a) $S^{\sf SF} ({\bf q})$ for $J_2 = 0$;   
(b) $S^{\sf SF} ({\bf q})$ for $J_2 = - 0.07 D$; and   
(c)$S^{\sf SF} ({\bf q})$ for  $J_2 = -0.09 D$.   
%
%
The sharp pinch--point structure, characteristic of classical spin ice, 
and visible in classical Monte Carlo simulations 
[cf. Fig.~\ref{fig:Sq.classical}(a)--(c) at, e.g.~${\bf Q} = (1,1,1)$], 
is eliminated by quantum fluctuations.~\cite{shannon12,benton12}
All simulations were carried out for a cubic cluster of 2000 sites,
with quantum tunnelling $g = 0.5 D$, and $J_{k \ne 2} \equiv 0$.   
$S^{\sf SF}({\bf q})$ is shown in the $[hhl]$ plane, in the 
spin-flip channel measured by Fennell {\it et al.} [Ref.~\onlinecite{fennell09}]. 
}
\label{fig:Sq.quantum}
\end{figure*} 


Simulations of cubic clusters of 128 and 1024 spins were performed using the 
zero-temperature Green's function Monte Carlo (GFMC) method described 
in Refs.~[\onlinecite{shannon12,benton12,sikora09,sikora11}] 
and Appendix~\ref{section:QMC-technical}.
Within this approach, only spin--configurations satisfying the ice--rules are considered, 
and $\mathcal{H}_{\sf QDSI}$~[Eq.~(\ref{eq:Hqdsi})] is taken to act on the space of all 
possible spin-ice ground states.
GFMC simulations were carried out for a range of values of quantum tunnelling 
$g/D$, for 2$^{\text{nd}}$--neighbour interaction $J_2/D$ spanning
all three classical ground states [cf. Fig.~\ref{fig:Ising-phase-diagram}].
All other exchange interactions $J_{k \ne 2}$ were set to zero.
The results of these simulations are summarised in Fig.~\ref{fig:phase.diagram.QMC}.

\subsection{Quantum spin liquid}
\label{subsection:QSL}

The zero--temperature quantum phase diagram is dominated by a QSL phase, shown 
in yellow in Fig.~\ref{fig:phase.diagram.QMC}.
The minimum value of quantum tunneling $g_c$ needed to stabilize a 
QSL for a given value of $J_2/D$, closely tracks the 
transition temperature $T_c$ for $g=0$ [cf. Fig.~\ref{fig:combined.phase.diagram}].  
Crucially, $g_c$ is always very small, being of order $g_c \sim 0.1 D$ for $J_2/D = 0$, 
and decreasing to a few percent of $D$ for $J_2/D \sim -0.06$.


Correlations within the QSL can once again be characterised by the equal--time structure 
factor $S(\bf q)$ [Eq.~(\ref{eq:def.Sq})].
While spin correlations in the CSL are dipolar 
leading to ``pinch--points'' in $S({\mathbf q})$ [cf.~Fig.~\ref{fig:Sq.classical}], 
spin correlations in the QSL decay as $1/r^4$ [\onlinecite{hermele04}], 
eliminating the pinch-points.~\cite{shannon12,benton12}.


Results for $S^{\sf SF}({\mathbf q})$ [Eq.~(\ref{eq:def.Sq.SF})], calculated using GFMC 
simulation, are shown in Fig.~\ref{fig:Sq.quantum}, for a range of values of $J_2/D$ 
spanning the phase diagram Fig.~\ref{fig:phase.diagram.QMC}, and $g = 0.5\ D > g_c$ 
within the QSL phase.  
As anticipated, the sharp zone--center pinch-points of the CSL are eliminated by 
quantum fluctuations [cf. Fig.~\ref{fig:Sq.classical}].
Correlations are also suppressed near to ${\mathbf q} = 0$ 
[cf. Ref.~\onlinecite{shannon12,benton12}].
All of these features are universal characteristics of the QSL, and therefore 
independent of the values of $J_2$ and $D$.


Correlations at short wave length, on the other hand, show a marked imprint 
of the long--range dipolar interactions, when compared with results for $D=0$ 
[Ref.~\onlinecite{shannon12,benton12}].   
These features are only weakly constrained by the structure of the QSL, and 
therefore depend strongly on the ratio of $J_2/D$ for which the simulations 
were carried out.


\begin{figure}[t]
\includegraphics[width=0.9\columnwidth]{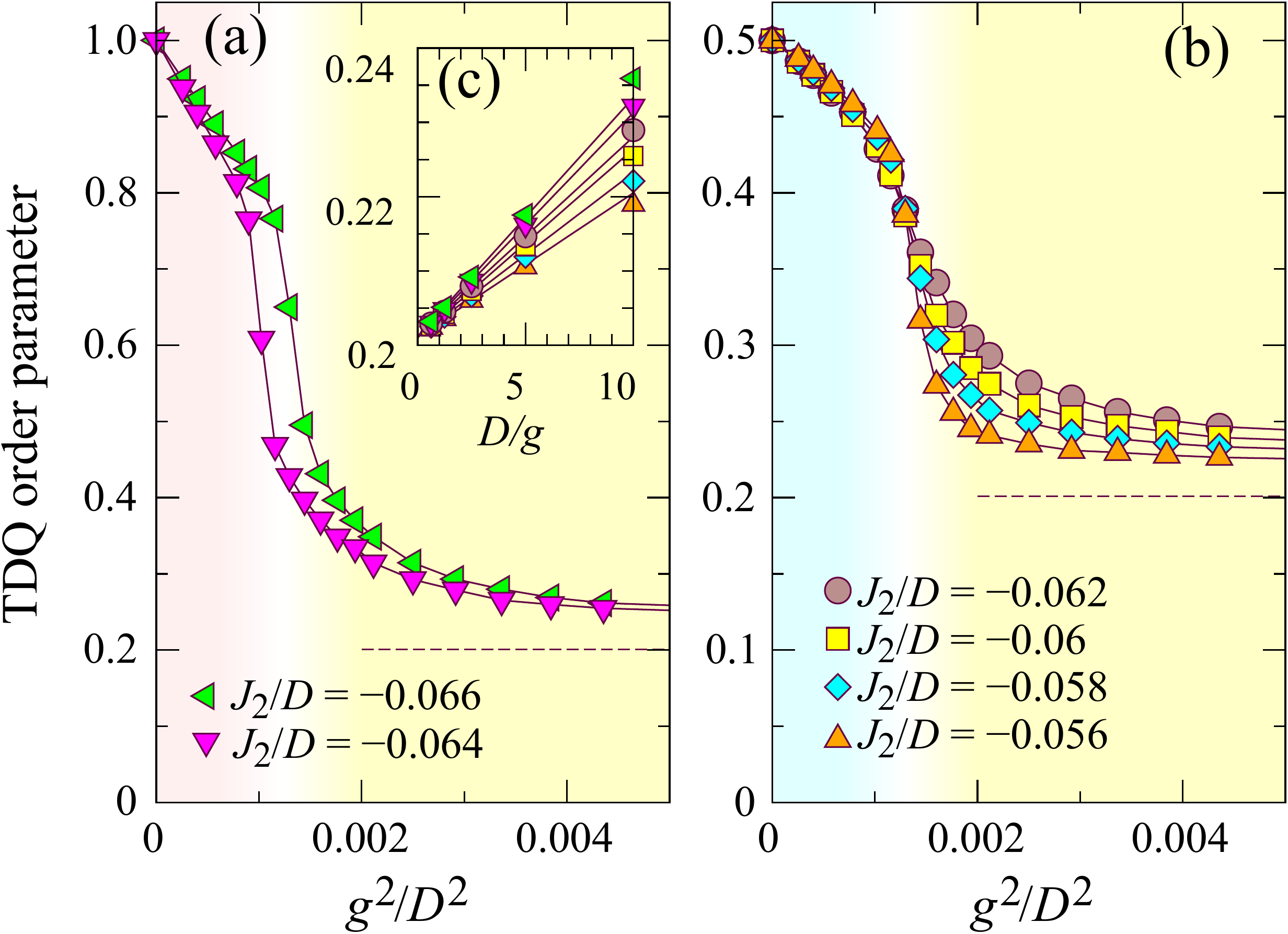}
\caption{(Color online)   
Evidence for the transition from the quantum spin liquid (QSL) 
into ordered tetragonal double--Q (TDQ) and orthorhombic zig-zag (OZZ) 
ground states, as determined by Green's function Monte Carlo (GFMC) 
simulation of $\mathcal{H}_{\sf QDSI}$~[Eq.~(\ref{eq:Hqdsi})].
(a) Results for the order parameter ${\mathcal O}_{\sf TDQ}$ [Eq.~(\ref{eq:TDQ.OP})], 
showing the transition between the Quantum spin liquid (QSL) and TDQ phases.  
%
(b) Equivalent results for the transition between the QSL and OZZ phases.    
%
(c) [Inset to (a)]  Scaling of order parameter within the QSL, showing how, 
for $D/g \to 0$, ${\mathcal O}_{\sf TDQ} \to 0.20(1)$, a finite-size 
value characteristic of the cluster simulated.
Dashed lines in (a) and (c) show the corresponding asymptote for $g/D \to \infty$.   
Simulations were carried out for a cubic cluster of 128 sites, with $J_{k \ne 2} = 0$.
Values of $J_2/D$ are shown on the legend within the figure.  
\label{fig:order.parameters.in.QMC}
}
\end{figure}

\subsection{Ordered ground states}
\label{subsection:OZZ}

For $g < g_c$, quantum fluctuations are not sufficient to stabilise a QSL, 
and the system orders.
For $g \to 0$ we find the same three, chain--based states discussed 
in Sec.~\ref{sec:classical.mean.field.theory} and Sec.~\ref{section:effective.Ising.model} 
--- a cubic antiferromagnet (CAF), a tetragonal double-Q (TDQ) state and 
a ferromagnet (FM).   
However quantum simulations also reveal a new ordered state, the 
orthorhombic zig--zag (OZZ) state shown in Fig.~\ref{fig:ordered.phases}(d).
The OZZ occurs at the boundary between the CAF and the TDQ, 
and is stabilised by quantum fluctuations at finite $g$. 
We consider each of these ordered states in turn, below.  


The FM and CAF are ``isolated states'', unconnected to other 
spin-ice configurations by matrix elements of $\mathcal{H}_{\sf tunneling}$~[Eq.~(\ref{eq:Htunneling})].
Quantum phase transitions between the QSL and the FM and CAF 
are therefore first-order, and can be determined by a simple comparison 
of ground state energies.
The corresponding values of $g_c$, as a function of $J_2$, are shown in 
Fig.~\ref{fig:phase.diagram.QMC}, for clusters of 128 and 1024 spins.
Finite-size effects are relatively small, at least in the range of $J_2$ 
for which is was possible to converge simulations for both clusters.


The TDQ state, in contrast, is directly connected with QSL by matrix elements 
of $\mathcal{H}_{\sf tunneling}$.
%
%
In this case $g_c$ was determined from a jump in the 
the order parameter of the TDQ state
\begin{align}
{\mathcal O}_{\sf TDQ}
    = \sum_{\eta=1}^{48} \sum_{j} 
    \left\langle 
       \left({\sf S}^{z,\text{\sf TDQ},\eta}_j 
       \sf{S}^z_{j} \right)^2 
    \right\rangle_{\text{\sf QMC}}
\label{eq:TDQ.OP}
\end{align}
where the spin configurations ${\sf S}^{z,\text{TDQ},\eta}_j$ 
are drawn from the 48 different TDQ ground states enumerated 
in \mbox{Table~\ref{table:ordered.states}}.  


Results for ${\mathcal O}_{\sf TDQ}$ within GFMC simulation 
are shown in Fig.~\ref{fig:order.parameters.in.QMC}(a), 
for parameters spanning the TDQ and QSL states.
An abrupt change in the order parameter marks the onset 
of TDQ order, with ${\mathcal O}_{\sf TDQ} \to 1$
in the fully ordered state.
In the spin liquid, for $g > g_c$, ${\mathcal O}_{\sf TDQ} \to 0.20$, 
a finite-size value determined by the cluster used in simulations
[cf. Fig.~\ref{fig:order.parameters.in.QMC}(b)].  


The OZZ is one of the many degenerate classical 
ground states found at the border between the CAF and TDQ phases 
[cf. Sec.~\ref{subsection:Ising.phase.diagram}].   
Unlike the CAF, the TDQ and OZZ both  contain ``flippable'' hexagonal plaquettes 
where $\mathcal{H}_{\sf tunneling}$~[Eq.~(\ref{eq:Htunneling})]
can act. 
As a result, both states gain energy from quantum fluctuations, 


Besides having flippable plaquettes, the OZZ is also directly connected 
with the QSL.
%
And, since the spin configurations in the one of the sets of 
parallel chains which make up the OZZ are identical 
to those of the TDQ (cf.~Table.~\ref{table:ordered.states}), it is also 
possible to use ${\mathcal O}_{\sf TDQ}$ [Eq.~(\ref{eq:TDQ.OP})] 
as an order parameter for the OZZ state.
Corresponding results for ${\mathcal O}_{\sf TDQ}$ are shown in 
Fig.~\ref{fig:order.parameters.in.QMC}(b), 
for parameters spanning the OZZ and QSL states.
We note that, in this case, \mbox{${\mathcal O}_{\sf TDQ} \to 0.5$} 
in the fully ordered state.  

\begin{figure}[t]
\includegraphics[width=0.9\columnwidth]{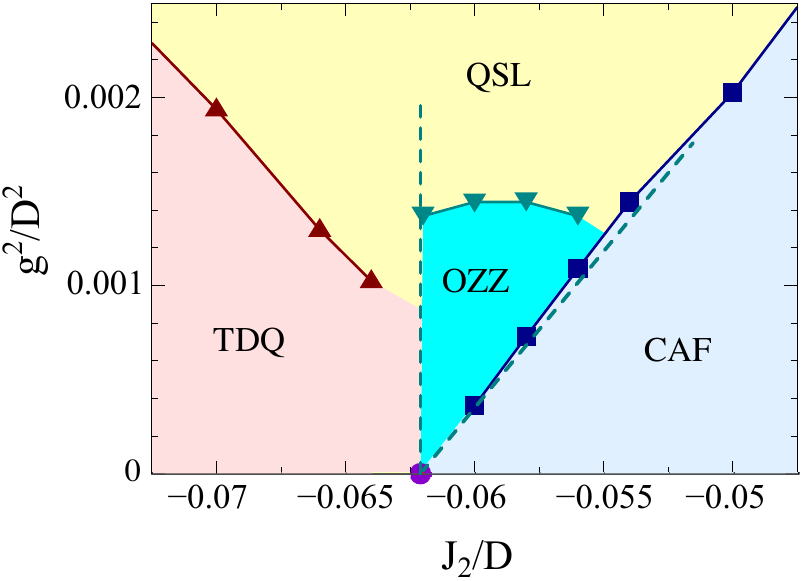}
\caption{(Color online)   
Detail of the ground-state phase diagram of a dipolar spin ice 
[Fig.~\ref{fig:phase.diagram.QMC}], showing how a small fan of 
orthorhombic zig-zag (OZZ) order opens between the cubic 
antiferromagnet (CAF) and tetragonal double--Q (TDQ) states. 
Filled symbols show the results of Green's function Monte Carlo (GFMC) 
simulation of $\mathcal{H}_{\sf QDSI}$~[Eq.~(\ref{eq:Hqdsi})], 
as described in the text.  
Dashed lines show the predictions of the degenerate perturbation theory
described in Appendix~\ref{appendix:2nd.order.perturbation.theory.in.g}.
Simulations were carried out for a cubic cluster of 128 sites, with 
$J_{k \ne 2} = 0$.
\label{fig:detail.of.quantum.phase.diagram}
}
\end{figure}

Colllecting all of these results, we obtain the quantum ground state 
phase shown in Fig.~\ref{fig:phase.diagram.QMC}.
We find that a small fan of OZZ order opens from the highly degenerate 
point $J_2/D = -0.0621$, $g/D=0$, at the expense of the CAF.
Detail of this highly frustrated region of the phase diagram is 
given in Fig.~\ref{fig:detail.of.quantum.phase.diagram}.  


It is possible to estimate the phase boundaries between the 
TDQ, OZZ and CAF states from 2$^{\text{nd}}$--order 
perturbation theory in $g$, as described in Appendix~\ref{appendix:2nd.order.perturbation.theory.in.g}.
The corresponding results are shown as dashed lines in 
Fig.~\ref{fig:phase.diagram.QMC} and Fig.~\ref{fig:detail.of.quantum.phase.diagram}.
In the case of the boundary between the CAF and OZZ states, 
it is possible to make direct comparison between this perturbation
theory and GFMC.
As shown in Fig.~\ref{fig:phase.diagram.QMC}, the agreement is excellent.   


While no new ordered states, besides the OZZ, were found for GFMC simulations of 
cubic clusters of 128 or 1024 states, it is interesting to speculate that 
quantum fluctuations might stabilise further new ordered state in the 
thermodynamic limit --- perhaps in the form of the ``fans'' found in 
classical anisotropic next-nearest neighbour Ising (ANNNI) models.~\cite{bak82,selke88}   
It is also plausible that thermal fluctuations might stabilise the OZZ, or 
some other state like it, in a more general model.

\section{Application to spin--ice materials}
\label{section:application.to.real.materials}

In this Article we have used a variety of numerical and analytic techniques 
to explore the nature of the equilibrium ground state of a dipolar spin ice 
with competing exchange interactions and quantum tunnelling between 
different spin--ice configurations, as described by $\mathcal{H_{\sf QDSI}}$ 
[Eq.~(\ref{eq:Hqdsi})].


A clear picture emerges from this analysis.
Long--range dipolar interactions, ${\mathcal H}_{\sf dipolar}$ [Eq.~(\ref{eq:H.dipolar})],
are minimised by states composed of ferromagnetically polarised chains of spins, within 
which they are exponentially screened.
Exchange interactions, ${\mathcal H}_{\sf exchange}$ [Eq.~(\ref{eq:H.exchange})]
act to select between such ``chain states'', and in the absence of quantum fluctuations
the ground state of a dipolar spin ice is one of the three ordered states, described 
in Table~\ref{table:ordered.states}.
Quantum tunnelling between different spin--ice configurations, 
$\mathcal{H}_{\sf tunneling}$ [Eq.~(\ref{eq:Htunneling})], can stabilise new forms
of chain--based order, and if sufficiently strong, will drive a quantum spin liquid 
ground state.
We now consider the implication of these results for real materials, paying particular
attention to the dipolar spin ice,  Dy$_2$Ti$_2$O$_7$.


Dy$_2$Ti$_2$O$_7$ is perhaps the best studied of spin--ice materials.
Pioneering measurements of the heat capacity of Dy$_2$Ti$_2$O$_7$ by 
Ramirez {\it et al.}~[\onlinecite{ramirez99}] provided the first thermodynamic evidence 
for the existence of an extensive ground--state degeneracy, as predicted by the 
``ice rules'' [\onlinecite{pauling35}].
These results are consistent with subsequent measurements of the heat--capacity 
of Dy$_2$Ti$_2$O$_7$ by other groups.~\cite{higashinaka02,hiroi03,klemke11}
And, significantly, none of these studies reported evidence for a transition 
into an ordered ground state at low temperatures, despite the expectation that 
a classical dipolar spin ice should have an ordered ground state.~\cite{melko01}


As the understanding of spin ice has improved, it has become clear 
that non--equilibrium effects play an important role, and that
the thermodynamic properties of materials like Dy$_2$Ti$_2$O$_7$ 
are consequently subject to extremely long equilibriation times.~\cite{castelnovo12}
In the light of this, the evolution of the low--temperature heat capacity of 
Dy$_2$Ti$_2$O$_7$ was recently revisted by Pomaranski  {\it et al.}~[\onlinecite{pomaranski13}], 
using an experimental setup designed to track the equilibration of the sample.
Their study reports equilibration times in excess of {\it 4 days}  
at $340\ \text{mK}$, and a dramatically revised profile for the low--temperature
specific heat.~\cite{pomaranski13}  
One of the most striking features of these results is an upturn in $\partial S/\partial T|_V = C_V/T$
below $T \approx 500 \text{mK}$, suggestive of an ordering transition, of the type
studied in Sec.~\ref{section:classical.monte.carlo}, or the emergence of a new 
(quantum) energy scale.  


\begin{figure}[t]
\includegraphics[width=0.9\columnwidth]{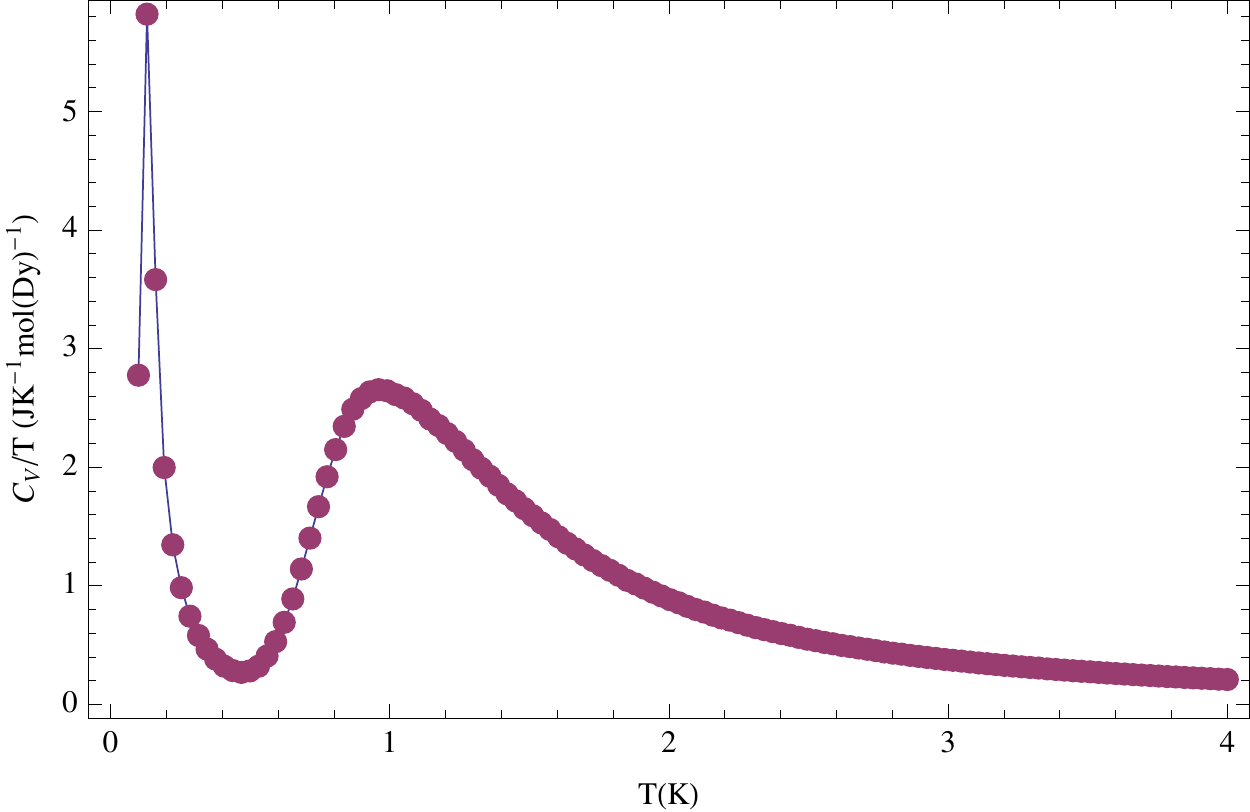}
\caption{(Color online) 
Plot of heat capacity $C_{\rm V}$ divided by temperature, $T$, 
for a dipolar spin ice described by $\mathcal{H_{\sf DSI}}$~[Eq.~(\ref{eq:Hdsi})], 
with parameters taken from fits to experiment on Dy$_2$Ti$_2$O$_7$ 
[\onlinecite{yavorskii08}], as specified in [Eq.~(\ref{eq:yavorskii.parameters})].
Results are taken from classical Monte Carlo simulation of a cluster of $128$ spins.
\label{fig:CVuponT}
}
\end{figure}


The results of Pomoranskii  {\it et al.}~[\onlinecite{pomaranski13}] clearly motivate 
a number of questions, including~: 
What is the origin of the upturn in $C_V/T$ ?
%
What is the nature of the ground state of Dy$_2$Ti$_2$O$_7$ ?
And, what is the reason for its extremely slow approach to equilibrium ?


These questions are most easily addressed within the well--established,
classical, dipolar spin--ice model $\mathcal{H_{\sf DSI}}$~[Eq.~(\ref{eq:Hdsi})].
As discussed in Sec.~\ref{subsection:Ising.phase.diagram}, the parameters 
reported by Yavorskii {\it et al.}~[\onlinecite{yavorskii08}], place the classical 
ground state of Dy$_2$Ti$_2$O$_7$ in the cubic antiferromagnetic (CAF) 
phase [cf. Fig.~\ref{fig:ordered.phases}(a)], previously investigated 
by Melko {\it et al.}~[\onlinecite{melko01}].
It is therefore natural to ask whether the upturn in $C_V/T$, observed 
in Dy$_2$Ti$_2$O$_7$ [\onlinecite{pomaranski13}], marks the onset of CAF order ?


At present, it is only possible to approach this question with reference to 
the heat capacity measurements of Pomaranskii  {\it et al.}~[\onlinecite{pomaranski13}].
To this end, in Fig.~\ref{fig:CVuponT} we show estimates of 
$C_V/T$ taken from classical Monte Carlo 
simulations of $\mathcal{H_{\sf DSI}}$~[Eq.~(\ref{eq:Hdsi})], for the parameters given 
by Yavorskii {\it et al.}~[\onlinecite{yavorskii08}] --- cf. Eq.~(\ref{eq:yavorskii.parameters}).
Simulations were carried out using the methods described
in Appendix~\ref{section:CMC.technical}, for a cubic cluster of 128 sites.
A complete comparison between experiment and simulation is not possible, 
since experimental data for $C_V/T$  is only available down to 
$T=340\ \text{mK}$ [\onlinecite{pomaranski13}].
However simulations correctly reproduce the measured peak in $C_V/T$ 
at  $T \approx 1\ \text{K}$, characteristic of the onset of spin-ice correlations, and 
exhibit a second peak at $T \approx 100\ \text{mK}$, associated with a first-order 
transition into the CAF ground state. 
For temperatures $340 < T \lessapprox 500\ \text{mK}$, simulations suggest 
an upturn in $C_V/T$ which is reminiscent of, but a little weaker than,
that observed in experiment.


At first sight, the comparison between simulation might seem good enough 
to justify a diagnosis of CAF order.
However the CAF is only one of the infinite family of chain--based ground states
described by the effective Ising model ${\mathcal H}^{\sf 2D}_{\sf Ising}$~[Eq.~(\ref{eq:HIsing})]  
--- cf. Sec.~\ref{section:effective.Ising.model}. 
And, since dipolar interactions are exponentially screened within these 
chain--states --- cf. Sec.~\ref{subsection:madlung.sum} --- the nature 
of the classical ground state is {\it extremely} sensitive to small differences in 
the exchange interactions ${\mathcal H}_{\sf exchange}$~[Eq.~(\ref{eq:H.exchange})].


For the specific set of parameters provided by Yavorskii {\it et al.}~[\onlinecite{yavorskii08}] --- 
Eq.~(\ref{eq:yavorskii.parameters}) --- the inter--chain interactions of 
${\mathcal H}^{\sf 2D}_{\sf Ising}$~[Eq.~(\ref{eq:HIsing})] take on the values 
\begin{eqnarray}
K_{(1,\sqrt{2})} &=& -35 \;\text{mK}   \, ,  \nonumber\\
K_{(2,0)} &=& \phantom{-}28 \;\text{mK}  \, , \qquad [\text{Dy$_2$Ti$_2$O$_7$}]  \\
K_{(0,2\sqrt{2})} &=& \phantom{-}1 \;\text{mK} \, .  \nonumber
\end{eqnarray}
These very weak interactions between chains should be compared with the 
uncertainty in exchange interactions, which is {\it at least} $10\ \text{mK}$ [\onlinecite{yavorskii08}].


It follows from definition of $K_{\boldsymbol{\delta}}$ [cf. Table~\ref{table:Kdelta}], 
that any change $\delta J_k$ in the value of exchange parameters leads directly 
to a change $\delta K_{\boldsymbol{\delta}}$ in the interactions between chains 
of spins
\begin{eqnarray}
\delta K_{(1,\sqrt{2})} &=& - \delta J_2/3 - \delta J_{3c} - \delta J_{3d}  \, ,  \nonumber\\
\delta K_{(2,0)} &=&  \delta J_{3d} \, . 
\end{eqnarray}
Since the parameters given by Yavorskii {\it et al.}~[\onlinecite{yavorskii08}] place
Dy$_2$Ti$_2$O$_7$ close to borders of CAF, tetragonal double Q (TDQ) 
and ferromagnetic (FM) phases --- cf. Fig.~\ref{fig:Ising-phase-diagram} --- 
an error as small as $\delta J \sim 7\, \text{mK}$ could be enough 
to convert the CAF into a TDQ ground state, while \mbox{$\delta J \sim 50\, \text{mK}$} 
could stabilize a FM.


This extreme sensitivity of the ground state of dipolar spin ice to small changes 
in exchange interactions makes very challenging to reliably predict the ground state 
in a real material from high-temperature estimates of model parameters.
However this challenge 
brings with it an opportunity~:
it seems entirely plausible that changes in $J_k$ of the scale 
$\delta J \sim 50\, \text{mK}$ could be achieved through the 
application hydrostatic pressure, or by chemical 
substitution,\cite{zhou12} allowing a spin ice to be tuned from 
one ground state to another.


The ``chain picture'' of ground--state order in a dipolar spin ice may also offer
some insight into the very slow equilibration of Dy$_2$Ti$_2$O$_7$ at low 
temperatures.~\cite{pomaranski13}
In order to achieve an ordered, equilibrium ground state, a dipolar spin ice 
must first select the low--energy chain--based states from the extensive set 
of states obeying the ice rules, and then single out the chain--state with 
the lowest energy.
At low temperatures, this thermal equilibration will be achieved through the motion 
of magnetic monopoles.
However, to connect one chain--state with another, a monopole would have to 
reverse all of the spins in chain.
This can only be achieved by the monopole traversing the entire length of a 
chain --- potentially the entire width of the sample.
Such dynamics would be activated, since it costs energy to make a pair of monopoles, 
and extremely slow.


The range of possible outcomes for the low--temperature physics of Dy$_2$Ti$_2$O$_7$ 
becomes much richer once quantum effects are taken into account.
One possibility is that quantum tunnelling, of the type described by 
${\mathcal H}_{\sf tunneling}$~[\ref{eq:Htunneling}] could stabilise a quantum 
spin--liquid (QSL) ground state, described by a quantum $U(1)$
lattice gauge theory [cf. Sec.~\ref{section:QMC}].  
In this case, the upturn in $C_V/T$ would signal the crossover between
the classical and a quantum spin liquid regimes.~\cite{benton12,kato-arXiv}


Another possibility, where exchange interactions place the system close to a classical 
phase boundary, is that quantum fluctuations could stabilise a new form of order, 
such as the orthorhombic zig--zag (OZZ) state studied in Sec.~\ref{subsection:OZZ}.  
%
Such a ground state could melt into a classical (or quantum) spin liquid at finite
temperature, leading to an upturn in $C_V/T$.


No reliable estimate is currently available for the strength of quantum tunneling 
in Dy$_2$Ti$_2$O$_7$.
And the uncertainty in published estimates of exchange 
interactions is also too great to assess how close it lies to a classical phase boundary.
For both reasons, it is difficult to draw any firm conclusions about 
the quantum or classical nature of its ground state.
\cite{iwahara15,rau-arXiv,tomasello-arXiv} 


However, one of the interesting consequences of chain--based order, 
and in particular of the exponential screening of dipolar interactions 
within chain states, is that quantum tunnelling does not need to be very 
strong to have a significant effect.
From Quantum Monte Carlo simulations for parameters 
similar to those proposed for Dy$_2$Ti$_2$O$_7$ [cf. Sec.~\ref{subsection:QSL}], 
we estimate that the value of quantum tunnelling $g$ needed to stabilize a QSL 
may be as little as \mbox{$g^{\text{Dy$_2$Ti$_2$O$_7$}}_c \approx \; 70\; \text{mK}$}.


Consequently --- and perhaps counter--intuitively --- a ``classical'' spin ice
like Dy$_2$Ti$_2$O$_7$, in equilibrium, may not be bad place to look for a QSL.
In this context it is interesting to note that the pinch--points observed
in Dy$_2$Ti$_2$O$_7$,~\cite{morris09} and its sister compound 
Ho$_2$Ti$_2$O$_7$,~\cite{fennell09} are 
somewhat reminiscent of the QSL at finite temperature.~\cite{benton12,kato-arXiv}

\section{Conclusions}
\label{section:conclusions}


In conclusion, determining the  zero-temperature, quantum, ground
state of a realistic model of a spin ice is an important challenge, 
motivated by recent experiments on Dy$_2$Ti$_2$O$_7$ \cite{pomaranski13}
and ongoing studies of quantum spin-ice 
materials.~\cite{thompson11,ross11-PRX1,chang12,fennell12,fennell14,kimura13}
In this Article, we have used a variety of numerical and analytic techniques
to address the question~: ``What determines the equilibrium ground state of 
spin ice, once quantum effects are taken into account ?''


In Sec.~\ref{sec:classical.mean.field.theory} and Sec.~\ref{section:effective.Ising.model}, 
we have shown how a new organisational principle emerges~: 
ordered ground states in a dipolar spin ice are built of 
alternating chains of 
spins, with net ferromagnetic polarisation.
These ``chain states'' minimise long--range dipolar interactions, and 
provide a natural explanation for the slow dynamics 
observed in Dy$_2$Ti$_2$O$_7$ [\onlinecite{pomaranski13}].
And, since dipolar interactions are exponentially screened within chain states, 
they can be described by an extended Ising model on an anisotropic triangular 
lattice, ${\mathcal H}^{\sf 2D}_{\sf Ising}$~[Eq.~(\ref{eq:HIsing})]. 


In Sec.~\ref{section:classical.monte.carlo} and Sec.~\ref{section:QMC}, 
using Monte Carlo simulation, we have determined both the quantum
and classical phase diagrams of ${\mathcal H}_{\sf QDSI}$~[Eq.~(\ref{eq:Hqdsi})],
as a function of quantum tunneling $g$, and temperature $T$.
We find that only a modest amount of quantum tunneling 
$g_c$ is needed to stabilize a quantum spin liquid (QSL), with deconfined fractional 
excitations,~\cite{hermele04,banerjee08,savary12,shannon12,benton12,lee12,savary13,moessner03}.
These results are summarized in Fig.~\ref{fig:combined.phase.diagram}.


We have also considered the implication of these results for real materials, 
concentrating on the spin ice Dy$_2$Ti$_2$O$_7$.
Based on published estimates of exchange parameters,\cite{yavorskii08} 
we find that an ordered ground state in Dy$_2$Ti$_2$O$_7$ would be a
cubic antiferromagnet (CAF).
However this state lies tantalisingly close in parameter space to other, competing ordered 
phases, and only a very small amount of quantum tunneling would be 
needed to convert it into a quantum spin liquid.  


While we have chosen to emphasize Dy$_2$Ti$_2$O$_7$, there are a great 
many rare-earth pyrochlore oxides,\cite{gardner10} in which to search for 
quantum spin ice, and other unusual forms of magnetism.\cite{savary12,lee12,yan-arXiv}  
In many of these materials, dipolar interactions will also play a role, and 
the small values of $g_c$ found in our simulations offer hope 
that quantum spin-liquids may be found in other materials at low temperature.

\section*{Acknowledgements}
\label{section:acknowledgements}


PM and OS contributed equally to this work.


The authors acknowledge helpful conversations with 
Owen Benton, Tom Fennell, and David Pomaranski, and thank 
Peter Fulde, Michel Gingras and Ludovic Jaubert for critical 
readings of the manuscript.  


This work was supported by the Okinawa Institute of Science 
and Technology Graduate University, by Hungarian OTKA Grant 
No. K106047, by EPSRC Grants No. EP/C539974/1 
and No. EP/G031460/ 1, and by the Helmholtz Virtual Institute 
``New States of Matter and their Excitations''.  
PM acknowledges an STFC Keeley-Rutherford fellowship held jointly 
with Wadham College, Oxford.
KP, PM, NS and OS and gratefully acknowledge support from the visitors 
program of MPI-PKS Dresden, where part of this work was carried out.


Since completing this work the authors have become aware of a parallel 
study of classical spin ice with long-range dipolar interactions and 
competing further-neighbour exchanges, by Henelius and 
coauthors~\cite{gingras-private-communication}.  

\begin{appendix} 

\section{Ewald summation of long-range dipolar interactions}
\label{appendix:ewald.sum}

The quantum and classical Monte Carlo simulations described in 
this Communication were carried out for cubic clusters of 
\mbox{$N = 16 \times L^3 = 128, \, 432, \, 1024, \, 2000$} spins, 
with periodic boundary conditions.
The long-range dipolar interactions ${\mathcal H}_{\sf dipolar}$ [Eq.~(\ref{eq:H.dipolar})], 
which cross the periodic boundaries of the cluster, were treated by Ewald summation.


Imposing periodic boundary conditions on a cubic cluster of dimension $L$, 
converts it into an infinitely-extended system, repeating with period $L$, 
for which the sum over long-range dipolar interactions is only 
conditionally convergent.
Within Ewald summation, this slowly converging sum, 
$U = {\mathcal H}_{\sf dipolar}/D$, is divided 
into two rapidly and absolutely convergent sums, $U^{({\rm R})}$ --- which is evaluated 
in real space, and $U^{({\rm G})}$ --- which is evaluated in reciprocal space.   
The rate of convergence of both sums is determined by a parameter 
$\alpha$, with dimension of inverse length, which determines the 
crossover between short-range interactions 
(treated in real space) and long-range interactions 
(treated in reciprocal space).
Since the system is periodic, the self-energy $U^{({\rm SE})}$ arising from a spin 
interacting with an infinite number of copies of itself must also be taken into account.
And since it is infinitely-extended, care must also be 
taken to impose an appropriate boundary condition at infinity.


Following [\onlinecite{wang01}], we impose boundary conditions  
through a macroscopic field term $U^{({\rm MF})}$, and write
\begin{equation}
{\mathcal H}_{\sf dipolar}/D =  U^{({\rm R})} + U^{({\rm G})} + U^{({\rm SE})} + U^{({\rm MF})} \, .
\end{equation}


\begin{widetext}
The sum evaluated in real space is given by 
\begin{eqnarray}
U^{({\rm R})} 
& = & \frac{1}{2} \sum_{i,j=1}^{N} \sum_{{\bf n}} '
\left\{  
\left( {\bf S}_i \cdot {\bf S}_j  \right)
  F_1( \vert {\bf R}_{ij} + {\bf n}  \vert  )  
- 
\left[{\bf S}_i \cdot  ( {\bf R}_{ij} + {\bf n} ) \right]
\left[{\bf S}_j \cdot  ( {\bf R}_{ij} + {\bf n} ) \right]
F_2( \vert {\bf R}_{ij} + {\bf n}  \vert  ) \right\} 
\; , \\
F_1(x) & = & \frac{1}{x^3} \left( {\rm erfc} \left( \alpha x\right) + \frac{2\alpha x}{\sqrt{\pi}} e^{-\alpha^2 x^2} \right) 
\;,  \\
F_2(x)  & = &  \frac{1}{x^5} \left( 3 \, {\rm erfc}\left( \alpha x\right) + \frac{2\alpha x}{\sqrt{\pi}} 
\left( 3 + 2\alpha^2 x^2 \right) e^{ -\alpha^2 x^2}  \right) \;,
\end{eqnarray}
%
%
where ${\bf n}\equiv (n_x,n_y,n_z)L$ 
with $n_a \in {\mathbb Z}$, and the prime on 
$
 \sum^\prime_{{\bf n}} 
$
indicates that the divergent terms arising for ${\bf n}=0$ and $\mathbf{R}_{ij}=0$ are omitted 
from the sum. 
The functions $F_1(x)$ and $F_2(x)$, which control the convergence of $U^{({\rm R})}$, 
are expressed in terms of the complementary error function ${\rm erfc}(z)$. 
The real space sum runs over all the periodic images of the cubic cluster of dipole moments. 


The sum to be evaluated in reciprocal space is a sum over the points 
${\bf G}\equiv (G_x,G_y,G_z)L$ (with $G_a \in {\mathbb Z}$) of the reciprocal lattice:
%
%
\begin{align}
U^{({\rm G})} = \frac{1}{2L^3} \sum_{\mathbf{G}\neq 0 } \frac{4\pi}{G^2} 
\exp\left[ -\left( \frac{\pi G}{\alpha L} \right)^2 \right] 
\sum_{i,j =1}^{N} \left( {\bf S}_i \cdot  {\bf G} \right) \left( {\bf S}_j \cdot  {\bf G} \right) 
\exp \left( \frac{ 2\pi i}{L} {\bf G}\cdot {\bf R}_{ij}  \right) \;.
\end{align}
\end{widetext}


The self-energy of spins is given by 
\begin{align}
U^{({\rm SE})} = - \frac{2\alpha^3}{3\sqrt{\pi}} \sum_{i=1}^{N}  {\bf S}^2_i \, .
\end{align}


The boundary conditions ``at infinity'' are imposed by the macroscopic 
field term
\begin{align}
U^{({\rm MF})} = \frac{2\pi}{\left( 2\epsilon + 1 \right) L^3} 
   \sum_{i=1}^{N}\sum_{j=1}^{N} {\bf S}_i \cdot {\bf S}_j \,, \label{eq:MF}
\end{align}
where the choice of boundary conditions is determined by the 
effective ``permitivity'' $\epsilon$.  


In this work we make the choice $\epsilon\rightarrow \infty$.
This is equivalent to embedding the periodic array of finite-size clusters in 
a medium which perfectly screens the net dipole moment of each cluster, 
so that the macroscopic field term $U^{({\rm MF})} \to 0$.   
The main justification for this choice of boundary condition comes
from the perfect quantitative agreement between the results of classical 
Monte Carlo simulation in the limit $T \to 0$, and the classical 
ground states determined through mapping onto an effective 
Ising model, as described in Section~\ref{section:effective.Ising.model}.
In real materials, phases with a net moment, such 
as the ferromagnet (FM), will form domains to screen the 
macroscopic field, and the effective boundary condition ``at infinity'' 
will also depend on the shape of the sample.

\section{Equivalence of exchange interactions within the spin--ice manifold}
\label{appendix:equivalence.J2.and.J3c}


\begin{figure}[t]
\includegraphics[width=0.85\columnwidth]{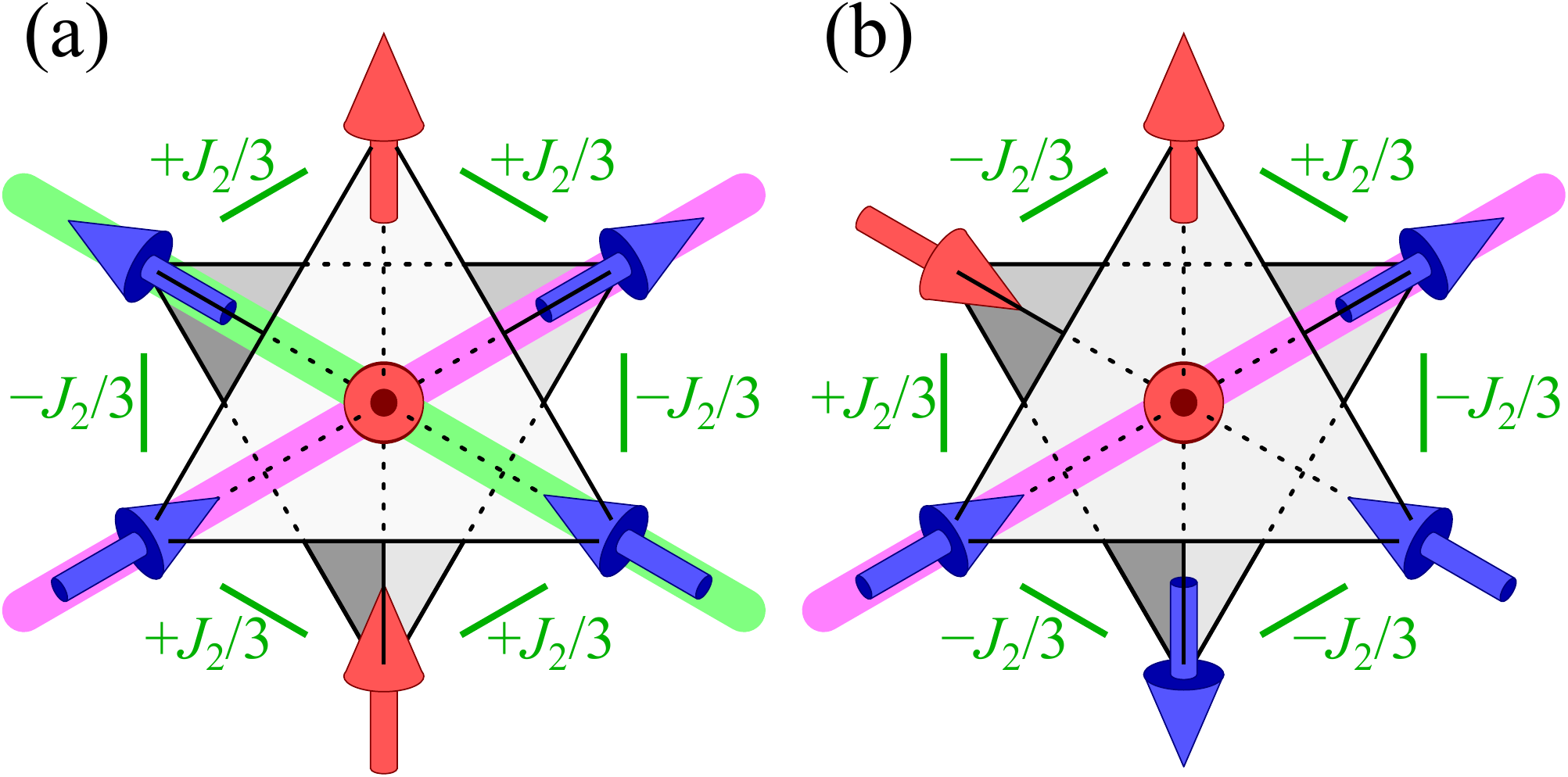}
\caption{(Color online)  
Equivalence of the exchange interactions $J_2$ and $J_{3c}$ within the 
manifold of spin-ice configurations.
All possible spin-ice states can be constructed from the two configurations 
(a) and (b), with energies $E_a$~[Eq.~(\ref{eq:Ea})] 
and $E_b$~[Eq.~(\ref{eq:Eb})].
In both cases, the energy is a function of $J_2 + 3 J_{3c}$, and the 
effect of the two exchange interactions is equivalent, up to a factor $3$.
Ferromagnetic chains of spins, which form the building-block for ordered states, 
are shown by thick magenta and green lines.
\label{fig:2tetrahedra}
}
\end{figure}


For spin--configurations obeying the ``ice rules'', a further simplification 
arises from the fact that second--neighbour exchange $J_2$, and the third--neighbour 
exchange in the direction of the $[110]$ chains, $J_{3c}$, are {\it no longer} 
independent parameters.


To understand how this works, we consider the two corner--sharing tetrahedra
shown in Fig.~\ref{fig:2tetrahedra}.  
The 2--in, 2--out  ``ice--rule'' reduces the number of possible spin-configurations
from $2^7=128$, to 18.
Each of these 18 configurations is equivalent to one of the two 
configurations shown in Fig.~\ref{fig:2tetrahedra}.
%
%
The energy of these spin configurations can be calculated by counting 
the number of satisfied and unsatisfied bonds of each type. 
Second--neighbour bonds (denoted by green lines) contribute
\begin{eqnarray}
\delta E_{J_2} = \pm\ J_2/3 \; .
\end{eqnarray}
Third--neighbour bonds, of the type  $J_{3c}$, meanwhile, contribute
\begin{eqnarray}
\delta E_{J_{3c}} = \pm J_{3c} \; .
\end{eqnarray}
Counting the relevant bonds, we find that the energies of the spin 
configurations shown in Fig.~\ref{fig:2tetrahedra}(a) 
and Fig.~\ref{fig:2tetrahedra}(b), are given by 
\begin{eqnarray}
E_{a} = \phantom{-}\frac{2}{3} J_2 + 3 J_{3c} &=& J_{3c} + \frac{2}{3} \left( J_2 + 3 J_{3c} \right) 
\label{eq:Ea} \;, \\
E_{b} = -\frac{2}{3} J_2 -J_{3c} &=& J_{3c} - \frac{2}{3} \left( J_2 + 3 J_{3c} \right) 
\label{eq:Eb}  \;.
\end{eqnarray}


Comparing the two results, we see that the interactions $J_2$ and $J_{3c}$ 
both have the same effect --- up to a factor~$\times 3$ --- when acting on any 
spin--configuration obeying the ice rules.
The constant shift $J_{3c}$, which appears in both $E_a$~[Eq.~(\ref{eq:Ea})]
and $E_b$~[Eq.~(\ref{eq:Eb})], is the same for {\it all} spin-ice configurations,
and so does not distinguish between different ordered or disordered states.


The physically relevant parameter, within a spin--ice, is therefore 
\begin{equation}
J_2 + 3 J_{3c}  \; , 
  \label{eq:j2_j3c_equivalence_appendix}
\end{equation}
as given in Section~\ref{subsection:model.parameters}.

\section{Classical Monte Carlo - technical details}
\label{section:CMC.technical}

Our classical Monte Carlo was carried out using cubic cells with periodic boundary 
conditions with $16\times L^3$ Ising spins with $L=2,3,4,5$, though for the phase 
diagram we chose cubic clusters with $128$ ($L=2$) and $1024$ ($L=4$), compatible 
with all three ordered phases. 
The long-ranged dipolar interaction was handled using a pre-tabulated Ewald summation 
(see Section~\ref{appendix:ewald.sum} and Ref.~[\onlinecite{mcclarty14}]). 
As is now standard for simulations of spin ice, the Monte Carlo allowed for single spin flips 
and worm updates.\cite{melko01}  
The worm updates allow for efficient sampling of spin-ice states with a short autocorrelation 
time compared to simulation time scales.  
We simulated up to $128$ temperatures simultaneously on the hydra cluster based 
in Garching with parallel tempering moves to assist equilibration.   
The highest temperature was taken below the heat capacity peak into the ice states.
The simulations for $L=4$ at low temperature were somewhat hampered by slow 
equilibration despite the presence of loop moves and parallel tempering.  
Whereas $L=2$ simulations were found to be independent of the starting configuration, 
this ceased to be the case for $L=4$.  
We therefore conducted simulations by starting from each of the three known ordered states 
and also from states that are degenerate at the phase boundaries --- for example the 
orthorhombic zigzag state (OZZ).  

\section{Quantum Monte Carlo - technical details}
\label{section:QMC-technical}

We have performed Green's function Monte Carlo (GFMC) simulations of 
$\mathcal{H}_{\sf QDSI}$~[\ref{eq:Hqdsi}], using methods previously developed 
to study the quantum dimer model on a diamond lattice,\cite{sikora09,sikora11}
and quantum spin ice in the absence of long-range dipolar 
interactions.~\cite{shannon12,benton12} 
GFMC is a form of zero-temperature Quantum Monte Carlo simulation, 
which is numerically exact where simulations converge.
%


Our implementation of GFMC closely parallels that of [\onlinecite{calandra-buonaura98}].
We work explicitly with spin--ice configurations and, starting from a given spin configuration,
use a population of ``walkers'' to  sample the space of other configurations connected by 
off-diagonal matrix elements of the Hamiltonian, $\mathcal{H}_{\sf QDSI}$~[\ref{eq:Hqdsi}].  
A guide wave function, optimised by a separate variational Monte Carlo simulation
is used to improve the convergence of simulations.
As such, GFMC can be thought of as a systematic way of improving upon 
a variational wave function.
A suitable variational wave function for a quantum spin ice, based on 
plaquette-plaquette correlations, is described in [\onlinecite{sikora11}].   
The number of variational parameters used in simulations, depended on the 
cluster, and was typically 20-40.
Populations of up to 1000 walkers were used in GFMC simulation.
The population of walkers was reconfigured after a typical period of 
45 steps, with simulations run for a few thousand consecutive 
reconfigurations.
The averages used in estimators for the ground state energy, etc., 
were calculated for sequences of 50-300 steps.


 We performed GFMC simulations for clusters of 
128, 1024, and 2000 sites, with the full cubic symmetry of the pyrochlore lattice.
Since not all of the ordered states considered are compatible with the 
2000-site cluster, this was used to explore the correlations $S(\mathbf{q})$ 
of the QSL phase, and not to determine the ground-state phase diagram.
To test the accuracy of the method, simulations of were also performed for 
a 80-site cluster with lower symmetry.
Exact diagonalization calculations were carried out for the same 80-site cluster,
and found to be in perfect numerical agreement with the results of GFMC.


Simulations  for ``large'' values of $g \gtrsim 0.1 D$, within the QSL, 
are relatively easy to converge, since all spin-ice configurations, apart for a 
tiny subset of ``isolated states'', are connected by matrix elements of 
$\mathcal{H}_{\sf QDSI}$~[\ref{eq:Hqdsi}], and all spin-ice configurations 
enter into the QSL ground state with comparable weight.
Simulations are relatively difficult to converge for large clusters and 
``small'' values of $g$, especially in the highly frustrated region 
$-0.08 \lesssim J_2/D \lesssim -0.06$, where the coupling between 
parallel ``chains'' is vanishingly small and many different  ground states compete.
Detail of this region of the phase diagram is given in 
Fig.~(\ref{fig:detail.of.quantum.phase.diagram}).


The Hilbert space of different possible spin--ice configurations, on which 
$\mathcal{H}_{\sf QDSI}$~[\ref{eq:Hqdsi}] acts, can be divided into distinct 
topological sectors, according to the net flux of spin moments 
through the boundaries of a cluster~\cite{shannon12,sikora11}.
Under the dynamics described by $\mathcal{H}_{\sf QDSI}$, these 
fluxes are conserved.
The QSL, and the CAF, TDQ and OZZ ground state all belong
to the zero--flux sector, while the FM has a finite value of flux.
We have GFMC performed simulations in a representative selection 
of flux sectors, and find no evidence of other competing ground 
states with finite values of flux.
We have also verified that the energies of the QSL in different flux sectors 
satisfies the expected scaling with flux at fixed system size, as 
described in Ref.~\onlinecite{shannon12}.
 

\begin{table}[t]
\caption{ 
Ground state energy $E_0$, excitation gap $\Delta$
and number of flippable plaquettes $N_{\text{flip}}$ 
for cubic clusters with $N=128$ and $N=1024$ sites, 
used in constructing the degenerate perturbation theory 
Eq.~(\ref{eq:degenerate-perturbation-theory}).
}
\centering
\begin{ruledtabular}
\begin{tabular}{c r c c r}
state & $N$ & $E_0/N$ & $\Delta_0$ & $N_{\text{flip}}$ \\ [0.5ex] 
\hline\vspace{-0.25cm}\\
CAF & 128 & $-1.94759 D - 2 J_2/3$  &  & $0$ \\
CAF & 1024 & $-1.94760 D - 2 J_2/3$ &  & $0$ \\
OZZ & 128 & $-1.92688 D - J_2/3$ & $0.2906 D + 8 J_2/3$ & 32\\
OZZ & 1024 & $-1.92687 D - J_2/3$ & $0.2919 D + 8 J_2/3$ & 256\\
TDQ & 128 & $-1.90617 D $ & $-0.3725 D - 8 J_2$ & 32\\
TDQ & 1024 & $-1.90613 D$ & $-0.3717 D - 8J_2$ & 256\\
\end{tabular}
\end{ruledtabular}
\label{table:pert2nd}
\end{table}

\section{2$^\text{nd}$ order perturbation theory in $g$}
\label{appendix:2nd.order.perturbation.theory.in.g}


We can use perturbation theory in $g$ 
to calculate the effect of the quantum fluctuations about the TDQ and OZZ ground states.  
To second order in $g$, the ground state energy is given by
\begin{equation}
 E^{(2)} = E^{(0)} - N_{\text{flip}} \frac{g^2}{\Delta_0}\,,
\label{eq:degenerate-perturbation-theory} 
\end{equation}
where $E^{(0)}$ is the classical ground state energy and $\Delta_0$ is the energy gap 
between the ground state and the excited state obtained by flipping the spins on a 
hexagon (where $N_{\text{flip}}$ is the number of such hexagons, and all the flippable hexagons are equivalent).   
 These numbers, found by the numerical enumeration of states, are presented for 
 the 128 and 1024 site cluster in Table~\ref{table:pert2nd}.


Comparing these energies close to the classical phase boundary where the TDQ, 
the OZZ, and the CAF are degenerate, we get that OZZ state has the lowest energy 
and is stabilized between the TDQ and CAF phases.  
The phase transition lines between the TDQ and OZZ phases are essentially 
independent of $g$:
\begin{align}
J_2/D &= -0.0621 \quad \text{(128 sites)}\,,\label{eq:pertbond128TDQ}\\ 
J_2/D &= -0.0622 \quad \text{(1024 sites)}\,.
\end{align}
In contrast, the phase boundaries between the CAF and OZZ depend on $g/D$ as
\begin{align}
 J_2/D &= -0.0621  + 6.01 (g/D)^2 \quad \text{(128 sites)}\,,\label{eq:pertbond128OZZ}\\
 J_2/D &= -0.0622  + 5.95 (g/D)^2 \quad \text{(1024 sites)}\,.
\end{align}
These phase boundaries are shown in Fig.~\ref{fig:phase.diagram.QMC} 
and Fig.~\ref{fig:detail.of.quantum.phase.diagram} as dashed lines 
(the finite--size effects are not discernible on the scale of the figure).

\end{appendix}




\begin{thebibliography}{99}

%
\bibitem{fazekas74}
P. Fazekas and P. W. Anderson, 
Phil. Mag. {\bf 30}, 423 (1974).

%
\bibitem{lee08}
Patrick A. Lee
Science {\bf 321}, 1306 (2008).

%
\bibitem{balents10}
Leon Balents, 
Nature, {\bf 464} 199 (2010).

%
\bibitem{bramwell01-Science294} 
S. T. Bramwell and M. J. P. Gingras, 
Science {\bf 294}, 1495 (2001).

%
\bibitem{castelnovo12}
C. Castelnovo, R. Moessner, and S. L. Sondhi, 
Annu. Rev. Condens. Matter Phys. {\bf 3}, 35-55 (2012).

%
\bibitem{powell11}
Stephen Powell, 
Phys. Rev. B {\bf 84}, 094437 (2011).

%
\bibitem{pomaranski13}
D. Pomaranski, L. R. Yaraskavitch, S. Meng, K. A. Ross, H. M. L. Noad, 
H. A. Dabkowska, B. D. Gaulin and J. B. Kycia, 
Nature Physics {\bf 9}, 353 (2013).

%
\bibitem{ramirez99}
A. P. Ramirez, A. Hayashi,  R. J. Cava, R. Siddharthan and B. S. Shastry, 
Nature {\bf 399}, 333 (1999).

%
%
\bibitem{klemke11}
B. Klemke, M. Meissner, P. Strehlow, K. Kiefer, S. A. Grigera and D. A. Tennant,  
J. Low Temp. Phys. {\bf 163}, 345 (2011).

%
\bibitem{moessner03}
R. Moessner and S. Sondhi,
Phys. Rev. B {\bf 68}, 184512 (2003)

%
\bibitem{hermele04} 
M.~Hermele, M.P.A.~Fisher, and L.~Balents, 
Phys. Rev. B {\bf 69}, 064404 (2004).

%
\bibitem{banerjee08} 
A. Banerjee, S. V. Isakov, K. Damle and Y. B. Kim,
Phys. Rev. Lett. {\bf 100}, 047208 (2008).

%
\bibitem{savary12}
L.~Savary and L.~Balents.
Phys. Rev. Lett. {\bf 108}, 037202, (2012).

%
\bibitem{shannon12}
N. Shannon, O. Sikora, F. Pollmann, K. Penc and P. Fulde, 
Phys. Rev. Lett. {\bf 108}, 067204 (2012).

%
\bibitem{benton12}
O. Benton, O Sikora and N. Shannon, 
Phys. Rev. B. {\bf 86}, 075154 (2012).

\bibitem{reimers91}
J. N. Reimers, A. J. Berlinsky, and A.-C. Shi,
Phys. Rev. B {\bf 43}, 865 (1991).

%
\bibitem{lee12}
S.-B. Lee, S. Onoda and L.~Balents, 
Phys. Rev. B {\bf 86}, 104412 (2012).

%
\bibitem{savary13}
L. Savary and L. Balents, 
Phys. Rev. B {\bf 87}, 205130 (2013).

%
\bibitem{gingras14}
M J P Gingras and P A McClarty,
Rep. Prog. Phys. {\bf 77}, 056501 (2014).

%
\bibitem{hao-arXiv}
Z-H, Hao, A. G. R. Day and M. J. P. Gingras
Phys. Rev. B {\bf 90}, 214430 (2014).

%
\bibitem{kato-arXiv}
Y. Kato and S. Onoda, 
arXiv:1411.1918

%
\bibitem{thompson11}
J.~D. Thompson, P.~A. McClarty, H.~M. Ronnow, L.~P. Regnault,  A. Sorge, and M.~J.~P. Gingras,
Phys. Rev. Lett. {\bf 106},  187202  (2011).

%
\bibitem{ross11-PRX1}
K. A. Ross, L. Savary, B. D. Gaulin and L. Balents, 
Phys. Rev. X {\bf 1}, 021002 (2011).

%
\bibitem{chang12}
L.~J. Chang, S. Onoda, Y. Su, Y.-J. Kao, K.-D. Tsuei, Y. Yasui, K. Kakurai and M.~R. Lees, 
Nature Commun. {\bf 3},  992  (2012).

%
\bibitem{molovian07}
H. R. Molavian, M. J. P. Gingras and Benjamin Canals, 
Phys. Rev. Lett. {\bf 98}, 157204 (2007)

%
\bibitem{fennell12}
T. Fennell, M. Kenzelmann, B. Roessli, M. K. Haas and R. J. Cava, 
Phys. Rev. Lett. {\bf 109}, 017201 (2012).

%
\bibitem{fennell14}
T. Fennell, M. Kenzelmann, B. Roessli, H. Mutka, J. Ollivier, 
M. Ruminy, U. Stuhr, O. Zaharko, L. Bovo, A. Cervellino, 
M. K. Haas and R. J. Cava, 
Phys. Rev. Lett. {\bf 112}, 017203 (2014).

%
\bibitem{kimura13}
K. Kimura, S. Nakatsuji, J.-J. Wen, C. Broholm, M.~B. Stone, E. Nishibori and H. Sawa,  
Nature Commun. {\bf 4},  1934  (2013).

%
\bibitem{yavorskii08}
Taras Yavors'kii, Tom Fennell, Michel J. P. Gingras and Steven T. Bramwell,
Phys. Rev. Lett. {\bf 101}, 037204 (2008).

%
\bibitem{siddharthan99} 
R. Siddharthan, B. S. Shastry, A. P. Ramirez, A. Hayashi, R. J. Cava and S. Rosenkranz,
Phys. Rev. Lett. {\bf 83}, 1854 (1999).

%
\bibitem{siddharthan-arXiv} 
R. Siddharthan, B. S. Shastry and A. P. Ramirez,
arXiv:cond-mat/0009265

%
\bibitem{denHertog00}
B. C. den Hertog and M. J. P. Gingras, 
Phys. Rev. Lett. {\bf 84}, 3430 (2000).

%
\bibitem{bramwell01-PRL87} 
%
S. T. Bramwell, M. J. Harris, B. C. den Hertog, M. J. P. Gingras, 
J. S. Gardner, D. F. McMorrow, A. R. Wildes, A. L. Cornelius, 
J. D. M. Champion, R. G. Melko and T. Fennell, 
Phys. Rev. Lett {\bf 87}, 047205 (2001).

%
\bibitem{melko01} 
R. G. Melko, B. C. den Hertog, and M. J. P. Gingras, 
Phys. Rev. Lett. {\bf 87} 067203 (2001).

%
\bibitem{supplemental}
See Supplemental Material at [URL will be inserted by publisher] 
for animated images of ordered states.

%
\bibitem{hiroi03}
Z. Hiroi, K. Matsuhira and M. Ogata, 
J. Phys. Soc. Jpn. {\bf 72}, 3045 (2003)

%
\bibitem{dublenych13}
Y. I. Dublenych, 
J. Phys.: Condens. Matter {\bf 25}, 406003 (2013).

%
\bibitem{harris97}
M. J. Harris {\it et al.}, 
Phys. Rev. Lett. {\bf 79}, 2554 (1997).

%
\bibitem{mcclarty14}
P. A. McClarty, A. O'Brien, and F. Pollmann, 
Phys. Rev. B {\bf 89}, 195123 (2014).

%
\bibitem{fennell09}
%
T. Fennell, P. P. Deen, A. R. Wildes, K. Schmalzl, D. Prabhakaran,
A. T. Boothroyd, R. J. Aldus, D. F. McMorrow, and S. T. Bramwell, 
Science {\bf 326}, 415 (2009).


\bibitem{morris09}
D. J. P. Morris, {\it et al.}, 
Science {\bf 326}, 411 (2009).

%
\bibitem{sikora09}  
O. Sikora, F. Pollmann, N. Shannon, K. Penc and P. Fulde, 
Phys. Rev. Lett. {\bf 103}, 247001 (2009).

%
\bibitem{sikora11}  
O. Sikora, N. Shannon, F. Pollmann, K. Penc and P. Fulde, 
Phys. Rev. B {\bf 84}, 115129 (2011).

%
\bibitem{bak82}
P. Bak,
Rep. Prog. Phys. {\bf 45},  587 (1982).

%
\bibitem{selke88}
W. Selke
Physics Reports {\bf 170}, 213 (1988).

%
\bibitem{zhou12}
H. D. Zhou, J. G. Cheng,  A. M. Hallas, C. R. Wiebe, G. Li, L. Balicas, 
J. S. Zhou, J. B. Goodenough, J. S. Gardner and E. S. Choi, 
Phys. Rev. Lett. {\bf 108}, 207206 (2012).

%
\bibitem{gardner10}
J. S. Gardner, M. J. P. Gingras and J. E. Greedan,
Rev. Mod. Phys. {\bf 82}, 53 (2010).

%
\bibitem{yan-arXiv}
H. Yan, O. Benton, L.~D.~C. Jaubert, and N. Shannon, 
arXiv:1311.3501

%
\bibitem{isakov04}
S.V. Isakov, K. Gregor, R. Moessner and S. L. Sondhi,
Phys. Rev. Lett {\bf 93}, 167204, (2004).

%
\bibitem{henley05}
C. L. Henley,
Phys. Rev. B {\bf 71}, 014424, (2005).

%
\bibitem{henley10}
C. L. Henley,
Annu. Rev. Condens. Matter Phys. {\bf 1}, 179, (2010).

%
\bibitem{enjalran04}
M. Enjalran and M. J. P. Gingras, 
Phys. Rev. B {\bf 70}, 174426 (2004).

%
\bibitem{savary12-PRL109}
Lucile Savary, Kate A. Ross, Bruce D. Gaulin, Jacob P. C. Ruff, and Leon Balents, 
Phys. Rev. Lett. {\bf 109}, 167201 (2012).

%
\bibitem{savary12-PRL108}
L.~Savary and L.~Balents.
Phys. Rev. Lett. {\bf 108}, 037202, (2012).

%
\bibitem{isakov05}
S. V. Isakov, R. Moessner, and S. L. Sondhi, 
Phys. Rev. Lett. {\bf 95}, 217201 (2005).

%
\bibitem{onoda11}
S. Onoda and Y. Tanaka,
Phys. Rev. B {\bf 83}, 094411 (2011)

\bibitem{luttinger46}  
J. M. Luttinger and L. Tisza, Phys. Rev. {\bf 70}, 954 (1946).

%
\bibitem{Wannier1950}
G. H. Wannier, 
Phys. Rev. {\bf 79}, 357 (1950).

%
\bibitem{anderson56} 
P. W.~Anderson, 
Phys. Rev. {\bf 102}, 1008 (1956).

%
\bibitem{castelnovo08}
C. Castelnovo, R. Moessner, and S. L. Sondhi, 
Nature {\bf 451}, 42 (2008).

\bibitem{gingras01}
M. J. P. Gingras and B. C. den Hertog, 
Can. J. Phys. {\bf 79}, 1339 (2001).

\bibitem{melko04}
R. G. Melko and M. J. P. Gingras, 
J. Phys. Condens. Matter {\bf 16},  R1277 (2004).

%
\bibitem{fukazawa02}
H. Fukazawa, R. G. Melko, R. Higashinaka, Y. Maeno, and M. J. P. Gingras,
Phys. Rev. B {\bf 65}, 054410 (2002)

%
\bibitem{calandra-buonaura98}
M. Calandra Buonaura and S. Sorella, 
Phys. Rev. B {\bf 57}, 11446 (1998).

\bibitem{wang01}
Z. Wang and C. Holm,
J. Chem. Phys. {\bf 115} 6277 (2001).

%
\bibitem{pauling35}
L. Pauling, 
J. Am. Chem. Soc. {\bf 57}, 2680 (1935)

%
\bibitem{higashinaka02}
R. Higashinakaa, H. Fukazawaa, D. Yanagishimaa and Y. Maeno, 
J. Chem. Phys. Solids {\bf 63}, 1043 (2002).

\bibitem{curnoe08}
S. Curnoe, Phys. Rev. B {\bf 78}, 094418 (2008).

%
\bibitem{iwahara15}
N. Iwahara and L. F. Chibotaru, 
Phys. Rev. B {\bf 91}, 174438 (2015).

%
\bibitem{tomasello-arXiv}
B. Tomasello, C. Castelnovo, R. Moessner and J. Quintanilla,
arXiv:1506.02672.

%
\bibitem{rau-arXiv}
J. G. Rau and  M. J. P. Gingras,
arXiv:1503.04808.

%
\bibitem{gingras-private-communication}
Michel Gingras, {\it private communication}.

\end{thebibliography}
\end{document}